\PassOptionsToPackage{dvipsnames}{xcolor}
\PassOptionsToPackage{oneside}{geometry}
\documentclass[twocolumn, twocolappendix, oneside]{aastex631}
\received{\today}
\shorttitle{}
\graphicspath{{figures/}}

\usepackage{tabularx}
\usepackage{lipsum}
\usepackage{physics}
\usepackage{multirow}
\usepackage{xspace}
\usepackage{natbib}
\usepackage{fontawesome5}
\usepackage{wrapfig}
\usepackage[figuresright]{rotating}
\usepackage{makecell}

\usepackage{rotating}

\usepackage[hang,flushmargin]{footmisc} 

\usepackage{styles/todo_notes, styles/abbreviations}

\makeatletter
\newcommand{\unit}[1]{%
    \,\mathrm{#1}\checknextarg}
\newcommand{\checknextarg}{\@ifnextchar\bgroup{\gobblenextarg}{}}
\newcommand{\gobblenextarg}[1]{\,\mathrm{#1}\@ifnextchar\bgroup{\gobblenextarg}{}}
\makeatother

\usetikzlibrary{shapes.geometric, arrows}

\tikzstyle{startstop} = [rectangle, rounded corners, minimum width=1.5cm, minimum height=1cm, text centered, draw=black, fill=gray!30]
\tikzstyle{process} = [rectangle, minimum width=2cm, minimum height=1cm, text centered, draw=black, fill=teal!30]
\tikzstyle{decision} = [diamond, aspect=1.5, minimum width=1cm, minimum height=1cm, text centered, draw=black, fill=BrickRed!30]
\tikzstyle{arrow} = [thick,->,>=stealth]

\newif\ifstartedinmathmode
\newcommand{\msun}{%
  \relax\ifmmode\startedinmathmodetrue\else\startedinmathmodefalse\fi
  {\ifstartedinmathmode\unit{M_{\odot}}\else$\unit{M_{\odot}}$\fi}\xspace%
}

\newif\ifstartedinmathmode
\newcommand{\rsun}{%
  \relax\ifmmode\startedinmathmodetrue\else\startedinmathmodefalse\fi
  {\ifstartedinmathmode\unit{R_{\odot}}\else$\unit{R_{\odot}}$\fi}\xspace%
}

\definecolor{dark}{rgb}{0.2,0.2,0.2}

\defcitealias{cogsworth}{Paper~I}
\defcitealias{DeDonder+2003:2003NewA....8..817D}{DD03}
\defcitealias{Zapartas+2017:2017AA...601A..29Z}{Z17}
\defcitealias{Eldridge+2011:2011MNRAS.414.3501E}{E11}
\defcitealias{Renzo+2019:2019A&A...624A..66R}{R19}

\makeatletter
\renewcommand\l@subsubsection[2]{}
\makeatother

\begin{document}


\title{Delayed and Displaced: The Impact of Binary Interactions on Core-collapse SN Feedback}

\newcommand{\UW}{\affiliation{Department of Astronomy, University of Washington, Seattle, WA, 98195}}
\newcommand{\cca}{\affiliation{Center for Computational Astrophysics, Flatiron Institute, 162 Fifth Ave, New York, NY, 10010, USA}}
\newcommand{\cmu}{\affiliation{McWilliams Center for Cosmology and Astrophysics, Department of Physics, Carnegie Mellon University, Pittsburgh, PA 15213, USA}}
\newcommand{\arizona}{\affiliation{University of Arizona, Department of Astronomy \& Steward Observatory, 933 N. Cherry Ave., Tucson, AZ 85721, USA}}
\newcommand{\princeton}{\affiliation{Department of Physics, Princeton University, Princeton, NJ 08544, USA}}
\newcommand{\rutgers}{\affiliation{Department of Physics and Astronomy, Rutgers, The State University of New Jersey, 136 Frelinghuysen Rd, Piscataway, NJ 08854, USA}}

\newcommand{\mpa}{\affiliation{Max-Planck-Institut f\"ur Astrophysik, Karl-Schwarzschild-Straße 1, 85741 Garching, Germany}}

\author[0000-0001-6147-5761]{Tom Wagg}
\UW{}
\cca{}

\author[0000-0002-1264-2006]{Julianne J.\ Dalcanton}
\cca{}
\UW{}

\author[0000-0002-6718-9472]{Mathieu Renzo}
\arizona{}

\author[0000-0001-5228-6598]{Katelyn Breivik}
\cmu{}

\author[0000-0003-1053-3081]{Matthew E. Orr}
\cca{}

\author[0000-0003-0872-7098]{Adrian~M.~Price-Whelan}
\cca{}

\author[0000-0001-7831-4892]{Akaxia Cruz}
\cca{}
\princeton{}

\author[0000-0002-0372-3736]{Alyson Brooks}
\rutgers{}
\cca{}

\author[0000-0001-8867-5026]{Ulrich P. Steinwandel}
\cca{}
\mpa{}

\author[0000-0001-8018-5348]{Eric C. Bellm}
\UW{}

\correspondingauthor{Tom Wagg}
\email{tomjwagg@gmail.com}

\begin{abstract}
    Core-collapse supernova feedback models in hydrodynamical simulations typically assume that all stars evolve as single stars. However, the majority of massive stars are formed in binaries and multiple systems, where interactions with a companion can affect stars' subsequent evolution and kinematics. We assess the impact of binary interactions on the timing and spatial distribution of core-collapse supernovae, using \cogsworth simulations to evolve binary star populations, and their subsequent galactic orbits, within state-of-the-art hydrodynamical zoom-in galaxy simulations.
    We show that binary interactions: (a) displace supernovae, 
    with ${\sim}13\%$ of all supernovae occurring more than $0.1\unit{kpc}$ from their parent cluster; and (b) produce delayed supernovae, such that ${\sim}25\%$ of all supernovae occur after the final supernova from a single star population.
    Delays are largest for low-mass merger products, which can explode more than 200 Myr after a star formation event. 
    We characterize our results as a function of: (1) initial binary population distributions, (2) binary physics parameters and evolutionary pathways, (3) birth cluster dissolution assumptions, and (4) galaxy models (which vary metallicity, star formation history, gravitational potential and simulation codes), and show that the overall timing and spatial distributions of supernovae are surprisingly insensitive to most of these variations.
    We provide metallicity-dependent analytic fits that can be substituted for single-star subgrid feedback prescriptions in hydrodynamical simulations, and discuss some of the possible implications for binary-driven feedback in galaxies, which may become particularly important at high redshift.
\end{abstract}



\section{Introduction}

Stellar feedback from massive stars plays a critical role in galaxy formation and evolution. Without this feedback, simulated galaxies tend to form too many stars and become too dense compared to observed galaxies \citep[e.g.,][]{Katz+1996:1996ApJS..105...19K, Hopkins+2018:2018MNRAS.477.1578H}. While several aspects of stellar evolution are likely to contribute to the regulation of star formation, supernovae (SNe) from massive stars are undoubtedly one of the most important \citep[e.g.,][]{Dekel+1986:1986ApJ...303...39D,Hopkins+2012:2012MNRAS.421.3522H,Somerville+2015:2015ARA&A..53...51S,Naab+2017:2017ARA&A..55...59N}. 

In the broadest picture of supernova (SN) feedback, the energy of the SN couples to the surrounding gas, driving it to lower densities and higher temperatures that are unfavourable for forming stars. However, the actual process is far more nuanced, with a large body of work clearly demonstrating that the timing and location of each SN has a significant impact on the efficacy of its feedback \citep[e.g.,][]{Walch+2015:2015MNRAS.454..238W,Girichidis+2016:2016MNRAS.456.3432G,Hu+2016:2016MNRAS.458.3528H,Hu+2017:2017MNRAS.471.2151H,Hu+2019:2019MNRAS.487.3252H,Smith+2021:2021MNRAS.506.3882S, Orr+2022:2022ApJ...932...88O}. In addition to a strong dependence on SN clustering, which affects how ``collectively'' SNe can impact the surrounding gas, much depends on the density of the surrounding gas where the SN explodes. SNe that occur near their birthplace are likely to directly disrupt their natal molecular clouds, rapidly shutting down star formation. In contrast, SNe that happen outside molecular clouds may have less direct effect on the efficiency within clouds, but have a greater chance of driving outflows \citep[e.g.,][]{Ceverino+2009:2009ApJ...695..292C,Ceverino+2014:2014MNRAS.442.1545C,Zolotov+2015:2015MNRAS.450.2327Z,Hu+2017:2017MNRAS.471.2151H,Andersson+2020:2020MNRAS.494.3328A,Steinwandel+2023:2023MNRAS.526.1408S}. At the extremes, massive runaway stars can be ejected far out of the plane of the galaxy \citep[e.g.,][]{Renzo+2019:2019A&A...624A..66R}, which are much more effective at boosting the momentum outflow rate of galaxies than SNe which are dispersed within the disc, since vertically displaced SN can more effectively couple to the hot phase of pre-existing wind outflows \citep{Steinwandel+2023:2023MNRAS.526.1408S}.

Simulations have additionally already demonstrated the importance of the timing of SNe for their relative impact on a galaxy \citep{Struck-Marcell+1987:1987ApJS...64...39S, Parravano+1996:1996ApJ...462..594P, Quillen+2008:2008MNRAS.386.2227Q}. In particular, more realistic stellar lifetime distributions, as opposed to fixed values for all stars, can prevent the formation of high density gas clouds and allow more feedback to occur in low density environments \citep{Kimm+2015:2015MNRAS.451.2900K}, as well as reduce the overall clustering of SNe \citep{Smith+2021:2021MNRAS.506.3882S, Hu+2023:2023ApJ...950..132H}.

One important mechanism for setting the timing and spatial distribution of SNe are interactions during binary stellar evolution. The vast majority of massive stars that reach core collapse and produce SNe are born in binaries and multiple systems \citep[e.g.,][]{Mason+2009, Almeida+2017, Moe+2017, Offner+2023:2023ASPC..534..275O}. A large subset of these stars will interact with their companion in their lifetime \citep[e.g,][]{Sana+2012:2012Sci...337..444S,deMink+2014}. 

The interactions during binary evolution can have a number of different impacts on the timing of core-collapse SNe. For example, adding or removing mass during mass transfer alters a star's evolutionary timescale and thus its time until core collapse, since more massive stars evolve more quickly. In the case of stellar mergers, two low mass stars that merge will reach core collapse much later than single stars of the same mass \citep{DeDonder+2003:2003NewA....8..817D,Zapartas+2017:2017AA...601A..29Z}. Rejuvenation of accretor stars can also have strong impacts on the timing of SNe, such that additional hydrogen is mixed into their cores, potentially altering their time to core collapse by ${\sim}$Myrs \citep[e.g.,][]{Neo+1977:1977PASJ...29..249N, Schneider+2016:2016MNRAS.457.2355S, Renzo+2023}, though this effect may be lower for different assumptions regarding convective boundary mixing \citep[e.g.,][]{Braun+1995}.

Binary interactions can also displace massive stars from their birth sites. First, the above changes in SN timing has a first-order effect on how close a SNe is to its birth place, where SNe delayed as a result of binary evolution will spend a longer time dispersing from their parent cluster \citep[e.g.,][]{Aghakhanloo+2017:2017MNRAS.472..591A}. Second, binary evolution can directly change the kinematics of individual stars. Secondary stars that were formed in binaries that are disrupted by a primary SN may travel large distances as runaway stars before a SN occurs \citep[e.g.,][]{Blaauw+1961,Boersma+1961,Renzo+2019:2019A&A...624A..66R}.

Despite the fact that binary interactions are likely important for the timing and locations of massive star SNe, most hydrodynamical zoom-in simulations currently implicitly assume all stars evolve without a companion \citep[e.g.,][]{Leitherer+1999:1999ApJS..123....3L,Leitherer+2014:2014ApJS..212...14L,Hopkins+2018:2018MNRAS.477.1578H,Hopkins+2023:2023MNRAS.519.3154H,Applebaum+2021:2021ApJ...906...96A,Smith+2021:2021MNRAS.506.3882S,Christensen+2023:2023arXiv231104975C}. As such, late SN feedback is not accounted for and the feedback is always assumed to occur at the position of the star particle.

While the timing of SNe and the ejection velocities of their progenitors can be addressed entirely within the context of binary evolution models, understanding the subsequent impact of the resulting SNe requires embedding the evolving binaries within realistic galactic potentials. Earlier work has established the likelihood that runaway stars ejected from binaries may end their lives far from their birth locations \citep[e.g.,][]{Eldridge+2011:2011MNRAS.414.3501E, Renzo+2019:2019A&A...624A..66R}, but without considering the restraining forces provided by a galactic potential. Similarly, previous studies have considered the impact of binary physics on the timing of SNe and the delays that are possible \citep{DeDonder+2003:2003NewA....8..817D, Zapartas+2017:2017AA...601A..29Z}.

We build upon these works by leveraging the capabilities of the new open-source code \cogsworth \citep{Wagg+2025:2025JOSS...10.7400W, Wagg+2025:2025ApJS..276...16W}, which provides a framework for performing self-consistent population synthesis and galactic dynamics simulations. In this way, one can evolve binary stars within their galactic context, integrating their orbits through the galaxy while accounting for the effects of SN natal kicks on a star's galactic trajectory. This enables us to track the precise time and galactic location of each SN in a galaxy. \cogsworth also takes advantage of the advances in numerical galaxy simulations to draw realistic populations in age, position and local potential.



This paper is structured as follows. In Section~\ref{sec:methods}, we outline our methods for simulating populations of SNe with \cogsworth and the settings chosen for our fiducial model. We demonstrate and explain the impact of binary interactions on SN location and timing in Section~\ref{sec:results}. We explore how our results depend on our choice of initial conditions, binary physics and galaxy parameters in Section~\ref{sec:variations}. In Section~\ref{sec:fits}, we present analytic fits to our distributions for the timing and distance of SNe. We discuss the implications and limitations of our work in Section~\ref{sec:discussion} and draw conclusions in Section~\ref{sec:conclusions}.  All code for producing our simulations, as well as the simulations themselves, is available on \href{https://github.com/TomWagg/supernova-feedback/}{GitHub}\footnote{\url{https://github.com/TomWagg/supernova-feedback/}} and \href{https://doi.org/10.5281/zenodo.15273993}{Zenodo}\footnote{\url{https://doi.org/10.5281/zenodo.15273993}}.

\section{Simulating binary stellar populations with \cogsworth}\label{sec:methods}

\begin{table*}
    \centering
    \begin{tabular}{l|ccc|>{\scriptsize}c}
        \hline\hline
        Parameter & Symbol & Fiducial model & Variations & References \\
        \hline
        \textbf{Initial conditions} &&&&\\
        \quad Initial mass function slope & $\alpha_{\rm IMF}$ & $-2.3$ & $[-1.9, -2.7]$ & \citet{Kroupa+2001:2001MNRAS.322..231K, Schneider+2018:2018Sci...359...69S, Schneider+2018:2018AA...618A..73S} \\
        \quad Mass ratio slope & $\kappa$ & 0 & [-1, 1] & \citet{Mazeh+1992:1992ApJ...401..265M,Sana+2012:2012Sci...337..444S} \\
        \quad Orbital period slope & $\pi$ & -0.55 & [-1, 0] & \citet{Sana+2012:2012Sci...337..444S,deMink+2015:2015ApJ...814...58D}\\
        \quad Initial orbital period limit & $P_{\rm 0, max}$ & $10^{5.5}\unit{d}$ & $10^3\unit{d}$ & \citet{deMink+2015:2015ApJ...814...58D} \\
        \quad Eccentricity slope & $\eta$ & -0.45 & - & \citet{Sana+2012:2012Sci...337..444S} \\
        \quad Binary fraction & $f_{\rm bin}$ & 1.0 & 0.0 & \\
        \quad Metallicity & $\bar{Z}$ & $\bar{Z}_{\rm m11h} \approx 1.2 Z_{\odot}$ & $[0.5, 0.2, 0.1, 0.05] \bar{Z}_{\rm m11h}$ & \citet{Wetzel2023} \\
        \textbf{Binary physics} &&&& \\
        \quad Mass transfer efficiency & $\beta$ & $\beta \propto 10 \tau_{\rm th, acc}$\tablenotemark{a} & [0, 0.5, 1] & \citet{Schneider+2015:2015ApJ...805...20S}\\
        \quad Case B critical mass ratio & $q_{\rm crit, B}$ & HW1987 & $[0, \infty]$ & \citet{Hjellming+1987:1987ApJ...318..794H}\\
        \quad Common-envelope efficiency & $\alpha_{\rm CE}$ & 1.0 & [0.1, 10.0] & \citet{Webbink+1984:1984ApJ...277..355W, deKool+1990:1990ApJ...358..189D}\\
        \quad CCSN natal kicks & $\sigma$ & $265 \unit{km}{s^{-1}}$ & $20 \unit{km}{s^{-1}}$ & \citet{Hobbs+2005:2005MNRAS.360..974H, Igoshev+2020:2020MNRAS.494.3663I}\\
        \quad ECSN/USSN natal kicks  & $\sigma_{\rm low}$ & $20 \unit{km}{s^{-1}}$ & $265 \unit{km}{s^{-1}}$ & \citet{Hobbs+2005:2005MNRAS.360..974H, Igoshev+2020:2020MNRAS.494.3663I}\\
        \quad Black hole kicks & - & Fallback limited & No rescaling & \citet{Fryer+2012:2012ApJ...749...91F} \\
        \textbf{Galaxy settings} &&&&\\
        \quad Hydrodynamical simulation & - & \fire \texttt{m11h} & \changa \texttt{r442} & {\citet{El-Badry+2018:2018MNRAS.473.1930E, Keith+2025:2025arXiv250116317K}}\\
        \quad Velocity dispersion & $v_{\rm disp}$ & $v_{\rm disp} \propto \avir$ & $[0.5, 5] \unit{km}{s^{-1}}$ & \citet{Bertoldi+1992:1992ApJ...395..140B}\\
    \end{tabular}
    \caption{A summary of the fiducial choices (and variations) of the settings for the simulations in this work.}
    \tablenotetext{a}{$\tau_{\rm th, acc}$ is the thermal (Kelvin-Helmholtz) timescale of the accretor.}
    \label{tab:simulations_overview}
\end{table*}

We use \cogsworth \citep{Wagg+2025:2025JOSS...10.7400W, Wagg+2025:2025ApJS..276...16W} to quantify the impacts of binary interactions on the timing and location of SNe. \cogsworth uses \texttt{COSMIC} \citep{Breivik2020} to rapidly synthesize populations of binary stars and \texttt{gala} \citep{gala_JOSS, gala_13377376} to integrate the subsequent orbits of the stars in model galactic potentials. A detailed description of \cogsworth can be found in \citet{Wagg+2025:2025ApJS..276...16W}, but we give a brief overview of our approach here; further details are presented below in Section~\ref{sec:fiducial}.

Here we use \cogsworth to simulate the most recent $150\unit{Myr}$ of star formation in a galaxy. We do so by replacing newly-born star particles in hydrodynamical galaxy simulations with an equivalent mass cluster of binary stars. We then rapidly evolve the individual binary stars while simultaneously integrating their orbits through the galaxy, using a galactic potential fit to the galactic mass distribution in the hydrodynamical simulation. We then record the time and location of each core-collapse SN in the galaxy, evolving each binary until $200 \unit{Myr}$ beyond present day, such that the oldest binaries are evolved for a total of $350 \unit{Myr}$; these timescales capture a reasonable fraction of the recent star formation whilst also allowing enough time for every star to reach core collapse. We evolve only binaries in which the initially more massive star has a mass of at least $4 \unit{M_\odot}$, since less massive binaries are unlikely to produce a core-collapse SN, even through mergers.

\subsection{Initial stellar distributions \& orbital evolution}

For our fiducial model, we use initial conditions from the \fire-2 galaxy \texttt{m11h} \citep{Hopkins+2018:2018MNRAS.477.1578H, El-Badry+2018:2018MNRAS.473.1930E}, which is a dwarf galaxy, similar in mass to the LMC ($M_\star = 4 \times 10^{9} \unit{M_\odot}$), with a strong disc component and a slightly super-solar typical metallicity of 0.017. The star formation history is reasonably uniform over the last $150 \unit{Myr}$, with ${\sim}0.5\unit{M_\odot}$ formed per year. We focus our study on \texttt{m11h} since earlier work has shown the effect of spatially distributed runaway star feedback is more prominent in dwarf galaxies \citep[e.g.,][]{Steinwandel+2023:2023MNRAS.526.1408S}. This simulation has a force resolution identical to its spatial resolution, which is fully adaptive and has a minimum length scale on the order 0.7~pc \citep{Hopkins+2018:2018MNRAS.480..800H}. However, the spatial scale resolved at the star formation density threshold is approximately 7~pc. We identify the initial position and kinematics of every star particle by reverse integrating it through the galactic potential.

When generating binary populations, we assume that each of the ${\sim}7000\msun$ star particles in the simulation represents a stellar cluster of radius $3 \unit{pc}$, which is typical of open clusters and similar to the radius of Orion Nebular Cluster \citep[e.g.,][]{Kroupa+2018:2018AA...612A..74K}. We use this radius to adopt a Gaussian spread in initial position for each binary relative to the location of its parent star particle. This is similar to methods used in previous works \cite[e.g.,][]{Sanderson+2020:2020ApJS..246....6S}.

We adopted an initial velocity dispersion of $1.7 \unit{km}{s^{-1}}$ for the binaries. This corresponds to a cluster of $10^4 \msun$ (the mass of a star particle in \texttt{m11h}) with virial parameter of $\avir = 1$, which means that the cluster is initially gravitationally bound \citep{Bertoldi+1992:1992ApJ...395..140B}. This velocity dispersion approximately follows measurements of the velocity dispersion of the Orion Nebula Cluster \citep{DaRio+2017:2017ApJ...845..105D,Kroupa+2018:2018AA...612A..74K, Kuhn+2019:2019ApJ...870...32K}.

When replacing star particles with collections of individual stars, we use \cogsworth to sample binary stellar populations that match the total mass, metallicity, and birth time of each star particle formed in the most recent $150 \unit{Myr}$ of the simulation. 

After formation, the positions of new stars are evolved within a galactic potential matching that of the original simulation. \cogsworth calculates the potential using the self-consistent field method implemented in \texttt{gala} based on \citet{Hernquist+1992:1992ApJ...386..375H} and \citet{Lowing+2011:2011MNRAS.416.2697L}, which fits the galactic mass distribution (accounting for the contributions from stars, gas and dark matter)
using a basis function expansion in spherical harmonics. For a more detailed explanation of the full implementation in \cogsworth, see Section 2.5 of \citet{Wagg+2025:2025ApJS..276...16W}.

Within the potential, and using the initial stellar kinematics, we integrate the galactic orbit of each single or binary system with an adaptive integration scheme \citep{dopri}, tracking individual stars if a binary is disrupted by a SN.

\subsection{Fiducial binary model} \label{sec:fiducial}

In this Section we describe our fiducial settings, which are also summarised in the third column of Table~\ref{tab:simulations_overview}. The fourth column lists the variations of these parameters, which we discuss in Section~\ref{sec:variations}.

\paragraph{Initial population sampling} We draw masses of the primary (initially more massive) star in each binary, $m_1$, following the broken power law initial mass function (IMF) from \citet{Kroupa+2001:2001MNRAS.322..231K}, such that $p(m_1) \propto m_1^{\alpha_{\rm IMF}}$ and sampling only stars with $m_1 > 4 \msun$. The high mass slope (for stars with $m_1 > 1 \msun$) of this IMF has $\alpha_{\rm IMF} = -2.3$. We sample mass ratios, $q \equiv m_2 / m_1$, uniformly in $[q_{\rm min}, 1]$, where $q_{\rm min}$ is set such that the pre-main sequence lifetime of the secondary is not longer than the full lifetime of the primary if it were to evolve as a single star \citep{Mazeh+1992:1992ApJ...401..265M, Goldberg+1994:1994A&A...282..801G}. Eccentricities, $e$, are sampled following \citet{Sana+2012:2012Sci...337..444S} such that $p(e) \propto e^{-0.45}$ with $e \in [0, 0.9]$, where the upper limit is chosen to avoid Roche lobe overflow at pericenter. For orbital periods, $P$, we follow the distribution of \citet{deMink+2015:2015ApJ...814...58D}, which is an extrapolation of \citet{Sana+2012:2012Sci...337..444S} to lower masses, such that $p(P) \propto (\log_{10}(P / \unit{days}))^{-0.55}$ with $\log_{10}(P / \unit{days}) \in [0.15, 5.5]$. We use a fixed binary fraction of 100\% since we simulate only massive stars which are almost all formed in binaries \citep[e.g.,][]{Offner+2023:2023ASPC..534..275O}. Additionally, though we do not include truly single stars, our upper orbital period limit allows for a fraction of stars to be effectively single, such that they have no interactions with companions. Metallicities and birth times are set based on the star particles from the galaxy that is postprocessed.

\begin{figure*}
    \centering
    \includegraphics[height=0.5\textwidth]{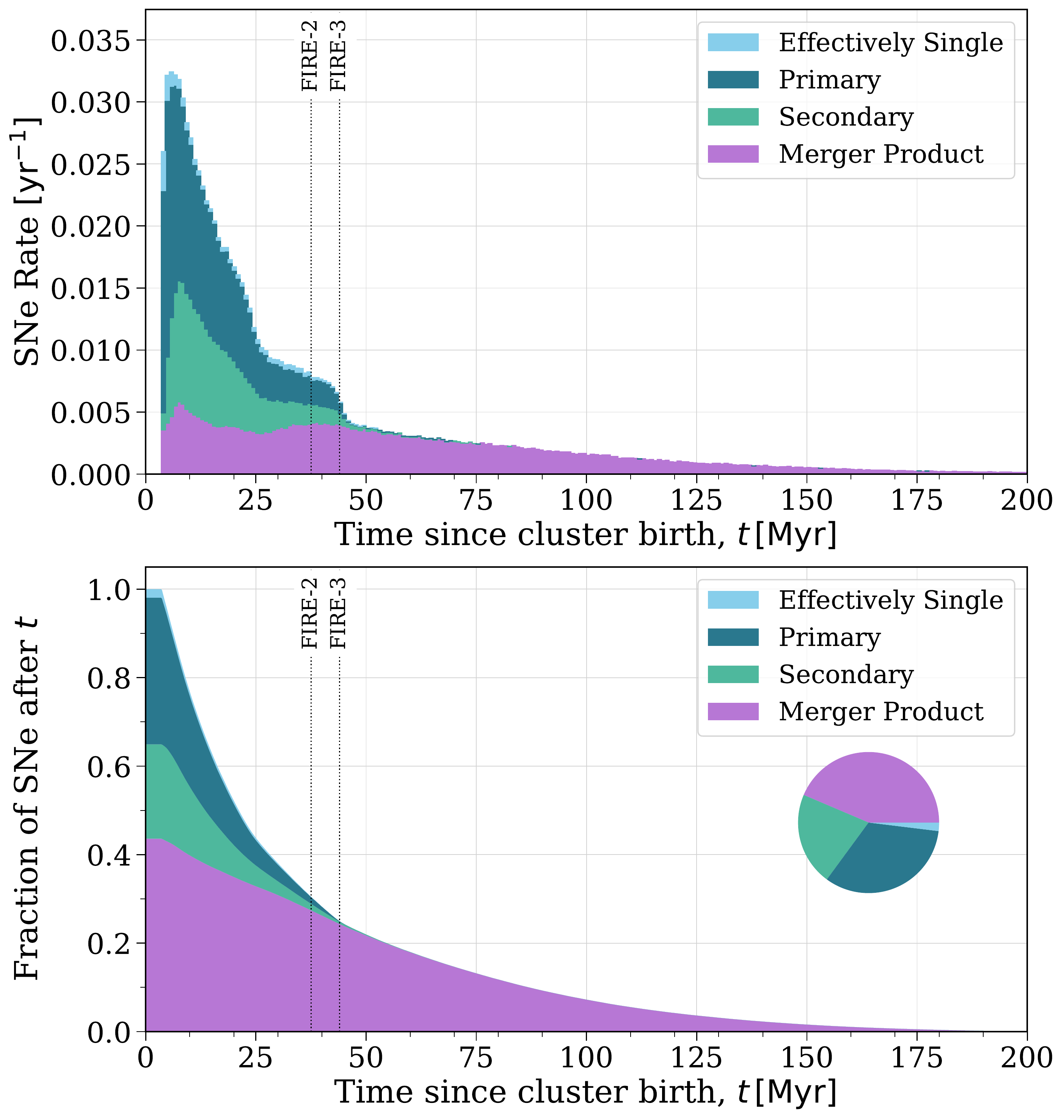}
    \includegraphics[height=0.5\textwidth]{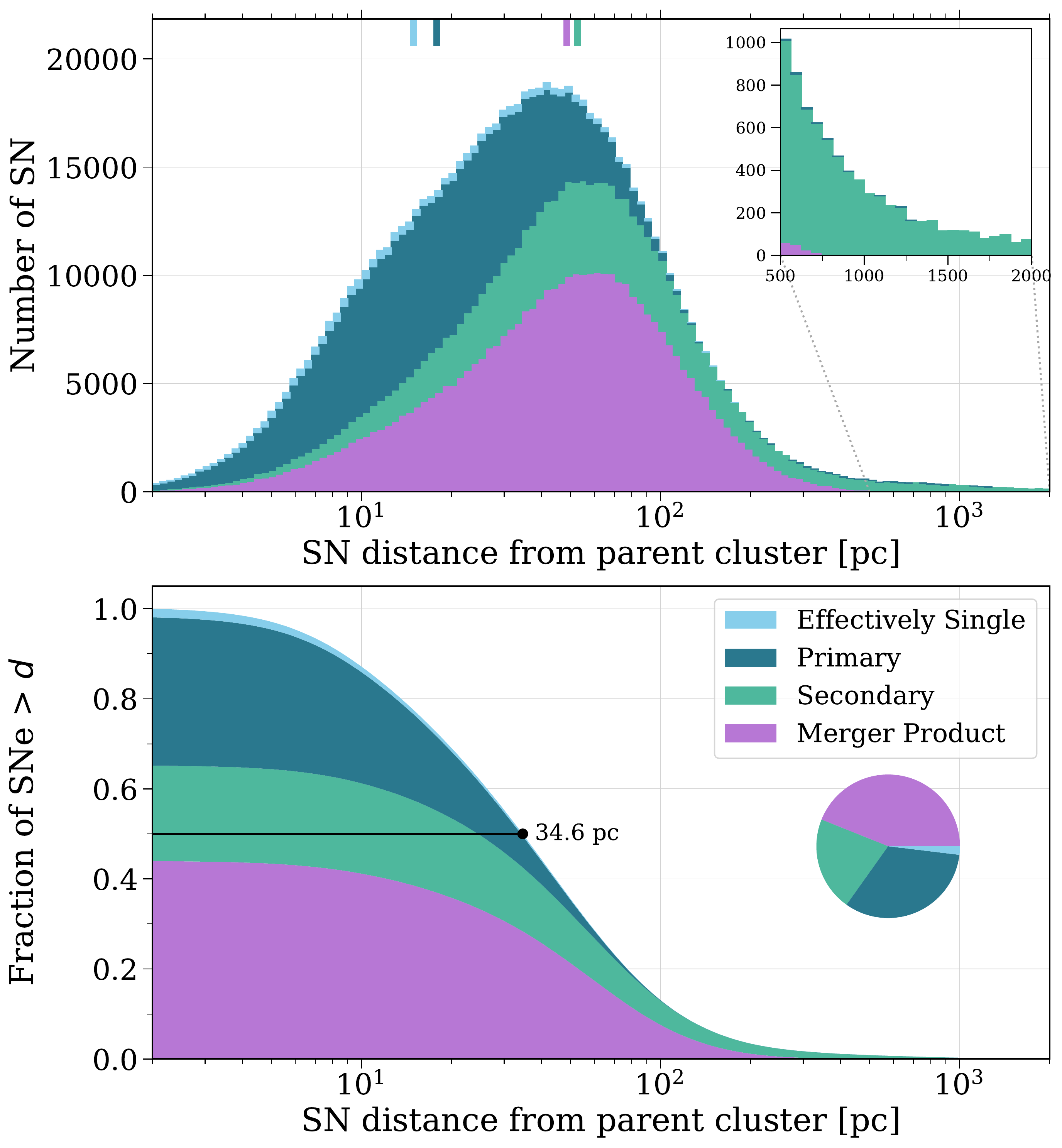}
    \caption{Binary interactions can result in delayed SNe and displace SNe far from their parent clusters and molecular clouds. \textbf{Top panels:} Stacked histograms separated by progenitor type. Effectively single stars had no binary interaction prior to SN (no Roche Lobe overflow and less than 5\% wind mass accretion). Primary (secondary) stars were the initially more (less) massive star in a binary that did not undergo a merger. Merger products are the result of a merger between two stars prior to SN in a binary. \textbf{Bottom panels:} Fractional complementary cumulative distribution also separated by progenitor type. Pie charts shows the relative contribution from each progenitor type. 
    \textbf{Left:} Histograms show the SN rate as a function of time since cluster birth. Dashed lines indicate time beyond which no SN occur in \fire simulations. \textbf{Right:} Histograms show the distance between a SN and its parent cluster at the time of the SN. The inset panel highlights the tail of SNe at extremely long distances. Solid markers on the top axis of the top panel indicate the median value for each progenitor type. Black circle on the bottom panel shows the overall median distance. (\href{https://www.tomwagg.com/html/interact/binary-supernova-feedback.html\#fig1}{\faLaptopCode{} Interactive figure available.})}
    \label{fig:sn-times-dists}
\end{figure*}

\paragraph{Binary physics} \cogsworth uses \texttt{COSMIC} for binary population synthesis. \texttt{COSMIC} is based on the \texttt{BSE} code \citep{Hurley+2000:2000MNRAS.315..543H, Hurley+2002:2002MNRAS.329..897H}, which uses fitting formulae from \citet{Tout+1997:1997MNRAS.291..732T} to the single star models of \citet{Pols+1998:1998MNRAS.298..525P}), but with extensive modifications and updated prescriptions based on more recent work \citep[see Section 3 of][]{Breivik2020}. Our simulations make use of \texttt{COSMIC} v3.4.16 and use the default settings for that version. In particular, we assume the efficiency of mass transfer, $\beta \equiv \Delta M_{\rm acc} / \Delta M_{\rm don}$, is such that the amount of mass accreted during Roche-lobe overflow is limited to 10x the thermal rate of the accretor for main sequence, Hertzsprung gap and core helium burning stars and unlimited for giant branch stars \citep[e.g.,][]{Kippenhahn+1967:1967ZA.....65..251K, Schneider+2015:2015ApJ...805...20S}. We assume that mass lost during Roche-lobe overflow is lost from the system via isotropic re-emission (as if it is a wind from the secondary) \citep[e.g.,][]{Massevitch+1975:1975MmSAI..46..217M}.

We determine the stability of mass transfer using the critical mass ratios defined in \citet{Hurley+2002:2002MNRAS.329..897H} for all stellar types except for giant branch stars, for which we use \citet{Hjellming+1987:1987ApJ...318..794H}. For unstable mass transfer we assume common-envelope events follow the $\alpha$-$\lambda$ prescription \citep{Webbink+1984:1984ApJ...277..355W, deKool+1990:1990ApJ...358..189D} and by default take $\alpha_{\rm CE} = 1$ and use the $\lambda$ prescription from \citet{Claeys+2014:2014A&A...563A..83C}. 

Asymmetries in SN explosions impart a natal kick on the compact object that is formed, which may unbind a binary orbit, thereby ejecting a secondary star. We assume SN natal kicks are distributed as a double Maxwellian distribution, with one component peaking at $\sigma_{\rm CC} = 265 \unit{km}{s^{-1}}$ for core-collapse SNe and the other at $\sigma_{\rm low} = 20 \unit{km}{s^{-1}}$ for electron-capture SNe (ECSN) and ultra-stripped SNe (USSN) \citep{Hobbs+2005:2005MNRAS.360..974H, Igoshev+2020:2020MNRAS.494.3663I}. By default in \cosmic ECSN are assumed to occur if the helium core mass is between $1.6$ and $2.25\msun$, whilst USSN occur for helium stars that undergo a common-envelope with a compact object companion. The magnitude of SN natal kicks for black holes is modulated based on the fallback mass following \citet{Fryer+2012:2012ApJ...749...91F}.

\section{Results I: Supernovae times and locations in Fiducial model}\label{sec:results}

In this section we consider the behaviour of the fiducial model described above.  Throughout, we separate the individual progenitor types into four groups: (1) effectively single stars that had no binary interaction prior to their SN, which we define as no Roche Lobe overflow and less than 5\% mass accretion from the stellar winds of a companion; (2) primary stars that were initially the most massive star in a binary that did not undergo a merger; (3) secondary stars that were the less massive partner in a non-merging binary; and (4) merger products that result from two stars merging in a binary prior to core collapse.

In the subsections below we discuss the trends in both timing and distances, based on the plots shown in Figure~\ref{fig:sn-times-dists} for our fiducial model. These plots show absolute (top row) and cumulative (bottom row) distributions for: the time between the formation of a stellar cluster and the eventual SNe of the member stars (left column); and the distance between the SN and the position that the original stellar cluster would have had at the time of the SN, had it not been replaced with a cluster of evolving binary stars (right column). Each distribution is plotted as a ``sandpile diagram'', with contributions from different SN precursors stacked on top of each other. Throughout the paper, we colour-code contributions from effectively single stars in light blue, primary stars in dark blue, secondary stars in green, and merger products in purple. 

\subsection{Supernova timing}\label{sec:sn_times}

For a single star population, there is a tight relationship between initial mass and time to core collapse, with some small scatter from metallicity effects \citep[e.g., Figure 5 of][]{Hurley+2000:2000MNRAS.315..543H}. Therefore, one would expect the SN rate to peak at early times and decline with a slope that is set by the convolution of the lifetime--mass relation and the initial mass function.

For a binary stellar population, mass transfer and mergers complicate this relationship. The addition or removal of mass during mass transfer will alter a star's evolutionary timescale. In the vast majority of cases, stars that lose mass to their companions will evolve more slowly and so have delayed SNe, with the opposite being true for those that accrete mass \citep[e.g.,][]{Pols+1994:1994A&A...290..119P}. Alternatively, two stars that were initially not massive enough to reach core collapse may merge after their (relatively) slow evolution and later explode, resulting in later SNe \citep[e.g.,][]{Zapartas+2017:2017AA...601A..29Z}.

In the left panels of Figure~\ref{fig:sn-times-dists} we show the distribution of SN times relative to cluster birth, separated by SN progenitor types. For a single star population, we would expect core-collapse SNe to cease after the lowest mass star that could reach SNe concludes its evolution. At the average metallicity of \texttt{m11h}, this corresponds to a star of ${\sim}7\msun$ and a time of ${\sim}44\unit{Myr}$. A lower mass star cannot reach core collapse with the default \cosmic settings and hence this is the latest time a single star could result in a core-collapse SN in our simulations. We note that the limits of \fire-2 and \fire-3 (shown as dotted lines on Figure~\ref{fig:sn-times-dists}) reproduce this limiting time well.

Critically, binary interactions allow for SNe to occur at later times. Similar to single star evolution models, the distribution in Figure~\ref{fig:sn-times-dists} shows that the earliest SNe occur around $3.7 \unit{Myr}$ and the overall distribution quickly peaks around $6 \unit{Myr}$ as the most massive stars reach core collapse. Unlike single star models, however, after ${\sim}44\unit{Myr}$ there is a long tail of SNe that is almost entirely from stars that merged before core collapse (purple). These merger product SNe occur later because the two less massive stars that merged evolve on slower timescales that massive stars that can reach core collapse alone.

\begin{figure}
    \centering
    \includegraphics[width=\columnwidth]{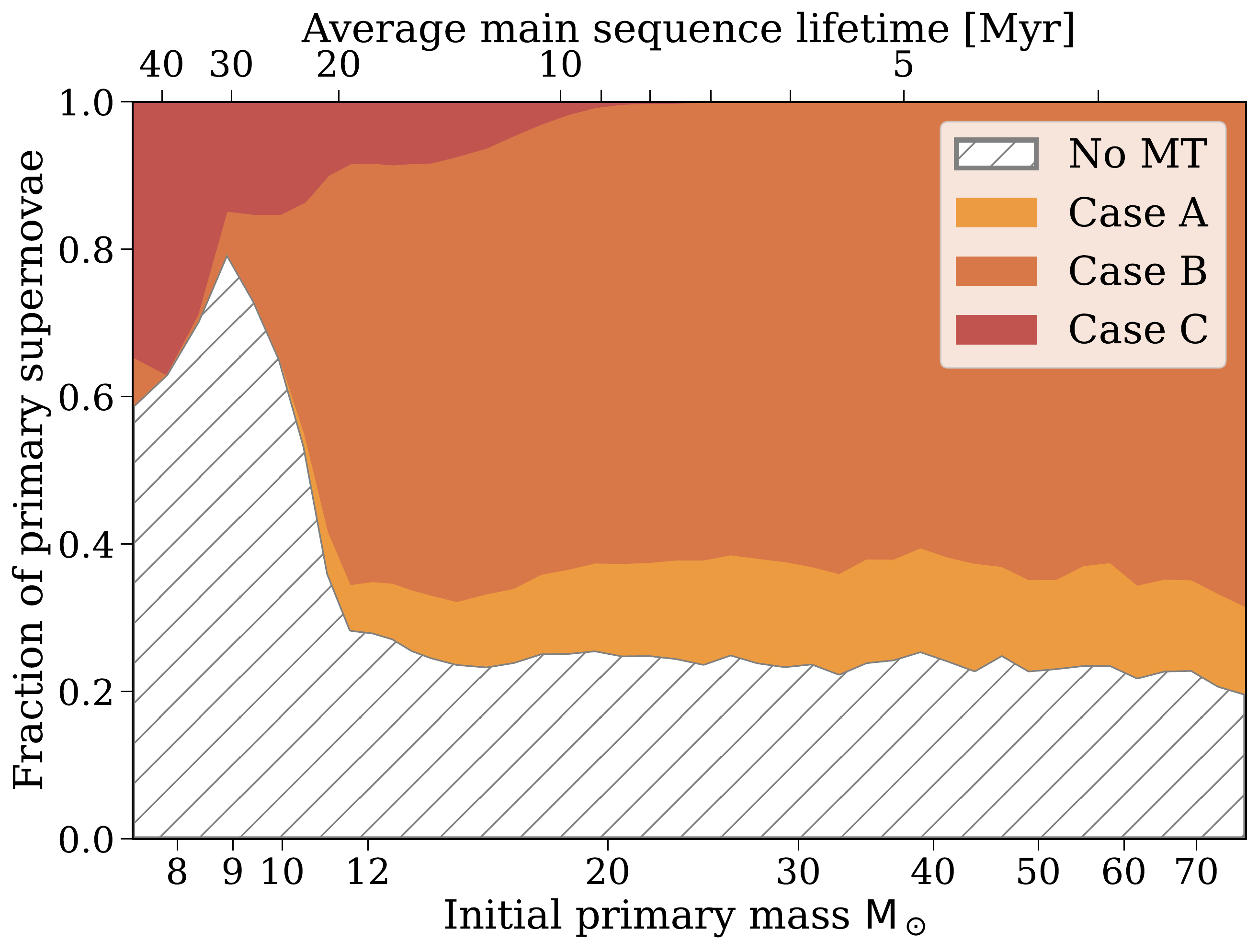}
    \caption{The type of mass transfer that a primary star first initiates prior to its core collapse is a function of its initial mass. The transition at ${\sim}10\msun$ leads to the knee in the distribution in the upper left panel of Figure~\ref{fig:sn-times-dists}. Mass transfer initiated on the donor star's main sequence is case A, while case B corresponds to expansion during the Hertzsprung gap, and case C occurs after the onset of core helium burning. Primary stars below ${\sim}10\msun$ mostly initiate case C mass transfer, whilst more massive stars initiate case A or B mass transfer. Top axis shows the main sequence lifetime corresponding to the initial primary mass for the average metallicity of \texttt{m11h} as calculated by \cosmic.}
    \label{fig:mt_type_by_mass}
\end{figure}

In addition to the late-time tail from mergers, the age distribution of SNe also shows a transition around ${\sim}25 \unit{Myr}$ (top left panel of Figure~\ref{fig:sn-times-dists}), which is an imprint of stable mass transfer on SN timing. The most massive stars expand significantly on the main sequence, and often transfer mass to their lower mass partner, delaying their eventual SNe. In contrast, lower mass stars rarely expand enough on their main sequence to initiate mass transfer \citep[e.g.,][]{deMink+2008:2008AIPC..990..230D, Burt_2025}. This transition appears in our simulations at roughly $10\unit{M_\odot}$ where the prevalence and mechanism of mass transfer changes, as shown in Figure~\ref{fig:mt_type_by_mass}. The transition leads to a break in the distribution at the lifetime corresponding to this mass, approximately ${\sim}25 \unit{Myr}$.

More specifically, this feature in the SN age distribution is driven by both the reduced likelihood of mass transfer at lower masses and the relative importance of case A, B, \& C mass transfer mechanisms. A primary star that donates a significant amount of mass during its main sequence (case A mass transfer), before developing a significant helium core, lengthens its nuclear timescale, delaying its SN. If the mass transfer occurs later in the star's life (case B or C mass transfer) then it has a lesser effect on the SN timing. We find that primary star SN progenitors with a lifetime longer than around $25\unit{Myr}$ do not undergo case A mass transfer, leading to the feature in the age distribution.
We note that the delays occurring as a result of case A mass transfer are strongly dependent on assumptions in population synthesis regarding core evolution and rejuvenation (see Section~\ref{sec:limitations}).

Due to these two effects --- mergers and mass transfer --- 25\% of SNe in our fiducial simulation occur beyond the cutoff used in the \fire-3 prescription for type II SNe (at $44 \unit{Myr}$), as shown in the lower left panel of Figure~\ref{fig:sn-times-dists}. The progenitors of these later SNe are primarily merger products, though with a small fraction of primaries and secondaries before $50 \unit{Myr}$.

\subsection{Supernova separation from parent clusters}\label{sec:sn_dists}

In our simulations, there are two mechanisms by which SNe can occur far from their birthplace. First, typical stars are born in stellar clusters, almost all of which rapidly dissolve into the field \citep[e.g.,][]{Lada+2003:2003ARA&A..41...57L, PortegiesZwart+2010:2010ARA&A..48..431P} after their natal gas clouds are expelled. This imprints stars with an initial velocity dispersion, leading them to have slightly different orbits through the galaxy, and thus to diverge from the cluster centre-of-mass over time. 

The second mechanism is through ejection from a binary. If a binary disrupts after the primary star reaches core collapse, the secondary star is ejected with approximately its orbital velocity \citep[e.g.,][]{ejection_vels_rnaas,Renzo+2019:2019A&A...624A..66R}. These secondaries can then travel large distances as runaway stars before reaching core collapse themselves.

In the right panels of Figure~\ref{fig:sn-times-dists}, we show the distance between each SN and its parent cluster (or simulation star particle)'s centre of mass at the moment of core collapse. SNe from effectively single stars and primary stars are typically located close to the parent cluster. This is expected given that they are short-lived with less time to disperse with their modest ${\sim}2\unit{km}{s^{-1}}$ initial velocity dispersion, and are almost never ejected from binaries. On average, SNe from effectively single or primary stars occur ${\sim}18\unit{pc}$ from their parent cluster and 90\% occur within $50\unit{pc}$.

In contrast, SNe from merger products, which constitute a third of all SNe in the fiducial simulation \citep[e.g.,][]{Sana+2012:2012Sci...337..444S}, are located further from their parent cluster at the time they explode. On average, they occur at a separation of ${\sim}49\unit{pc}$, with 17\% occurring beyond $100\unit{pc}$. This is a result of SNe from merger products happening later (see Section~\ref{sec:sn_times}), giving them more time to disperse from their parent cluster as a result of their initial velocity dispersion \citep[e.g.,][]{Aghakhanloo+2017:2017MNRAS.472..591A}.

The spatial separations are even further extended for SNe from secondary stars, which constitute nearly a quarter of SNe in the fiducial case. These SNe have an additional contribution from their high ejection velocities, and are located at an average separation of ${\sim}54\unit{pc}$, with 25\% occurring beyond $100\unit{pc}$ and 4\% beyond $500\unit{pc}$.

Overall, the average distance between a SNe and its parent cluster is ${\sim}35\unit{pc}$, which is within the resolution element of \fire simulations at the average ISM density (\fire's spatial resolution is $\approx 85\unit{pc}$ for an ISM density of $1\unit{cm^{-3}}$). However, the distribution has a significant tail, with ${\sim}13\%$ of SNe occurring at a separation of more than $100\unit{pc}$.

\subsection{The joint distribution of SN time delay and separation, for single and for binary stars}

\begin{figure}[htb]
    \centering
    \includegraphics[width=\columnwidth]{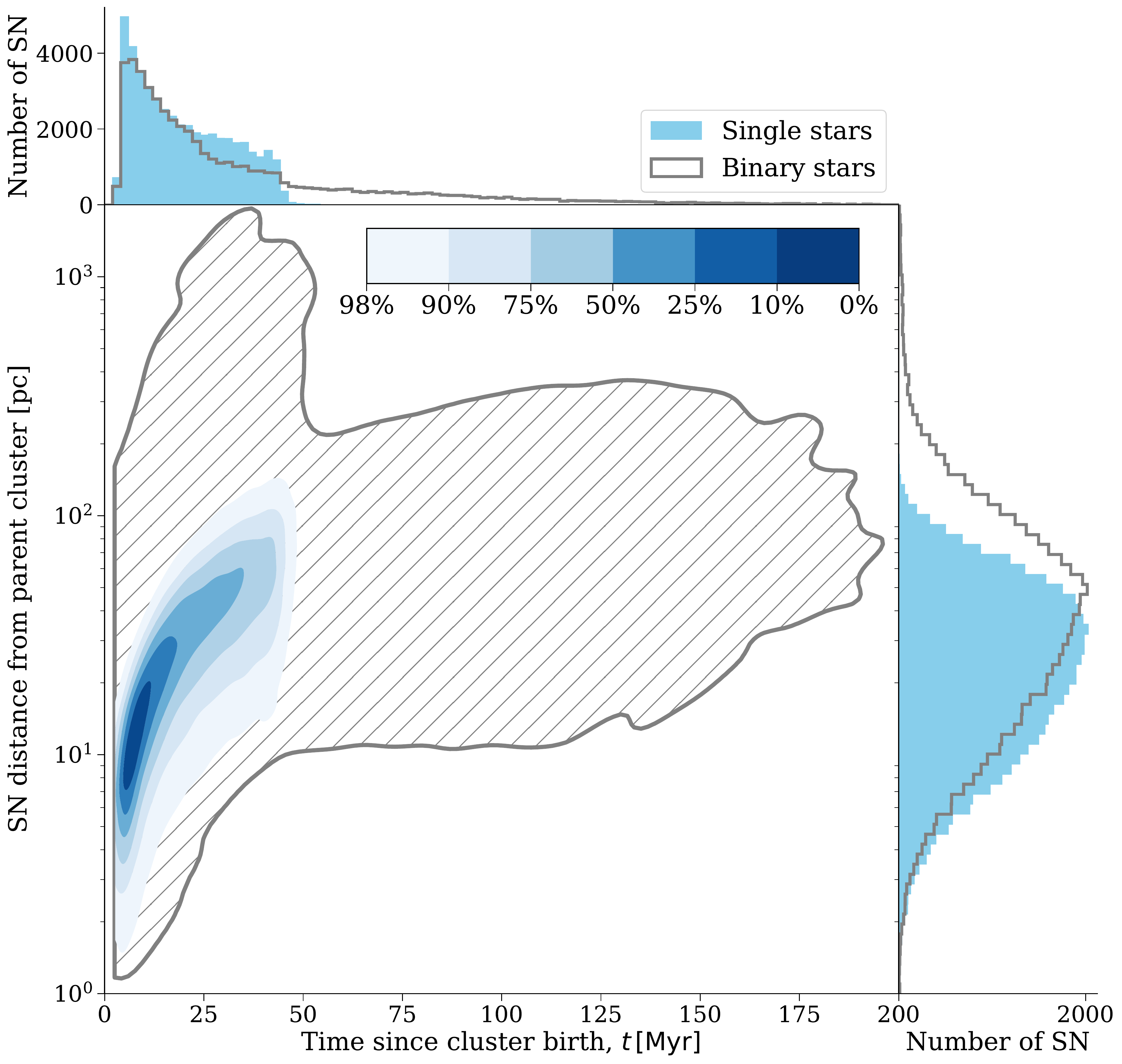}
    \includegraphics[width=\columnwidth]{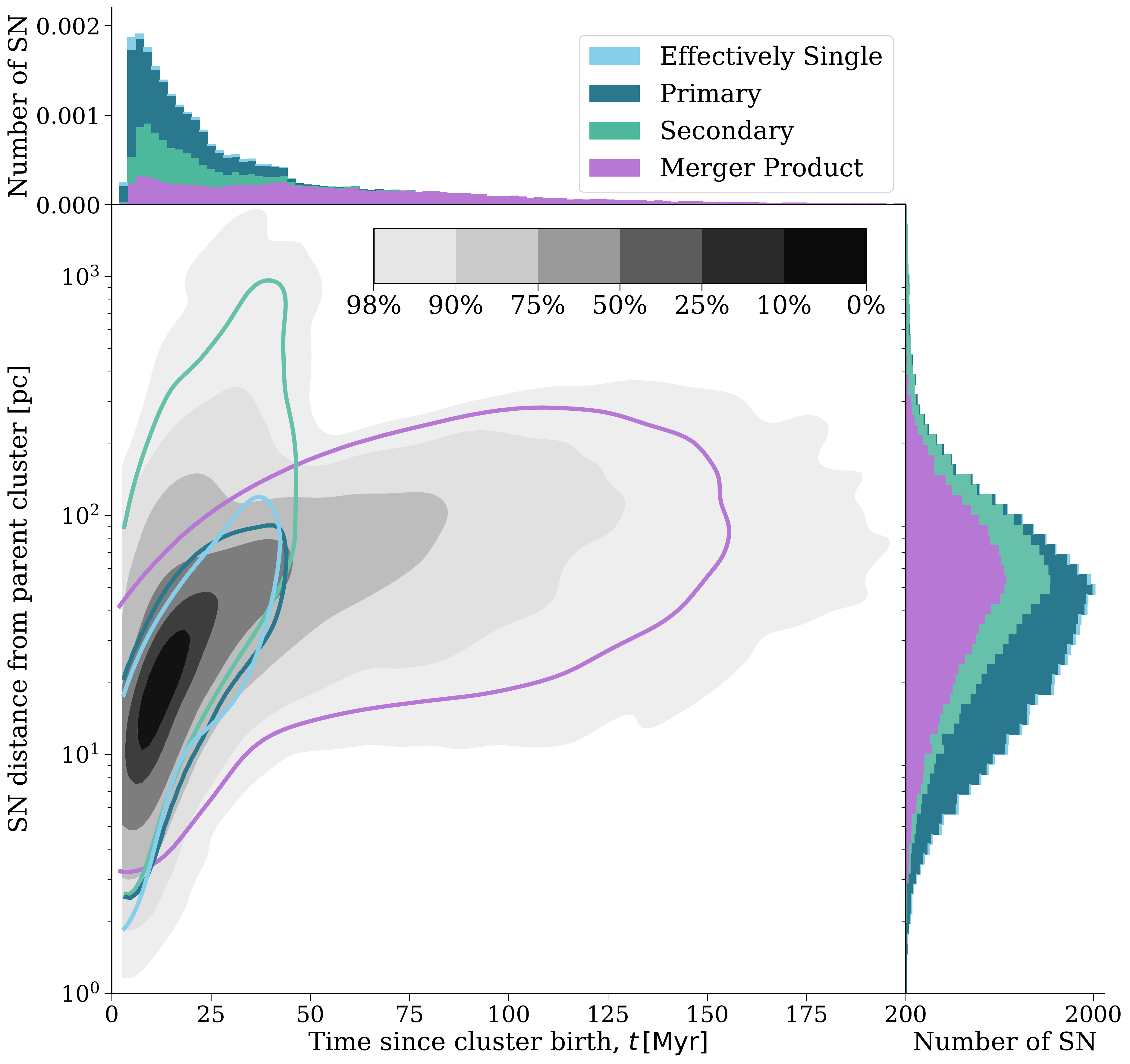}
    \caption{\textbf{Top:} Binary stars produce a significantly different distribution of SNe times and locations than single stars. Main panel shows a 2D kernel density estimation of the distribution of SNe for single star population in blue. The grey hatched contour shows the 98\% region for a binary star population. Marginal histograms are shown in each side panel. \textbf{Bottom}: The timing and location of SNe relative to their parent cluster are strongly correlated and a function of their progenitor type. Grey filled contours show the full 2D density distribution, whilst coloured contours indicate the region within which 90\% of each subpopulation is contained. Marginal histograms are the same distributions as the top panels of Figure~\ref{fig:sn-times-dists}.}
    \label{fig:2d-dists}
\end{figure}

In the above discussion of Figure~\ref{fig:sn-times-dists}, we treat SN timing and spatial distributions as independent. In practice, however, these two distributions are coupled; extended SN timescales allow more time for precursor stars to disperse, and specific binary evolutionary pathways can change both timing and velocities, due to mass transfer and kicks from evolving higher-mass partners.

In this section, we look at the joint distributions of SN timing and position, as a function of binary evolution channel. We also compare these distributions to expectations for purely single star populations. We generate this comparison set by using identical initial conditions to our fiducial model, except we set each binary's initial orbital period to effectively infinitely wide orbits, such that they never interact. This ensures a consistent normalisation of the two populations and that differences are entirely a result of binary interactions.


We show the resulting joint distributions in Figure~\ref{fig:2d-dists}. In the top panel, the blue distribution shows the joint distribution for single stars. As expected, there is a correlation between time and distance, which is entirely driven by the dissolution of stellar clusters, with individual progenitor stars' orbits diverging from the centre of mass after being perturbed by a velocity consistent with the assumed fiducial velocity dispersion (Table~\ref{tab:simulations_overview}).

Also included in the upper panel is a cross-hatched contour that contains 98\% of the fiducial binary population. This distribution is both more extended and more complex than the single star distribution. Binary interactions clearly result in significantly longer tails in both time and distance for core-collapse SN feedback when comparing to a population of single stars.

We note that in addition to the change in the joint distribution, the binary evolution channels also increase the overall number of SNe. As a result, binary interactions produce \textit{additional} feedback compared to single star evolution, with ${\sim}11\%$ more SNe occurring in our fiducial model than in the single star model. This change is mainly a result of merger products allowing two less massive stars to reach core collapse.

In the lower panel of Figure~\ref{fig:2d-dists}, we explore in further detail the evolutionary pathways and progenitor types that lead to the structure in the binary population. To first order, the distribution of primary (dark blue) and effectively single (light blue) stars are quite similar to the distribution of single stars in the the upper panel. This similarity is expected, as these progenitor types are not displaced by companions and primaries experience only small delays as a result of stable mass transfer.

More significant differences are seen for the secondaries (green) and merger products (purple). The secondaries experience significant velocity kicks when their higher mass primary companions reach core collapse. These kicks lead the stars to eventually explode at much larger distances than single stars.

The merger products in Figure~\ref{fig:2d-dists} are significantly extended in both time and distance. As discussed above, the long tail in ages is a result of the slower evolution of lower mass stars, which eventually merge and reach core collapse. This also means that the stars have much longer to disperse from their parent cluster. However, because the coalescence does not significantly change the kinematics of the system, the spatial distribution of the merger products continues the trend established by cluster dissolution, as seen for the single stars.

We also note that the joint distribution offers a way to help isolate different evolutionary pathways. 
Supernovae found at large separations from star-forming regions, but with young ages are very likely to be secondary stars, whilst older remnants are more likely to be merger products.

\section{Results II: Robustness of results to model variations}\label{sec:variations}

In this Section, we assess the robustness of the predicted temporal and spatial distributions to variations in the initial conditions of our binary stellar population, binary physics assumptions, and galactic orbit evolution parameters. We focus these variations on settings that have the largest potential to influence the results of our fiducial model. 

As discussed in Section~\ref{sec:sn_times}, the timing of SNe is altered by mass transfer and stellar mergers. Therefore, we vary both the efficiency and stability of mass transfer. Moreover, changes to the initial population, such as the initial mass function and metallicity, also strongly influence this distribution by varying the number of massive stars and the minimum mass that can reach core collapse.

The locations of SNe are driven by a combination of cluster dissolution and stars ejected from their binaries with their orbital velocity. Therefore, we vary parameters that (i) change the rate of cluster dissolution and (ii) change the orbital velocity of the secondary at the moment of the primary SN. We also consider how using a different galactic potential may change these distances. We first outline the specific variations below and summarise them in the fourth column of Table~\ref{tab:simulations_overview}. We then discuss the impacts of these variations in Section~\ref{sec:variations}.

For variations that alter our assumptions regarding binary physics, we use an identical initial population, such that the same evolution of the same binary can be compared for different settings. This approach relies on \cosmic's ability to record the precise initial conditions (and random seeds) of sampled populations, making them easily reproducible with different evolution settings \citep{Breivik2020}. For variations that alter galactic orbits, we use populations with identical initial populations and stellar evolution, such that the direction, magnitude and timing of SN kicks are consistent across variations.

\subsection{Variation descriptions}\label{sec:variation-descriptions}

\paragraph{Common-envelope events} A common-envelope (CE) event occurs when mass transfer becomes dynamically unstable in a runaway process. In such an event, the donor continues to expand faster and further overflow its Roche lobe as it transfers mass. Eventually the cores of both the donor star and the accretor are engulfed within the envelope material of the donor. These events can significantly shrink the binary orbit of the massive cores, and in some cases lead to a stellar merger. 

Given the uncertainty of these events, we vary the efficiency of common-envelope events significantly, from to $\alpha_{\rm CE} = 0.1$ and $10.0$. This efficiency changes the fraction of orbital energy that is available to unbind the envelope \citep{Webbink+1984:1984ApJ...277..355W, deKool+1990:1990ApJ...358..189D}. Thus $\alpha_{\rm CE}$ affects both the fraction of binaries that survive a CE without merging, and the post-CE orbital separations of those that survive as a binary, which later impacts the ejection velocity of the secondary if a disruption occurs. 

We also vary our assumptions of the critical mass ratio, $q_{\rm crit}$, necessary for stable case B mass transfer, which is the most common type of mass transfer in our simulations \citep[in agreement with][]{vandenHeuvel+1969:1969AJ.....74.1095V}. The stability of mass transfer is determined by comparing the mass ratio of the binary at the moment of Roche-lobe overflow ($q_{\rm MT} \equiv m_{\rm donor}/m_{\rm accretor}$) to the critical mass ratio, with unstable mass transfer occurring when $q_{\rm MT} > q_{\rm crit}$. We calculate the two possible extremes: $q_{\rm crit,caseB} = 0$ (all case B mass transfer is unstable) and $q_{\rm crit,caseB} = \infty$ (all case B mass transfer is stable).

\paragraph{Stable mass transfer} The efficiency of mass transfer, $\beta$, changes the fraction of mass transferred from the primary star that is accreted by a companion. For a larger value of $\beta$, more mass is accreted by a companion, which could lead to stronger rejuvenation and changes to the time at which an accretor's SN occurs. Additionally, decreasing $\beta$ allows more angular momentum to leave the system and decrease orbital separations, thereby increasing the ejection velocities of ejected secondary stars.

In our fiducial model, our default assumption is that the efficiency is based on the thermal timescale of the accretor. This choice results in an average efficiency of $\beta\approx0.77$ in our fiducial simulations. We vary $\beta$ to three fixed values: $0.0$ (non-conservative), $0.5$ and $1.0$ (conservative). 

\paragraph{Supernova natal kicks} Binary orbits are frequently disrupted by SN natal kicks, resulting in the ejection of secondary stars \citep[e.g.,][]{Renzo+2019:2019A&A...624A..66R}.
In rare cases where a binary remains bound these SN kicks can also determine how far a secondary star travels before its own core collapse. We vary the strength of SN natal kicks in three different ways. By default, we assume a double maxwellian distribution for kick velocities, with a low velocity component from electron-capture and ultra-stripped SNe, and a higher velocity component from core-collapse SNe. Two of the variations change this distribution to a single maxwellian, the first assuming all remnants are in the low kick component of the fiducial double maxwellian and the other assuming all are in the high kick component. In the third variation we no longer reduce black hole kicks based on the amount of fallback mass.

\paragraph{Initial distributions} We vary the assumed power law distribution of initial primary masses, orbital periods and mass ratios. The initial mass function affects the fraction of stars that reach core collapse, as well as the number of lower mass stars available for mergers and later SNe. We vary the slope of the high mass end of the IMF from our fiducial assumption of $\alpha_{\rm IMF} = -2.3$ \citep{Kroupa+2001:2001MNRAS.322..231K} to $\alpha_{\rm IMF} = -1.9$ based on the finding for the 30 Doradus starburst \citep{Schneider+2015:2015ApJ...805...20S, Schneider+2018:2018AA...618A..73S} and additionally $\alpha_{\rm IMF} = -2.7$ to bracket our fiducial assumption.

The initial orbital period can change the evolutionary phase during which mass transfer occurs (or whether it occurs entirely), as well as the orbital separation of the binary at the moment of the first SN. We vary the power law slope, $\pi$, to $\pi = -1$ and $\pi = 0$, bounding the fiducial assumption of $-0.55$, based on the uncertainties in \citet{Sana+2012:2012Sci...337..444S} and compatible with several observational constraints \citep[e.g.,][]{Opik+1924:1924PTarO..25f...1O,Kobulnicky+2007:2007ApJ...670..747K,Moe+2017}. We additionally alter our assumed upper limit for the initial orbital period, changing our fiducial value of $10^{5.5}$ days to $10^3$ days. The lower maximum initial period is often assumed in earlier works, based on the limitations of previous spectroscopic surveys. These surveys could only observe binaries wider than ${\sim}10^{3}\unit{days}$, however interferometric studies have found wider binaries \citep[e.g.,][]{Sana+2014:2014ApJS..215...15S}. \citet{deMink+2015:2015ApJ...814...58D} introduced a higher limit of $10^{5.5}\unit{days}$ based on these results and argued that there is no physical reason for the binary fraction to sharply drop at $10^{3}\unit{days}$. For these reasons we adopt the same limit in our fiducial model.

The initial mass ratio can change the stability of mass transfer, because more unequal mass ratio binaries are more likely to experience unstable mass transfer, as well as the time between primary and secondary SNe. We vary the power law slope, $\kappa$, to $\kappa = -1$ and $\kappa = 1$, bounding the fiducial assumption of a uniform mass ratio distribution, based on the uncertainties in \citet{Sana+2012:2012Sci...337..444S} and compatible with several observational constraints \citep[e.g.,][]{Kobulnicky+2007:2007ApJ...670..747K, Mason+2009, Moe+2017}.

\paragraph{Metallicity} Stars at a lower metallicity lose less mass via stellar winds and lower radial expansion \citep{Leitherer+1992:1992ApJ...401..596L, Brunish+1982:1982ApJS...49..447B, Baraffe+1991:1991A&A...245..548B}. This means that they retain more mass and have different radii during their evolution and thus may experience mass transfer during different evolutionary phases. We lower the metallicity, $Z$, of each star in our initial population by constant factors of $0.5, 0.2, 0.1$, and $0.05$. This preserves the distribution of metallicites within the binary population, but shifts the distribution to a systematically lower mean metallicity, $\bar{Z}$.

\paragraph{Cluster velocity dispersion} The initial velocity dispersion of a cluster determines how quickly its member stars will disperse, changing the distance between the eventual SNe and the centre of the parent cluster. Our fiducial assumption is a velocity dispersion of $1.7 \unit{km}{s^{-1}}$ (based on a virial parameter of $\avir = 1$), which approximately follows measurements of the velocity dispersion of the Orion Nebula Cluster \citep{DaRio+2017:2017ApJ...845..105D,Kroupa+2018:2018AA...612A..74K, Kuhn+2019:2019ApJ...870...32K}. We consider the impact of initial velocity dispersion of the cluster in our simulation by using a lower choice of $v_{\rm disp} = 0.5 \unit{km}{s^{-1}}$ and higher choice of $v_{\rm disp} = 5 \unit{km}{s^{-1}}$, which bracket the range of dispersions found in \citet{Kuhn+2019:2019ApJ...870...32K}.

\paragraph{Hydrodynamical zoom-in simulation} Our fiducial model is calculated within the context of a specific simulated galaxy. Although the majority of our results are shaped by our choice of binary physics and cluster velocity dispersion, there may be additional effects that can be traced to the specifics of the simulated galaxy. In particular, how quickly cluster stars disperse can be shaped by the gravitational potential the cluster forms and evolves within. There may also be effects due to the metallicity distribution of the population of stellar clusters, which can change the binary evolution. To assess some of these possible effects, we repeat our calculations for the \changa \texttt{r442} galaxy, to compare with the fiducial \fire-2 \texttt{m11h} galaxy, with and without controlling for metallicity. 

\subsection{Summary plots of parameter variations}\label{sec:variation-plot-explained}

\begin{figure*}
    \centering
    \includegraphics[width=0.9\textwidth]{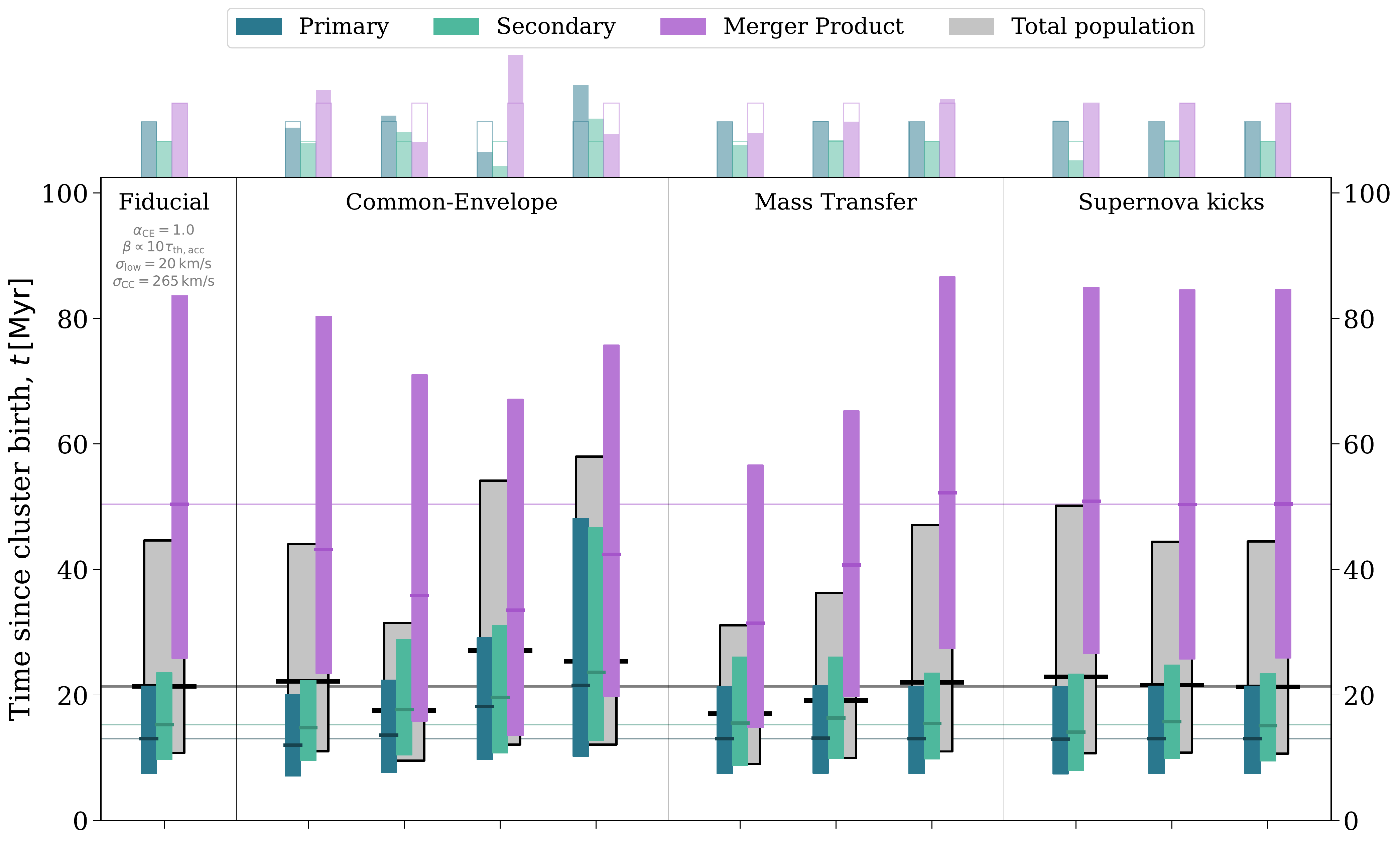}
    \includegraphics[width=0.9\textwidth]{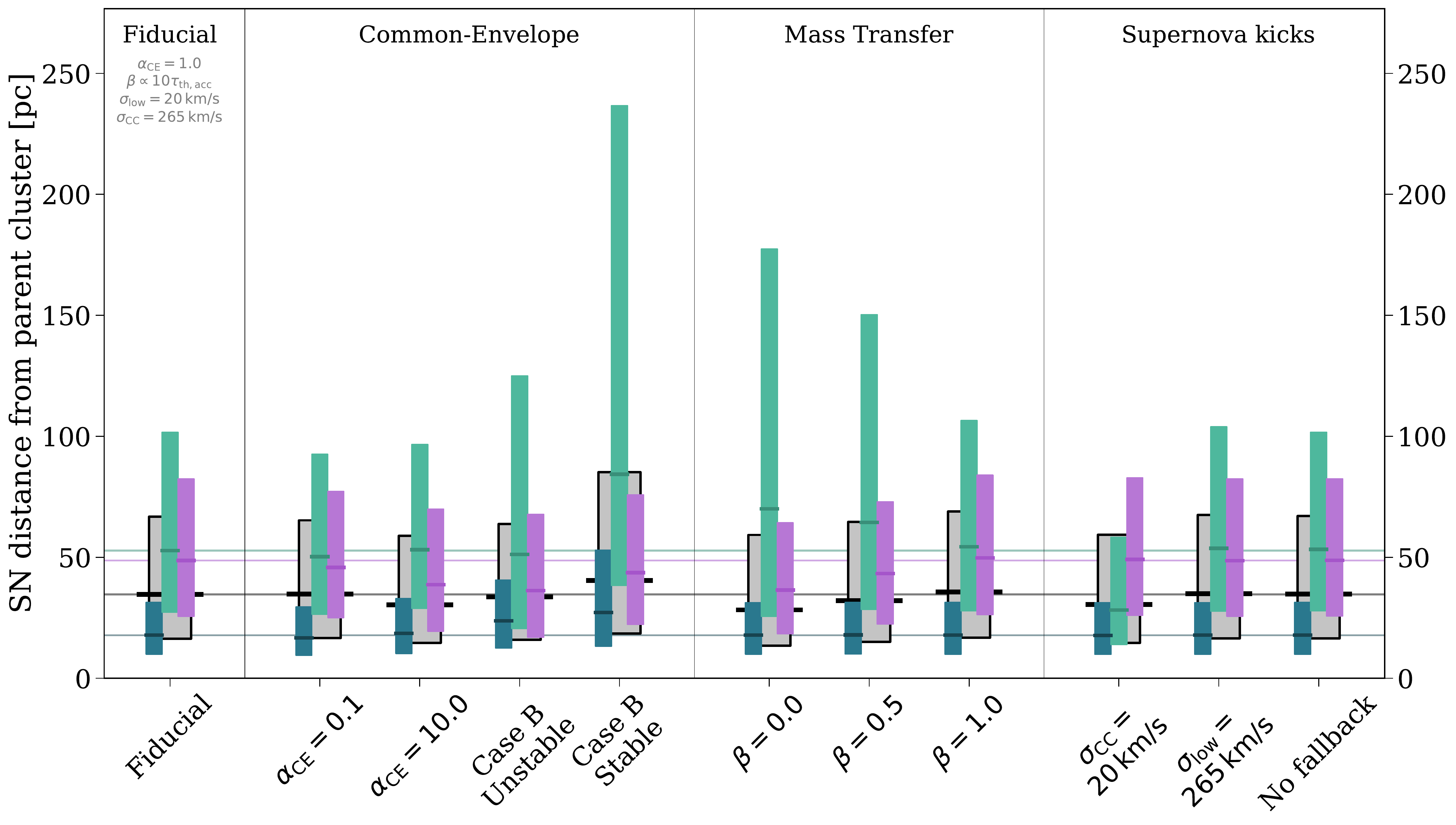}
    \caption{Comparison of the impact of binary physics variations on the timing (upper panel) and location (lower panel) of SNe. Each group of bars in the main panels corresponds to a different choice of binary physics. Coloured bars show the interquartile range for each subpopulation (labelled in legend) with darker lines at the median. The grey bars show the same information for the overall distribution. Full distributions for the fiducial model are shown in Figure~\ref{fig:sn-times-dists}. Thin horizontal lines indicate the fiducial median values for comparison. Fiducial parameter are annotated in the upper left corner. The small bars at the top indicate the number of SNe per $100\unit{M_\odot}$, with the fiducial values shown in a thin outline. Section~\ref{sec:variation-plot-explained} explains how to interpret this figure (and Figures~\ref{fig:var-ini}--\ref{fig:var-gal}) in further detail. (\href{https://www.tomwagg.com/html/interact/binary-supernova-feedback.html\#fig4-7}{\faLaptopCode{} Interactive figure available.})}
    \label{fig:var-bin}
\end{figure*}

In Figures~\ref{fig:var-bin}--\ref{fig:var-gal} we show comparisons of the distribution of SN times and locations across an array of parameter variations. Rather than reproducing the full distributions in Figure~\ref{fig:sn-times-dists} for every parameter value, we have chosen to make condensed summary plots that characterize the statistical properties of the distributions. We use vertical bars to indicate the interquartile ranges and medians (horizontal marks within the bars) of supernova timing (top panel) and distance from the parent cluster centre at the time of explosion (bottom panel) for the primary (blue), secondary (green), merger products (purple), and total (grey) populations.

We note that changes in the distributions can come from a combination of (a) a shifting of the times and locations of the original population in the fiducial model or (b) the creation or elimination of subpopulations which cause shifts. Therefore, we include bars at the top of each figure that indicate the total number of SNe that occur per $100\unit{M_\odot}$ of star formation. These can be compared to the thin outlined bars, which indicate the fiducial values.

In every plot, the fiducial model is reproduced in the leftmost column for reference, with the relevant fiducial settings annotated at the top of the column. Its median values are also propagated throughout the plot as thin horizontal lines. The parameters that are being varied are shown along the bottom of the plot, and are grouped by the physics that the parameters control. These parameter variations may have significant impacts on the distributions of particular evolutionary pathways (i.e., indicated by moving the coloured bars up or down, and/or changing their relative widths), which can be a useful diagnostic of the controlling physics.

In addition to these plots, throughout the following sections we will use a series of summary statistics to show how variations change the tails of our distributions. These are summarised in Table~\ref{tab:totals_and_stats} and displayed in Figure~\ref{fig:trends-tails}, where \flate is the fraction of SNe that occur ``late'' (beyond $44\unit{Myr}$, the \fire-3 cutoff for type II SN feedback), and \ffar and \fdistant are the fractions of SNe that occur more than $100 \unit{pc}$ and $500 \unit{pc}$ from their parent clusters respectively.

In Appendix~\ref{app:supple}, we include tables of data for reproducing the total population distributions from the comparison plots. Table~\ref{tab:percentiles} details the percentiles of both the timing and distance distributions, whilst Table~\ref{tab:totals_and_stats} contains the total number of SNe in each subpopulation in each variation.

\subsection{Binary physics variation trends}

In Figure~\ref{fig:var-bin}, we focus on the impact of varying parameters that directly control binary evolution -- common-envelope evolution, the efficiency of mass transfer, and the strength of SNe kicks. As described in Section~\ref{sec:variation-descriptions}, we vary these parameters over a wide range of possible values, many of which are quite extreme.

\subsubsection{Common-envelope efficiency}

Approximately 35\% of stars that reach core collapse experienced a common-envelope event during their evolution in our fiducial model. Therefore changing the efficiency of this phase can impact a large fraction of our population.

Figure~\ref{fig:var-bin} shows that decreasing the efficiency of common-envelopes (i.e., lower values of $\ace$), results in more stellar mergers. A lower value of $\ace$ reduces the fraction of orbital energy that is available to unbind the envelope and so stars need to get much closer before releasing enough energy to unbind the common envelope, making a merger more likely. This change is also notable in the bar charts on top of Figure~\ref{fig:var-bin}, which show that the number of mergers increase by $18\%$ for $\ace = 0.1$, such that their overall contribution to the total SNe population is $50\%$. A subset of these additional mergers come from stars that would have reached core collapse alone, and hence the number of primary and secondary SNe each decrease slightly by ${\sim}5\%$.

The increase in stellar mergers at lower $\ace$ reduces the overall median of the SN time distribution for merger products. The median decreases because more higher mass stars, which would have exploded as primary or secondary SNe in the fiducial model, instead merge and evolve to core collapse faster than an average merger product. Nevertheless, the overall distribution of times is relatively unaffected, because merger products represent a larger fraction of the population. Therefore, though the average merger product SNe is earlier, the average SNe of \textit{any} progenitor is relatively unchanged. Similarly, the total distance distribution is not significantly affected, though merger distances slightly decrease as a result of decreased times to core collapse giving them less time to disperse.

Following the same trend as lower $\ace$, the variation with a higher value of $\ace$ decreases the number of merger product SNe by $53\%$, such that merger products produce only $24\%$ of all SNe in the $\ace = 10.0$ variation. However, counterintuitively this variation \textit{also} reduces the average merger product SNe time even more significantly than the $\ace = 0.1$ variation. In this case, the increased efficiency allows more stars to avoid a merger during a common-envelope. However, unlike in the $\ace = 0.1$, these stars don't shift to a different progenitor channel, but instead often avoid core collapse entirely. In this way a large fraction of the subpopulation that occupy the late SN tail is removed, shifting the distribution to smaller times, such that $\flate = 14\%$ in this case. Additionally, the distance distribution decreases because there are fewer late SNe that have longer to disperse from their cluster.

\subsubsection{Case B mass transfer stability}

In our fiducial model, ${\sim}85\%$ of stars that reach core collapse have are involved in Roche-lobe overflow, either as a donor or an accretor. Of these stars, ${\sim}66\%$ experience case B mass transfer (i.e.,\ the subset of Roche-lobe overflow events that are initiated during expansion on the Hertzsprung gap). Given that ``case B'' is the most common type of mass transfer, we explore how changing our criteria for its stability affects our results. These variations are shown in the common-envelope section of Figure~\ref{fig:var-bin} for the two extremes of case B behaviour.

At one extreme, when all case B mass transfer is unstable, $\sim$75\% SN progenitors undergo a merger prior to core collapse. This increase in mergers is a result of failed common-envelope events in which the binary fails to eject its envelope and instead merges. Furthermore, the same process leads a greater fraction of merger product SNe to come from higher mass stars with shorter evolutionary timescales, which in turn decreases the typical SN times for merger products, compared to the fiducial model (in which these stars would have avoided merging).

While the evolution of merger products is accelerated with unstable case B mass transfer, the \textit{overall} median SN delay time of the population as a whole has the opposite behavior, and significantly increases to longer times. Mergers become more common, making the tail of late-time (${>}\,44\unit{Myr}$) SNe that they produce even more prominent (see Figure~\ref{fig:sn-times-dists}). Thus, despite the lower average time of the merger product subpopulation, the median time of the \textit{total} population increases by ${\sim}30\%$, such that $\flate = 31\%$, due to merged SNe progenitors representing a much greater fraction of the population. The secondary distance distribution also appears to show a slightly extended tail, but we caution that this distribution is very weakly populated (as shown in the bars at the top of the figure) and thus not well characterized or important to the overall behaviour of the population.

In the other extreme, we consider models that force all case B mass transfer to be stable. This change \textit{also} increases the average core-collapse SN delay time, but for different reasons. In this variation, the number of merger product SNe decreases significantly because many binaries avoid common-envelopes entirely. These binaries instead reach core collapse as primary or secondary star progenitors. As a result, many low-mass stars are able to stably accrete enough matter from their companions to reach core collapse without requiring a merger. This means that the times of primary and secondary SNe are significantly later in this variation, shifting the overall SNe distribution to later times.

Another feature of the ``always stable'' case B variation is that it shows the strongest increase in SN distances among all the binary physics variations that we considered, particularly for the secondary SNe. The additional population of distant secondary SNe are a result of low-mass stars that could not have reached core collapse without accretion. The stable mass transfer variation allows these stars to avoid mergers and thus remain in binaries, enabling them to later be ejected at high velocities when the primary star explodes. Furthermore, these no-longer-merging stars are primarily lower mass, which means that they travel further before exploding and hence increase the tail of the distribution. Overall, this increases the fractions of distant SNe to $\ffar = 21.1\%$ and $\fdistant = 2.9\%$, which are $1.6$x and $3$x the fiducial values respectively.

\subsubsection{Mass transfer efficiency}

The efficiency of mass transfer, $\beta \equiv \Delta M_{\rm acc} /\Delta M_{\rm donor}$, is defined as the fraction of material transferred by the donor ($\Delta M_{\rm donor}$) which is successfully accreted by the companion star ($\Delta M_{\rm acc})$, rather than being removed by winds. In our fiducial model, the efficiency is not set directly, but instead dictated by the thermal timescale of the accretor, such that the maximum accretion rate is given by
\begin{equation}
    \dot{M}_{\rm acc, max} = \begin{cases} 10 \frac{M_{\rm acc}}{\tau_{\rm KH}} & \text{Has radiative envelope},\\\infty & \text{else}, \end{cases}
\end{equation}
where $M_{\rm acc}$ is the mass of the accretor, $\tau_{\rm KH}$ is the Kelvin-Helmholtz timescale of the accretor star and this limit is only applied for stars with radiative envelopes \citep{Hurley+2002:2002MNRAS.329..897H}. We calculate the fraction of mass that's accreted in each instance of stable mass transfer in our fiducial model and find that this criterion results in an average of $\beta \approx 0.77$. We show three variations using fixed values for $\beta$ in Figure~\ref{fig:var-bin}.

The distributions of primary stars are almost entirely unaffected by these variations. Changes in $\beta$ only affect the amount of mass accreted, not the mass lost. Therefore, primary stars would only be affected by these variations in a rare case in which the accretion went the opposite way during a late phase of a binary's evolution, such that a primary star accreted a significant amount mass from an evolving companion before the primary's core mass is determined.

We do see overall changes in the timing of SN from variations in mass transfer efficiency. There is a clear trend that a lower value of $\beta$ results in a weaker tail of SN at late times (see Figure~\ref{fig:trends-tails}), primarily because there are fewer late merger-product SNe. In our fiducial model, these late SNe most often follow a formation pathway in which a merger occurs after unstable mass transfer from a secondary star onto a primary star \citep[e.g.,][]{Zapartas+2017:2017AA...601A..29Z}. A lower mass transfer efficiency results in a less massive secondary star and therefore fewer scenarios in which a secondary star expands sufficiently during its evolution to initiate a common-envelope at close separations. As a result, the fraction of merger product SNe from this pathway decreases from $50\%$, to $35\%$ to $17\%$ for $\beta = 1.0$ to $0.5$ to $0.0$, thus reducing the late SN rate. 

The distance distributions show two main trends: an overall decrease in the median distance of SNe, but an increase in the distances that secondary stars travel. The overall decrease follows from the same reasoning as above, fewer late merger product SNe mean a larger fraction of the SNe happen close to their parent cluster. The increase in secondary distances for low $\beta$ occurs for two reasons.
First, lower $\beta$ values mean secondary stars accrete less material, potentially delaying their SN explosions and thus giving them more time to travel after their primary explodes. Second, more angular momentum is lost from the system for lower $\beta$ values, and thus binary separations tend to be tighter prior to their disruption, leading them to be ejected with higher velocities.  Quantitatively, we find that the average ejection velocity for secondary stars changes from $14.6 \unit{km}{s^{-1}}$ to $16.5 \unit{km}{s^{-1}}$ to $22.1 \unit{km}{s^{-1}}$ as $\beta$ decreases from $1.0$ to $0.5$ to $0.0$. At the same time, secondary stars on average live for an additional $1.5\unit{Myr}$ after ejection from a binary when $\beta = 0.0$ compared to $\beta = 1.0$. For these reasons, secondary stars tend to reach core collapse further from their parent clusters when $\beta$ is lower.

\subsubsection{Supernova natal kick magnitude}

The magnitude of the SN natal kicks primarily alter our distributions by changing the fraction of binaries that are disrupted. Even in our fiducial model ${\sim}85\%$ of binaries are disrupted by their first SN, in agreement with earlier works \citep{DeDonder+2003:2003NewA....8..817D,Eldridge+2011:2011MNRAS.414.3501E,Renzo+2019:2019A&A...624A..66R}. We emphasise that, for natal kicks that already disrupt a binary, increasing the magnitude of the kick has almost no effect on the ejection velocity of secondary stars \citep[e.g.,][]{ejection_vels_rnaas}. The variations that increase the strength of the kicks only disrupt an additional 5\% of binaries, hence they have negligible effects on our distributions, as shown in the supernova kicks section of Figure~\ref{fig:var-bin}.

While the changes in the overall distributions do not change dramatically with variations in the SNe kick magnitudes, there are some subtle variations that are worth explaining. Most of these effects will be driven by the population of secondary stars; primary stars and merger products are unaffected as both cases represent the first SN in a binary system and hence their prior evolution has not been affected by SN kicks.

The first place one might expect to see the effects of SNe kicks are in the distance distributions, which are indeed skewed to lower values when using weaker core collapse SN kicks. However, the origin of this effect is not due to directly reducing the ejection velocities of secondary stars, as these velocities are entirely set by the pre-supernova orbital velocity \citep{ejection_vels_rnaas}. Instead, the reduction in the typical cluster distance is due to having fewer ejected secondary stars. The number of secondary stars that reach core collapse in this variation decreases by a factor of 2 compared to the fiducial, such that the total number of SNe decreases by ${\sim}12\%$. The decrease in secondary SNe is because the weaker primary SN kicks disrupt only ${\sim}35\%$ of binaries, compared to $85\%$ in the fiducial model. These weaker kicks cannot unbind tight binaries, such that the only ejected secondaries come from the subset of weakly-bound binaries that have wide orbits with lower orbital velocities.

In contrast, SN delay times are slightly later in the variation in which core-collapse SN natal kicks are weaker. This again is the result of a much smaller fraction of binaries being disrupted. The surviving binaries can then often experience a subsequent unstable mass transfer event, which leads to a merger between the compact object and secondary star and prevents a subsequent core-collapse SN. The decrease in the number of secondary SNe means that the SN times are generally later because merger products represent a greater fraction of the population.

\subsubsection{Summary of binary physics trends}

The story told by Figure~\ref{fig:var-bin} is that the expected distributions of SNe time delays and distances from cluster centres are surprisingly robust to even extreme variations in binary physics. Focusing only on the medians of the distributions (dark horizontal lines across the grey bars), there is not much change across the entire plot, for either the timing or distance distributions; these vary by no more than 25\% and 18\%, respectively.

The effects of the variations are more visible in the tails (Figure~\ref{fig:trends-tails}), and in the number and type of SNe precursors.  However, in all cases there is a long tail of late SNe from merger products, which is not currently accounted for in hydrodynamical simulations, and a long tail of distant SNe from secondary stars, which is particularly sensitive to changes to the stability of mass transfer.

\subsection{Initial condition variation trends}

\begin{figure}
    \centering
    \includegraphics[width=0.5\textwidth]{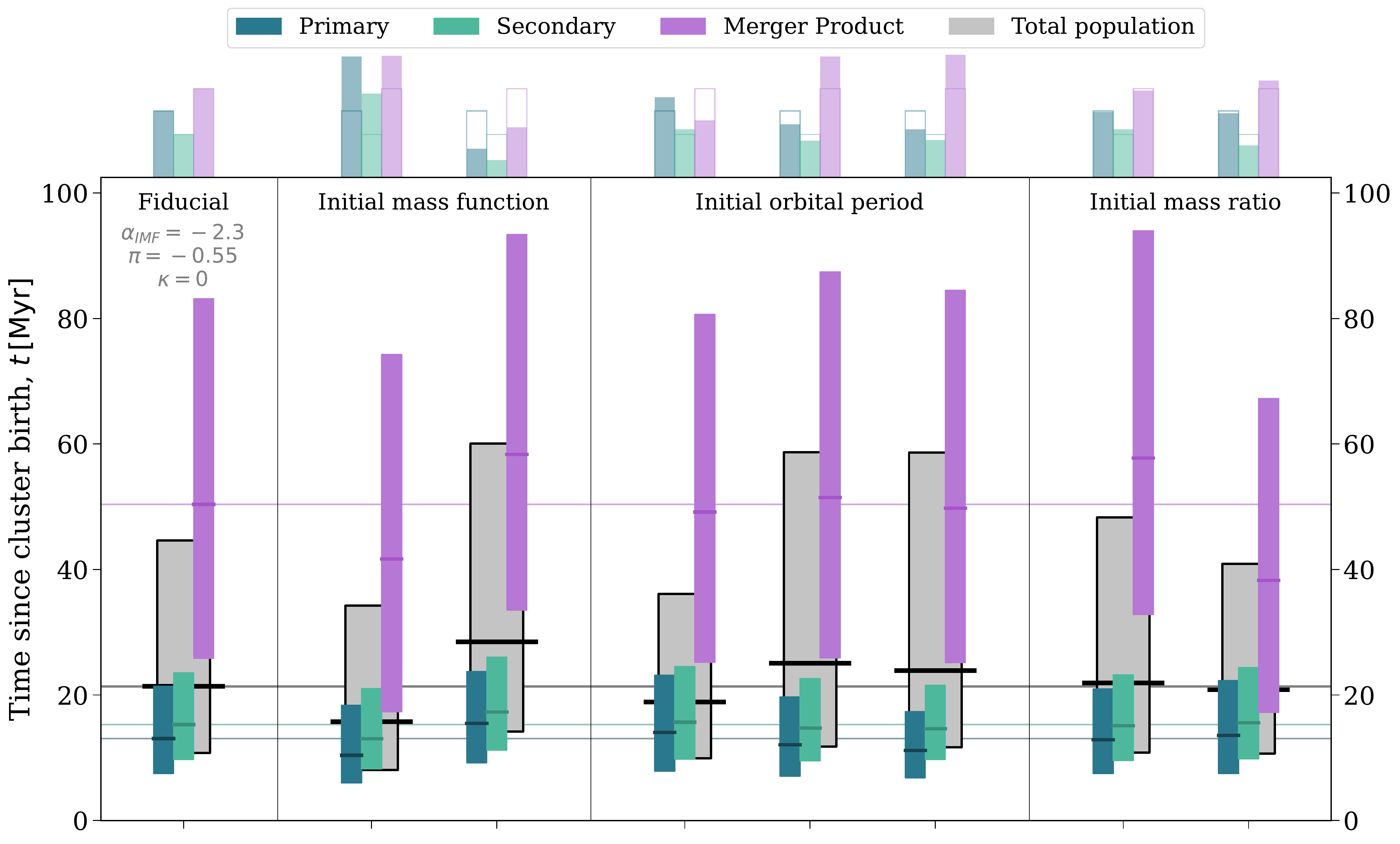}
    \includegraphics[width=0.5\textwidth]{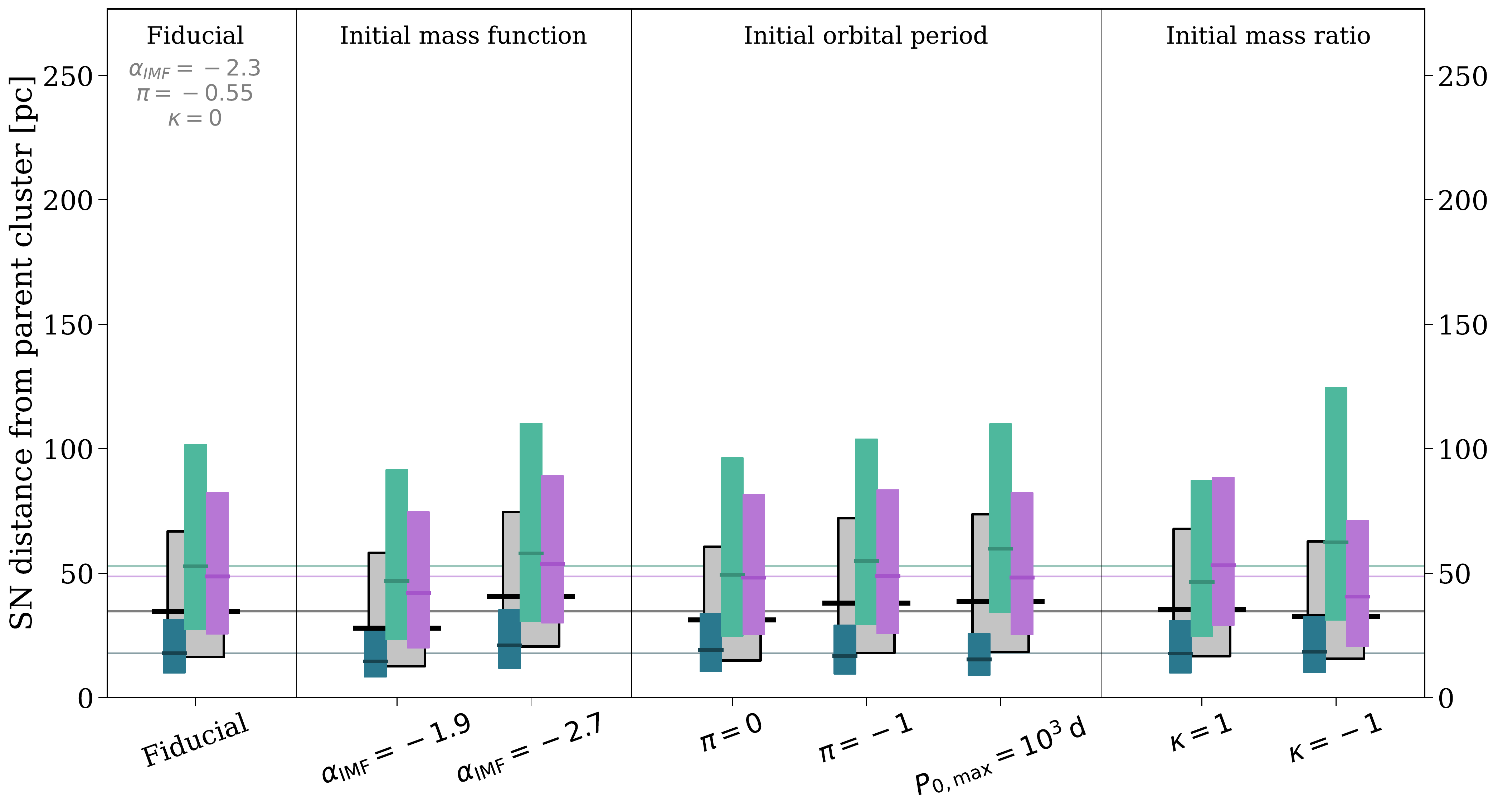}
    \caption{As Figure~\ref{fig:var-bin}, except varying different choices of initial conditions. Variations of the initial mass function and initial orbital period distribution can affect SNe timing, but initial conditions do not strongly impact SN spatial locations. (\href{https://www.tomwagg.com/html/interact/binary-supernova-feedback.html\#fig4-7}{\faLaptopCode{} Interactive figure available.})}
    \label{fig:var-ini}
\end{figure}

The initial conditions of binary stellar populations, such as the initial primary mass, mass ratio and orbital period distributions, are not strongly constrained and may differ from our fiducial assumptions. In this section, we test the sensitivity of our results to variations in these distributions and display the results in Figure~\ref{fig:var-ini}. 

\subsubsection{Initial mass function}

We consider two variations of the slope, $\alpha_{\rm IMF}$, of the high mass end of the IMF (for primary masses of $m_1 > 1\msun$). The $\alpha_{\rm IMF} = -1.9$ variation is a top-heavy IMF that favours more massive stars, while $\alpha_{\rm IMF} = -2.7$ is contrastingly a bottom-heavy IMF; the optimal value remains a major observational and theoretical uncertainty in any calculation of SNe rates. We keep the low mass ($m_1 < 1\msun$) IMF the same as the Kroupa IMF, while ensuring continuity at $m_1$. The impact of these variations are shown in Figure~\ref{fig:var-ini}.

As expected for any model (binary or otherwise), the overall total number of SNe is strongly affected by the IMF \citep{deMink+2015:2015ApJ...814...58D, Zapartas+2017:2017AA...601A..29Z}. Varying the IMF slope has a first-order effect simply through changing the fraction of stars that have high enough masses to be candidates for eventual core collapse. Here, the top-heavy IMF variation produces significantly more massive stars, resulting in twice as many primary and secondary SNe in our simulations. 

There is a more nuanced effect on the number of merger product SNe, which increases by a lesser extent, of only 40\% rather than a factor of 2. Many of these merger product SNe are formed from binaries in which the primary star is not massive enough to reach core collapse alone, but, these binaries constitute a lower fraction of the population in a top-heavy IMF. Quantitatively, roughly half (52\%) of the stars in our fiducial Kroupa IMF simulation have $m_1 < 7 \msun$, whilst for the top-heavy IMF the same fraction drops to only $35\%$ (noting that we only simulate primary masses of $m_1 > 4 \msun$; see Section~\ref{sec:fiducial}). This shift to higher masses reduces the fraction of the population that requires a merger in order to reach core collapse. 

All of the trends above are reversed for the bottom-heavy IMF, with overall decreases in the total SNe rate that are more drastic for primary and secondary SNe (a 55\% decrease) and less so for merger products (a 40\% decrease).

Although a more top-heavy IMF produces a greater \textit{number} of SNe, it decreases the fraction of these that occur at late times. The SN times of each progenitor type decrease for the top-heavy $\alpha_{\rm IMF} = -1.9$ case, because more stars form at high masses and reach core collapse in a shorter time. As such, only $\flate = 17.9\%$ of SNe occur after $44\unit{Myr}$ in the case of a top-heavy IMF. However, 
though this variation has a lower \textit{fraction} of the late SNe, there are still ${\sim}17\%$ more late SNe than the fiducial model because so many more stars reach core collapse. In contrast, the bottom-heavy IMF variation has the one of the largest fractions of late SNe of all variations we consider ($\flate = 32.9\%$), but with a total absolute rate that is only $64\%$ of the fiducial model.

These changes to the timing distribution lead to slight variations in the distances that stars travel before reaching core collapse. A top-heavy IMF results in a slightly more concentrated distribution of SNe, whilst a bottom-heavy IMF has a slight increase to all distances.

\subsubsection{Initial orbital period}\label{sec:ini_orb_period}

We explore the impact of changing our distribution of initial orbital periods in two variations from our fiducial model, which uses a power law slope of $\pi = -0.55$ (see Section~\ref{sec:fiducial}). In the $\pi = 0$ simulation, binaries are generally placed at wider orbital periods, whilst in $\pi = -1$ they are concentrated on closer initial orbits. These wide- and tight-orbit variations are shown in the third section of Figure~\ref{fig:var-ini}.

The strongest impact of these variations is to change the fraction of SNe from merger products. Binaries with tighter initial orbital periods are more likely to merge during their evolution. For the wide-binary case ($\pi = 0$) there are ${\sim}54\%$ fewer merger product SNe than the fiducial model with $\pi=-0.55$, whilst for $\pi = -1$ the total increases by ${\sim}38\%$.

The merger fraction is the dominant driver of late-time SNe, and thus the increase in mergers for the tight-binary $\pi = -1$ case leads directly to a more prominent tail to late times. In fact, the $\pi = -1$ variation in particular produces the second-highest\footnote{We note that the overall \textit{rate} of late SNe is highest in this variation, because the bottom-heavy IMF suppresses SNe.} fraction of late SNe of all of our variations, with $\flate = 32.7\%$, in agreement with rates found by \citet{Zapartas+2017:2017AA...601A..29Z}). In contrast, the overall distribution of SN times for the wide-binary $\pi = 0$ case decreases the late-time tail significantly.

Besides the change in merger frequencies, initially 
wider
binaries result in slower ejection velocities and thus less distant SNe. In the wide-binary ($\pi = 0$) variation, orbital speeds are typically slower, 
reducing the average ejection velocity of secondary stars when the primary star explodes. The lower ejection velocities result in slightly fewer distant SNe in this variation ($\ffar = 11.6\%$) compared to our fiducial model ($\ffar = 13.4\%$) and the tight-binary $\pi = -1$ variation ($\ffar =14.7\%$).

We additionally include a variation in which we reduce the upper limit on initial orbital periods to $P_{0,{\rm max}}=10^3\unit{days}$. This lower limit suppresses the number of wide, effectively-single binaries. Although there is no reason for there to be a strict upper limit at this period \citep[e.g.,][]{deMink+2015:2015ApJ...814...58D}, this stricter limit is more typical of early work on binary populations. Similar to the $\pi = -1$ variation, this skews the distribution of initial binaries to tighter orbits. For the same reasons as $\pi = -1$, this produces a high fraction of late SNe ($\flate = 33.3\%$) and moderately high fraction of distant SNe ($\ffar = 15.3\%$).

\subsubsection{Initial mass ratio}

We change the power-law slope, $\kappa$, of our initial mass-ratio distribution in two variations. The $\kappa = -1$ variation produces more unequal-mass binaries, whilst $\kappa = 1$ favours more equal-mass binaries. These variations are shown in the final section of Figure~\ref{fig:var-ini}.

Merger product SNe generally occur earlier in the variation with more unequal mass ratios ($\kappa = -1$). Generally, this timescale is set by the lifetime of the (more massive) primary star, for the subset of ``forward mergers'' --- those occurring when the primary star overflows its Roche lobe). More unequal mass ratios tend to produce more forward mergers ($73\%$ of mergers for $\kappa = -1$, compared to $52\%$ in our fiducial model), because the larger mass difference reduces the chance that both the primary and secondary will be evolving significantly at the same time. As a result, the times of merger product SNe are generally earlier in this variation. The inverse reasoning applies to the $\kappa = 1$ variation and explains its later merger times.

The earlier merger SNe times for more unequal mass ratios ($\kappa = -1$) also shifts the overall distance distribution to lower values, because those SNe have less time to disperse from the cluster. Since merger products represent the largest progenitor population (see histograms in the top row of Figure~\ref{fig:var-ini}), the overall distance distribution is skewed to lower values.

While the changes in merger-progenitor SNe times shifts the overall distribution to shorter times and smaller separations when the initial mass-ratio is more uneven, there is an additional (albeit smaller) change driven by changes to the secondaries, which have larger distance separations on average. The increased distances stem from secondary stars in binaries with more unequal mass ratios taking longer to evolve than their primary star. This delay means that there is more time between the disruption of a binary (after the primary SN) and the eventual secondary SN, allowing the ejected secondary companion star to travel further before exploding. In particular, we find that secondary stars in disrupted binaries travel for $2.5\unit{Myr}$ longer before exploding in the $\kappa = -1$ variation than in the $\kappa = 1$ variation. This means that though $\ffar$ slightly decreases to $11.9\%$ (from $13.4\%$) in the $\kappa = -1$ variation, $\fdistant$ increases to $1.2\%$ (from $0.9\%$).

\subsection{Metallicity}\label{sec:metallicity_variation}

\begin{figure}
    \centering
    \includegraphics[width=0.5\textwidth]{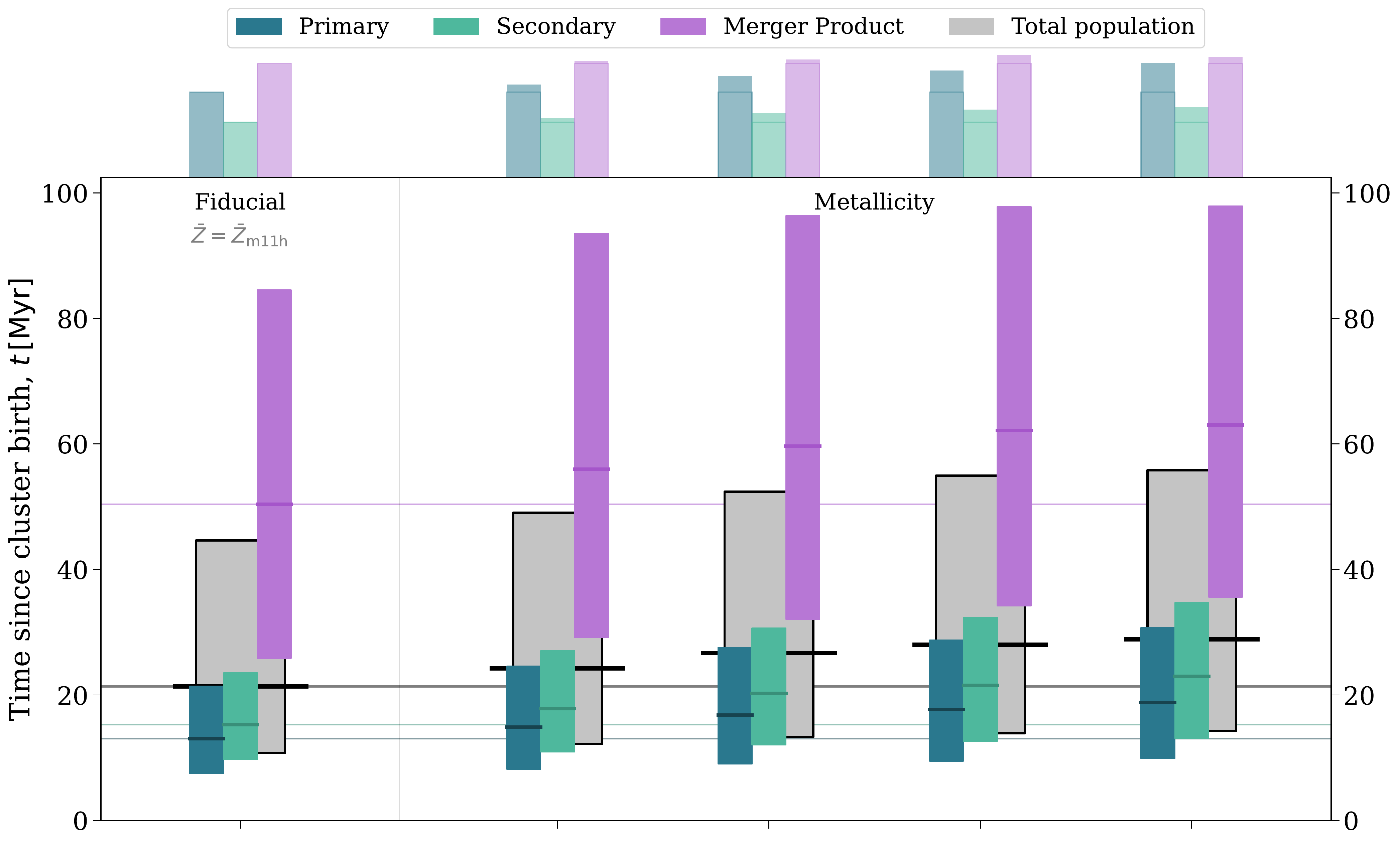}
    \includegraphics[width=0.5\textwidth]{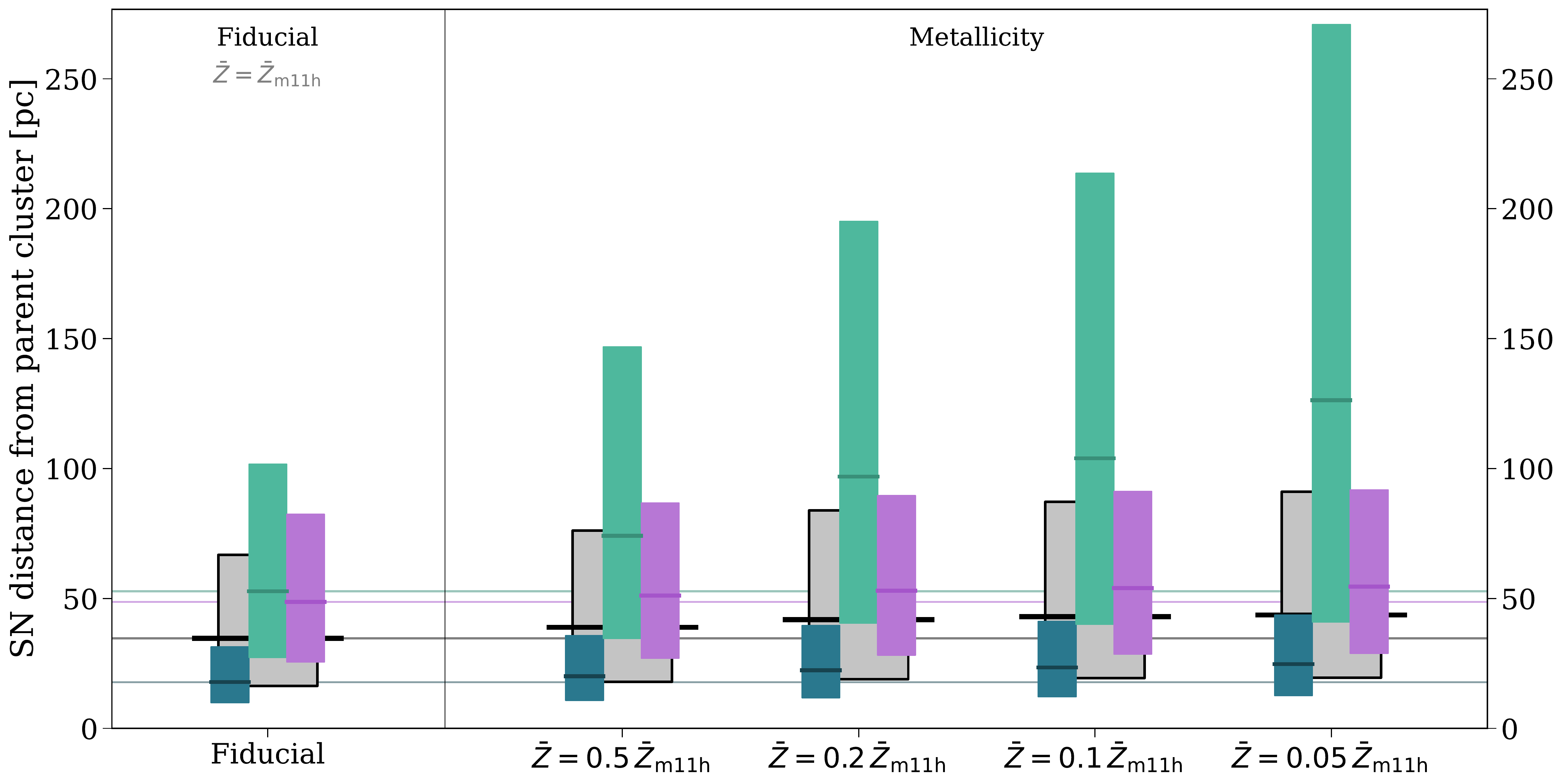}
    \caption{As Figure~\ref{fig:var-bin}, except varying average population metallicity. Decreasing the metallicity of simulated binary stars leads to a systematic increase in both SNe times and locations. (\href{https://www.tomwagg.com/html/interact/binary-supernova-feedback.html\#fig4-7}{\faLaptopCode{} Interactive figure available.})}
    \label{fig:var-z}
\end{figure}

In our fiducial model, there is an intrinsic distribution of stellar metallicities, set by the specific history of star formation and gas flows within a single model galaxy. However, metallicity is a well-known driver of stellar evolution, changing the internal structure, radius, and time-evolution of single stars, all of which will have downstream effects when those stars are evolving within binaries. We therefore now consider the impact of metallicity variations. 

Our fiducial simulation has a relatively narrow, near-solar metallicity distribution ($\bar{Z}_{\rm m11h}=0.017 \pm 0.001$, where the uncertainty indicates the interquartile range). We explore the impact of metallicity by decreasing the metallicity of each binary in the fiducial model simulation by a constant factor in a series of variations. This process leaves the overall shape of the metallicity distribution unchanged, while decreasing the average metallicity, $\bar{Z}$. We not explore higher metallicity variations as the highest metallicity that can be evolved in \cosmic is $Z = 0.03$ \citep{Breivik2020} and \texttt{m11h} already has star particles with $Z = 0.0223$, so any significant increase would require changing the overall metallicity distribution to avoid exceeding $Z = 0.03$. The resulting SNe distributions for these metallicity variations are shown in Figure~\ref{fig:var-z}, for a variety of increasingly sub-solar metallicities.

As metallicity decreases, the amount of feedback increases. In the $Z = 0.1 Z_{\rm m11h}$ variation, the total number of SNe increases by ${\sim}17\%$ compared to the fiducial near-solar model (as shown by the top histograms in Figure~\ref{fig:var-z}). At lower metallicities, opacity and the strength of wind mass loss is decreased. As a result, the lower limit for the initial mass of a single star that can reach core collapse is decreased \citep{Pols+1998:1998MNRAS.298..525P}. We find that the lowest mass star that accreted no mass from its companion and still reached core collapse is ${\sim}7.0\msun$ in our fiducial model, but ${\sim}6.2\msun$ in the $Z = 0.1 Z_{\rm m11h}$ variation. The ${\sim}17\%$ increase in SNe is therefore exactly as expected from integrating the initial mass function from this decreased lower mass limit.

The addition of this subpopulation of lower mass stars increases the times at which all subtypes of SNe occur. The main sequence lifetimes of these stars are longer than the rest of the population and thus their addition skews the average SN time to later values. For the lowest metallicity variation $\flate = 34.4\%$, producing the largest fraction of late SNe across all variations we consider.

The addition of more late SNe with decreasing metallicity also increases the overall SN distance distribution for each subtype. While some fraction of this effect is simply allowing more time for cluster dissolution, a particularly dramatic change is in the behaviour of SNe from secondary stars.

Secondary stars have strongly metallicity-dependent evolution, due to changes in 
the radial expansion of stars, which is decreased at low metallicities \citep[e.g.,][]{Brunish+1982:1982ApJS...49..447B, Xin+2022:2022MNRAS.516.5816X, Klencki+2022:2022A&A...662A..56K}. In our simulations, the median of primary stars' maximum radial extent decreases from ${\sim}1200\rsun$ in our fiducial model, to ${\sim}350\rsun$ in the lowest metallicity variation. The smaller stellar radius allows binaries to detach after Roche-lobe overflow at much smaller separations. As a result, binaries are typically tighter prior to the primary SN and thus the ejection velocities of secondary stars are increased, in agreement with \citet{Renzo+2019:2019A&A...624A..66R}. In our models the average ejection velocity of secondaries from disrupted binaries increases from $15 \unit{km}{s^{-1}}$ in the fiducial model to $27 \unit{km}{s^{-1}}$ at the lowest metallicity. As a result, the very low metallicity $\bar{Z} = 0.05 \bar{Z}_{\rm m11h}$ variation has one of the strongest distance SNe tails, with $\ffar = 22.3\%$ and $\fdistant = 2.4\%$ (see Figure~\ref{fig:trends-tails}).

While the metallicity clearly has a significant effect on the tails of the distance and age distributions, the bulk of the distribution experiences actually much more modest changes. The median age and distance change by only 34\% and 25\%, respectively, at the most extreme variations. The overall numbers of SNe also only changes by no more than 20\% (see Figure~\ref{fig:trends-totals}). 

\subsection{Galaxy parameter variation trends}

\begin{figure}
    \centering
    \includegraphics[width=0.5\textwidth]{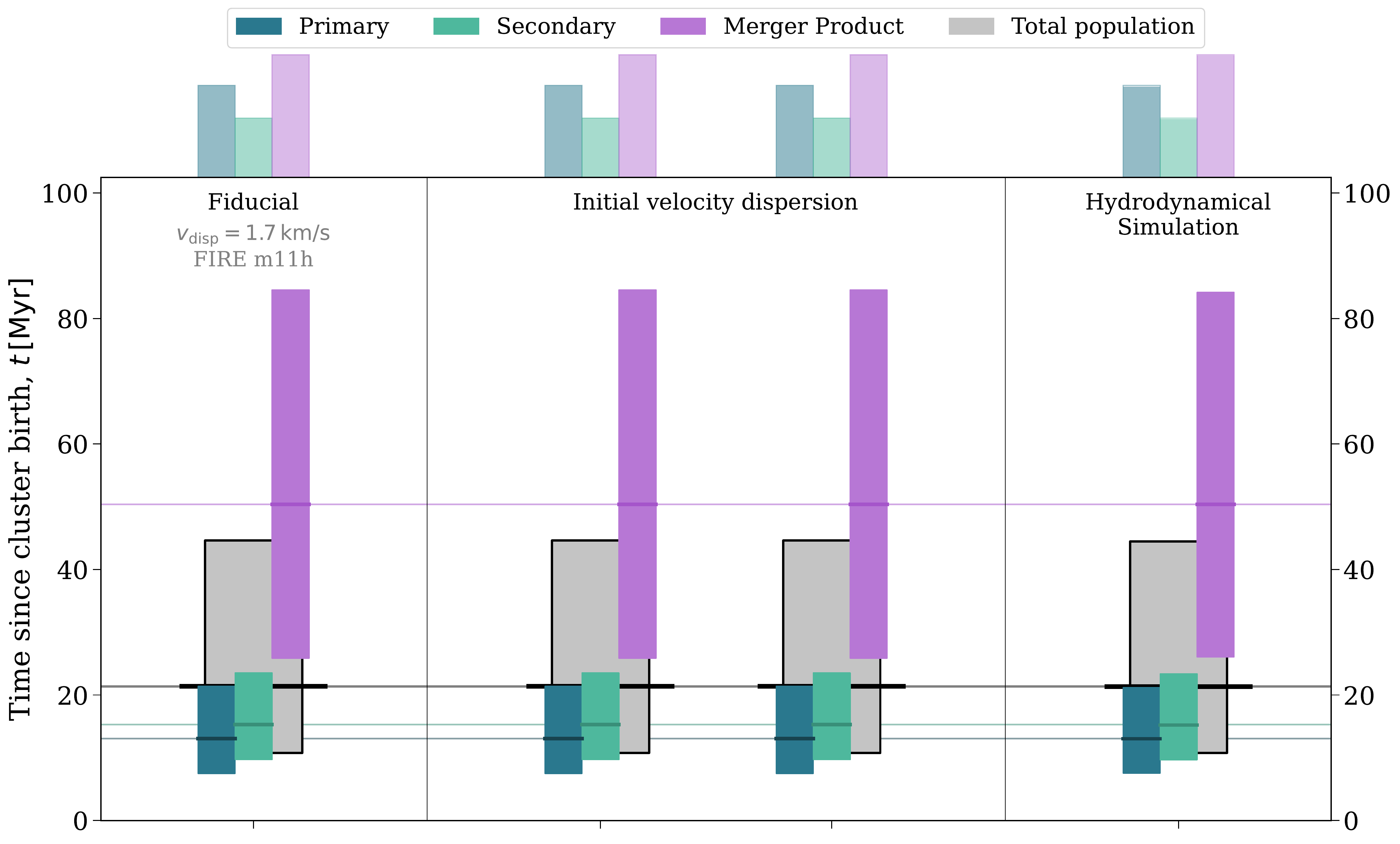}
    \includegraphics[width=0.5\textwidth]{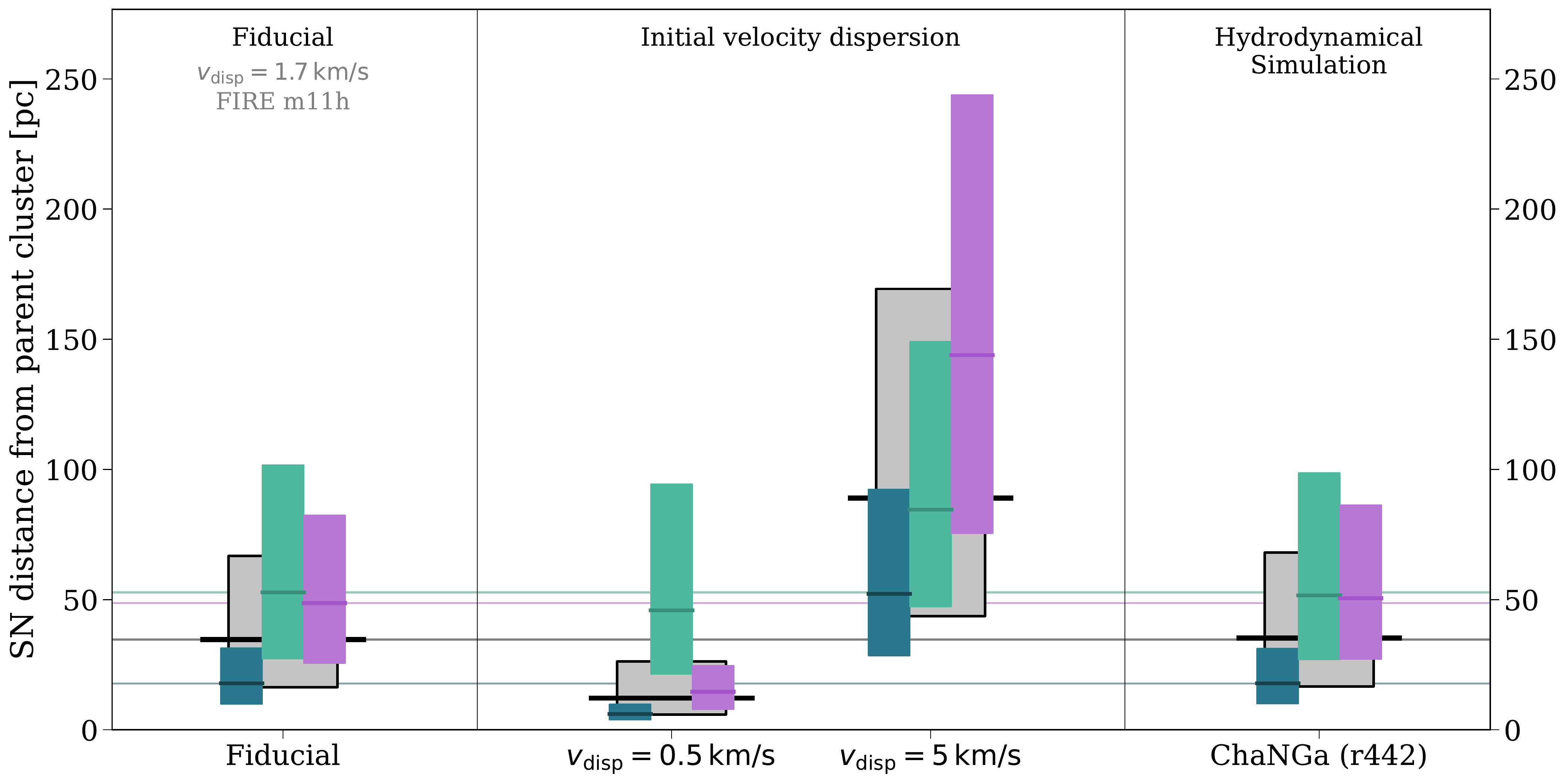}
    \caption{As Figure~\ref{fig:var-bin}, except varying different choices of galaxy settings. SNe timing is unaffected by galaxy settings, but SNe location is strongly dependent on the choice of initial cluster velocity dispersion (though a tail of distant ejected secondary SNe still exists with a low dispersion).\\ (\href{https://www.tomwagg.com/html/interact/binary-supernova-feedback.html\#fig4-7}{\faLaptopCode{} Interactive figure available.})}
    \label{fig:var-gal}
\end{figure}

There are two aspects of the work above that rest on assumptions we have made in connecting the \cogsworth infrastructure to a simulated galaxy. The first is how individual star particles in the simulation are turned into a cluster of binary stars. The chosen statistics of the binary populations are discussed above, but the assumed velocity dispersion of the stellar population is also a factor in setting the eventual orbital diffusion of the stars. The second aspect to be considered is the choice of simulated galaxy. Different evolutionary pathways can lead to different metallicities, whose impacts are discussed in Section~\ref{sec:metallicity_variation}. However, there are other potential impacts on the orbital evolution of young stars due to varying the gravitational potential and the spatial distribution of young stars. 

In this section we consider the impacts of both of these effects, by varying the assumed initial velocity dispersion of the stellar clusters that replace star particles, and by changing the galaxy simulation used for our calculations.

\subsubsection{Cluster initial velocity dispersion}

Observational evidence suggests that nearly all young stellar clusters become unbound after expulsion of their natal gas cloud \citep[e.g.,][and references therein]{Hills+1980:1980ApJ...235..986H,Krumholz+2019:2019ARA&A..57..227K, Hennebelle+2024:2024ARA&A..62...63H}. Thus, even in the absence of binary evolution, the stars in young stellar clusters will naturally drift apart due to their initial velocity dispersion. The amplitude of this velocity dispersion will directly impact the average separation of member stars from the cluster's guiding centre, at the time the stars undergo core collapse. 

There are currently no strong observational constraints on the appropriate value of the initial cluster velocity dispersion. We have made a ``typical'' choice for our fidicual model ($v_{\rm disp} = 1.7 \unit{km}{s^{-1}}$), but given the uncertainty, we also vary the initial cluster velocity dispersion to fixed values of $0.5$ and $5 \unit{km}{s^{-1}}$. We plot the impact of these variations in Figure~\ref{fig:var-gal}, but note that the timing distributions (top panel) are unchanged, because the velocity dispersion does not change any aspect of stellar evolution.

In contrast to the timing distributions, the initial velocity dispersion has a strong impact on the distances at which SNe occur (bottom panel). Indeed, these variations produce the strongest change in the median of the overall distance distribution among all of our variations. 

While the means do change, the tail of distant SNe from secondary stars is relatively unchanged at lower velocity dispersions. For these secondaries, the distances are mostly driven by binary physics rather than initial velocities. Therefore even for a very low choice of the initial velocity dispersion, some SNe still occur far from their parent cluster in low density environments. For $v_{\rm disp} = 0.5 \unit{km}{s^{-1}}$, the fraction of SNe beyond $100\unit{pc}$ decreases to $\ffar = 5.2\%$, but the most distant tail is almost unchanged from our fiducial model with $\fdistant = 0.8\%$.

\subsubsection{Hydrodynamical simulation codes}

So far, all of the calculations presented here have assumed that a binary population is evolved from the recent star formation in a specific simulated galaxy (\texttt{m11h} from \fire). This choice can affect the resulting distributions by changing the gravitational potentials and stellar kinematics of the young stars, as well as their metallicities (see Section~\ref{sec:metallicity_variation} above).  

We explore the possible impact of the choice of simulation here, by considering instead the galaxy \texttt{r442} from the Massive Dwarfs suite of simulations run with \changa \citep{Menon+2015:2015ComAC...2....1M}.
\texttt{r442} is a similar \citep{Keith+2025:2025arXiv250116317K, Ruan2025arXiv250316607R} but moderately larger galaxy, with a total mass approximately 1.5 times higher than \texttt{m11h}. Its recent star formation rate is similar, such that the total stellar mass formed in the most recent 150 Myr is only ${\sim}6\%$ higher than in the fiducial simulation \texttt{m11h}. Additionally, the disk of \texttt{r442} is approximately $20\%$ less massive which is important for the galactic potential.

\begin{figure}
    \centering
    \includegraphics[width=\columnwidth]{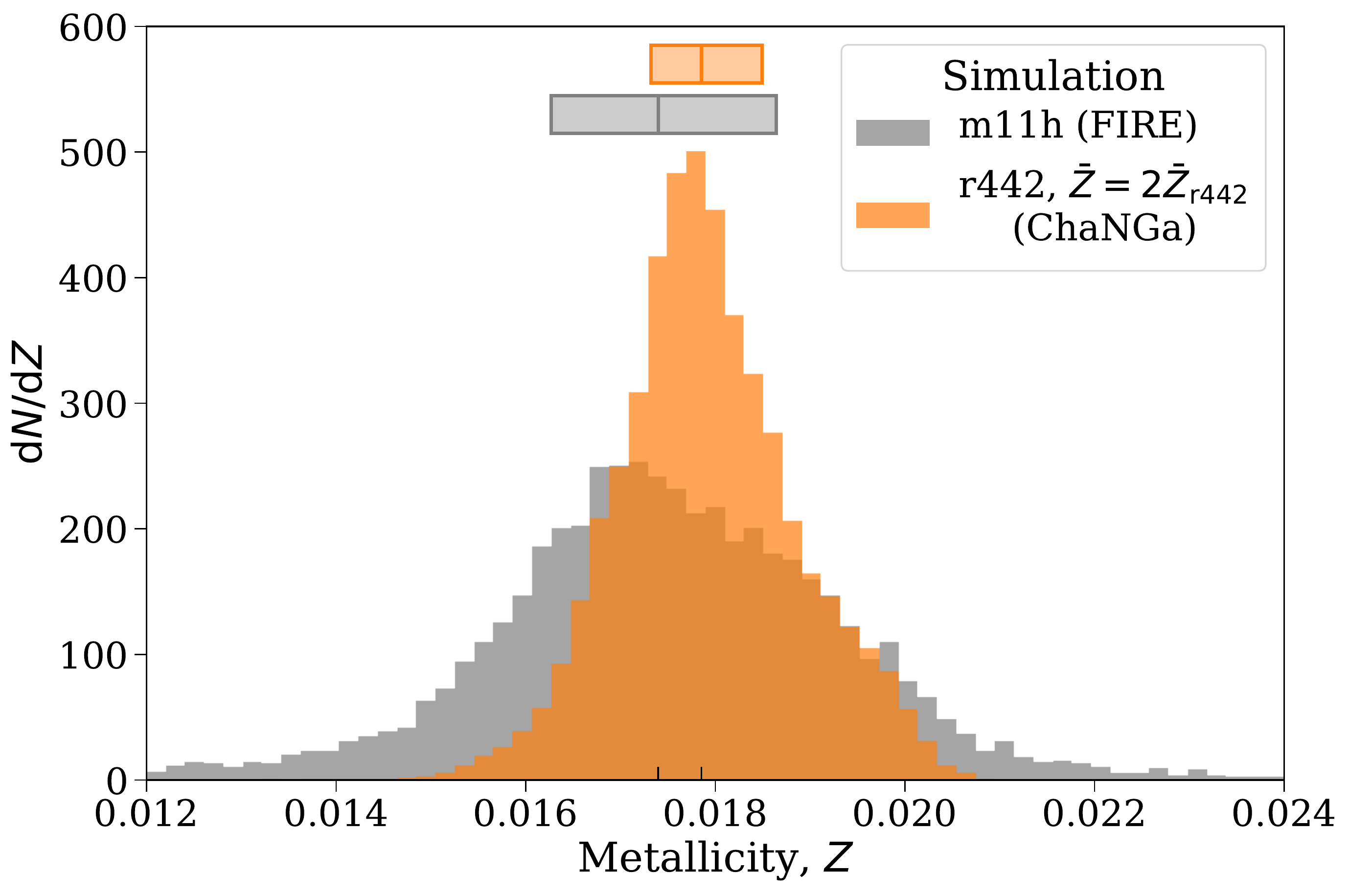}
    \caption{The distribution of metallicities of stars in our fiducial \texttt{m11h} simulation, compared to those in the \changa \texttt{r442} simulation with a factor 2 increased metallicity. The boxes at the top indicate the interquartile range, with the median shown with a line in the box. As noted in Section~\ref{sec:metallicity_variation}, the average metallicity is very similar between the simulations, with a slight difference in the width of the distribution.}
    \label{fig:metallicity-dists}
\end{figure}

In addition to changes in the mass and internal structure, the average metallicity of \texttt{r442} is a factor of ${\sim}2$ lower than \texttt{m11h}, which as discussed in Section~\ref{sec:metallicity_variation}, will strongly affect the binary evolution. We therefore increase the metallicity of each star particle in \texttt{r442} by a factor of two, such that the median metallicity of the stellar populations seeded from each galaxy are now very similar. As a result we can better isolate the effects of changing galaxy simulations. We show the resulting metallicity distributions in Figure~\ref{fig:metallicity-dists}. We note that the width of the metallicity distribution for \texttt{r442} is still slightly wider in \texttt{m11h}, which may contribute to slight differences between the populations. 

In the final panel of Figure~\ref{fig:var-gal} we show the timing and distance distribution of SNe for an \texttt{r442} simulation with a factor of 2 increase in metallicity. We find that the distributions of SN times between the two simulations are statistically indistinguishable, consistent with the lack of expected other impacts on binary populations once metallicity is controlled for. 

The same similarity is found for the distance distributions. Overall, our results show no strong dependence on the hydrodynamical simulation code used to seed star formation. Moreover, our results indicate that a change in the mass of the gravitational potential by a factor of three do not strongly affect the spatial distribution of SNe.


\section{Analytic model for feedback}\label{sec:fits}

The results in Section~\ref{sec:variations} suggest that there is enough consistency in the behaviour of SNe populations that models of stellar feedback should be able to include the effects of binary evolution without being overly sensitive to choice of parameters. Hydrodynamical simulations typically use ``subgrid models'' to implement stellar feedback. These models are currently based on single-star stellar evolution, and are implemented as simple parameterised analytic models \citep[e.g.,][]{Hopkins+2023:2023MNRAS.519.3154H}.

In this section, we develop an analytic model appropriate to evolving binary star populations. We fit both the rate of core-collapse SNe over time, and the velocities of SN progenitors, including the effects of metallicity. This model can be used in future hydrodynamical simulations to more consistently model the times and locations of core-collapse SN feedback.

We make a sampling routine for our analytic model available as a simple Python script. This routine can be used to rapidly generate an array of times and velocities for all SNe associated with a given star particle in a hydrodynamical simulation. The routine is available on \href{https://github.com/TomWagg/supernova-feedback/blob/main/analytic_feedback_model.py}{the GitHub repository}\footnote{\url{https://github.com/TomWagg/supernova-feedback/blob/main/analytic_feedback_model.py}} for this paper.

Our general procedure for sampling SNe resulting from a starburst of a given metallicity, $Z$, is as follows. We first use our model to sample the time, $t_{\rm SN}$, at which each SN occurs (Eq.~\ref{eq:time_fit}). Based on this time, we sample a progenitor velocity, $v_{\rm SN}$ (Eq.~\ref{eq:vel-fit}). The distribution for this velocity changes based on (a) whether the progenitor was ejected from its binary (Eq.~\ref{eq:f-eject}) and, if so, (b) what type of mass transfer it experienced mass transfer before doing so (Eq.~\ref{eq:f_noMT}--\ref{eq:f-CE}). This procedure is illustrated in a flowchart in Figure~\ref{fig:flowchart}. In the following subsections we outline our model for each of these distributions and assess their goodness-of-fit to our simulations.

\subsection{Core-collapse SN rate}

\begin{figure}
    \centering
    \includegraphics[width=\columnwidth]{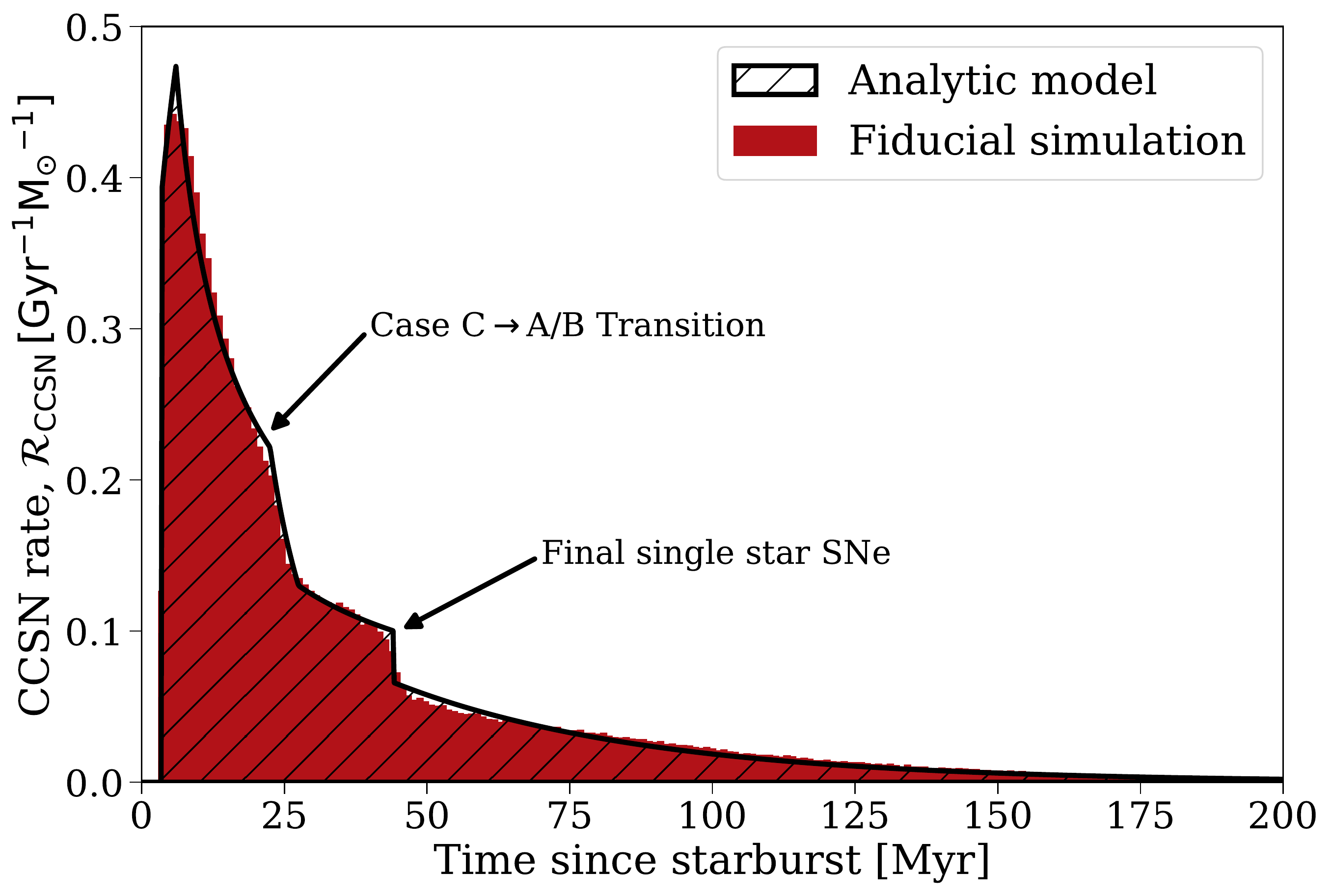}
    \includegraphics[width=\columnwidth]{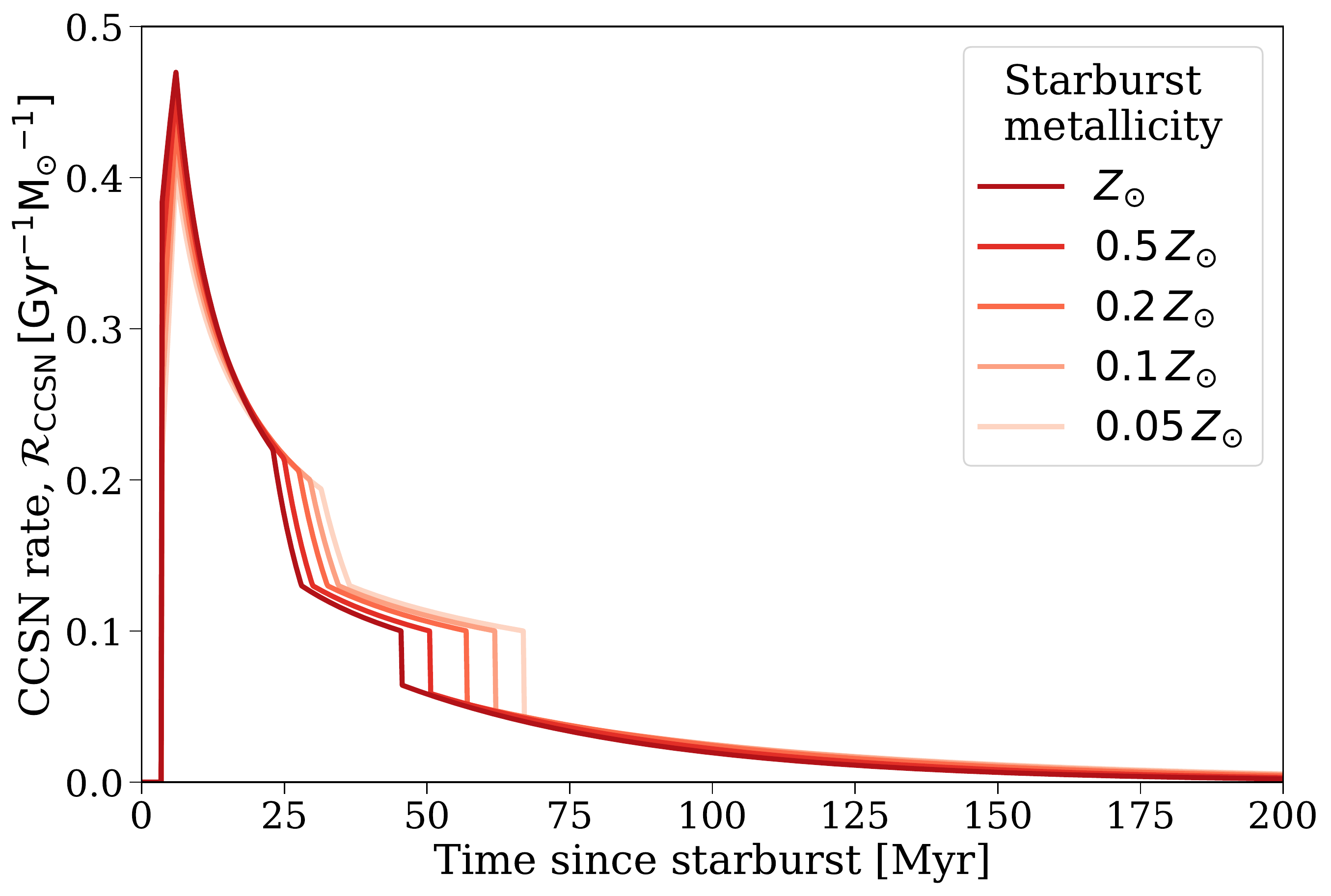}
    \caption{\textbf{Top:} A comparison of our analytic model (Eq.~\ref{eq:time_fit}, shown in black) to the SNe delay time distribution to our fiducial simulation (shown in red). Transition points are annotated with the physical process driving them. The model reproduces the late SN rate and overall normalisation to within $0.5\%$. \textbf{Bottom:} The metallicity dependence of our analytic model to the SN time distribution. Lower metallicity star formation events produce more SNe that typically occur later. Our analytic model reproduces the late SNe rates and normalisations of our lower metallicity simulations to within $1\%$. (\href{https://www.tomwagg.com/html/interact/binary-supernova-feedback.html\#fig9-10}{\faLaptopCode{} Interactive figure available.})}
    \label{fig:time-fit}
\end{figure}

In the top panel of Figure~\ref{fig:time-fit} we plot in red the distribution of SNe explosion times for our fiducial model.  This distribution has a number of clear features, including a rapid rise after $3.5\unit{Myr}$ to a peak at $6\unit{Myr}$, the distribution then declines, with a knee around $25\unit{Myr}$, which is produced by the transition from binaries that experience case A or case B mass transfer to those that experience case C mass transfer (see Section~\ref{sec:sn_times} for a discussion of this feature). After this decline there is a sharp drop at $44\unit{Myr}$ associated with the maximum age of a single-star that can still reach core-collapse, followed finally by a tail to long times due to the subset of SNe from merger products. Existing SNe prescriptions capture some of these features, but all lack the tail beyond $44\unit{Myr}$, which contains 25\% of the entire SNe population.

We develop a fitting formula for this distribution that captures these various transitions. We first define the rate of core-collapse SNe, $\mathcal{R}_{\rm CSSN}(t)$, as a piecewise power-law, with an exponential tail, as a function of time
\begin{equation}\label{eq:time_fit}
    \frac{\mathcal{R}_{\rm CSSN}(t / \unit{Myr})}{\unit{Gyr}{M_\odot}} \equiv \begin{cases}
        0 & t < t_1, t > t_6,\\
        a_{i} (t / t_{i})^{\psi_{i}} & t_{i} \le t < t_{i + 1}, i < 5,\\
        a_6 e^{-t / t_5} & t_5 \le t < t_6
    \end{cases}
\end{equation}
where
\begin{equation}
    \begin{split}
    a_i = [&0.38 + 0.13 \feh, 0.47 + 0.05 \feh,\\&0.22 + 0.02 \feh, 0.13, 0.1, 0.175 + 0.05 \feh],\end{split}
\end{equation}
\begin{equation}\label{eq:time-fit-times}
    \begin{split}
    t_i = [&3.5, 6, 23 - 6.5 \feh,\\&28 - 6.5 \feh, 45.5 - 16.5 \feh, 200],\end{split}
\end{equation}
the subscript $i$ ranges from 1 to 6 and represents the different physically-motivated transition points that we discussed above, $\psi_i = \ln(a_{i + 1} / a_i) / \ln(t_{i + 1} / t_i)$, and$\feh \approx \log_{10}(Z / Z_\odot)$ is the metallicity of the starburst relative to solar, where we assume $Z_\odot = 0.0142$.

Given that some of the features in the distribution in Figure~\ref{fig:time-fit} are metallicity-dependent (see arguments in Section~\ref{sec:metallicity_variation}), we include metallicity-dependent transition points. The transition from binaries that experience case A or case B mass transfer to those that experience case C mass transfer occurs around $25\unit{Myr}$ at the metallicity of \texttt{m11h} and moves to later times at lower metallicity (by $2 \unit{Myr}$ for every factor of 2 decrease in metallicity). The final single star SN occurs $44 \unit{Myr}$ at the metallicity of \texttt{m11h} and increases by $5 \unit{Myr}$ for every factor of 2 decrease in metallicity. 

The analytic model described above closely reproduces our simulated population, accounting for both metallicity and effects from binarity, shown as the black line in the top panel of Figure~\ref{fig:time-fit}.
The model reproduces the late SN rate to $0.5\%$ and overall normalisation to $0.3\%$. This model produces approximately 1.2 SNe for every 100 solar masses of star formation, which is in good agreement with the current \fire-3 model \citep{Hopkins+2023:2023MNRAS.519.3154H}. In the lower panel of Figure~\ref{fig:time-fit}, we show the metallicity dependence of the model for each of the metallicities that we consider in the variations in Section~\ref{sec:metallicity_variation}. In each case the model reproduces the late SNe rates and normalisations of our lower metallicity simulations to within $1\%$.

\subsection{Spatial distribution of core-collapse SNe}

The variations presented in Section~\ref{sec:variations} show that the SN distance distribution is dependent on the assumed initial cluster velocity dispersion, and on the metallicity and galactic potential of the simulated galaxy. An analytic model of the distance distribution in our fiducial model may therefore not always be applicable other simulations. Therefore, we instead chose to model the velocity distribution at which supernova progenitors travel away from their parent cluster, which can more easily be adapted to other contexts.

There are two components to the velocity distribution of SNe precursors: (1) the velocities of primary stars,  merger products, and unejected secondaries, which are all dominated by cluster dissolution through the initial cluster velocity distribution $v_{\rm disp}$; and (2) the velocities of progenitors that are ejected from binaries after a companion's SN. Binary physics and metallicity will shape the second of these distributions, and set the fraction of SNe that go through the latter channel.

We adopt a definition of a progenitor being considered ejected as a walkaway or runaway star if the ejection velocity is greater than $5\unit{km}{s^{-1}}$ \citep[following e.g.,][]{Eldridge+2011:2011MNRAS.414.3501E, Renzo+2019:2019A&A...624A..66R}. Given this definition, we estimate that the fraction of core-collapse SNe that originate from an ejected progenitor is approximately,
\begin{equation}\label{eq:f-eject}
    f_{\rm eject}(t) = \begin{cases}
        0.24 & 5 \le t_{\rm SN} / \unit{Myr} < t_5(Z) ,\\
        0.1 & t_5(Z) \le t_{\rm SN} / \unit{Myr} < 60,\\
        0.0 & {\rm else}
    \end{cases}
\end{equation}
where $t_{\rm SN}$ is the time at which the SN occurs and $t_5$ is the time of the last single-star SN, defined in Eq.~\ref{eq:time-fit-times}. We find that this relation does not have a significant metallicity dependence and varies by less than 1\% even when decreasing the metallicity by a factor of 20.

\paragraph{Unejected progenitor velocity distribution} For most SNe progenitors, the velocity at which they move away from the cluster is determined solely by the initial cluster velocity dispersion\footnote{The galactic potential can also alter these velocities, producing accelerations or decelerations based on the trajectory of an individual star, but this effect is generally secondary to the velocity dispersion.}. Under the assumption that each star moves isotropically away from the cluster following a 3D Gaussian with a dispersion velocity of $v_{\rm disp}$, the probability that a SN progenitor moved away with a given 1D velocity follows a Maxwellian as
\begin{equation}\label{eq:vel-fit-cluster}
    p_{\rm unejected}(v) = \frac{1}{\sigma^3} \sqrt{\frac{2}{\pi}} v^2 e^{-v^2 / (2\sigma^2)},
\end{equation}
where $v$ is the velocity in $\unit{km}{s^{-1}}$, $\sigma = v_{\rm disp} / \sqrt{3}$ and $v_{\rm disp}$ is the initial velocity dispersion of clusters.

\begin{figure}
    \centering
    \includegraphics[width=\columnwidth]{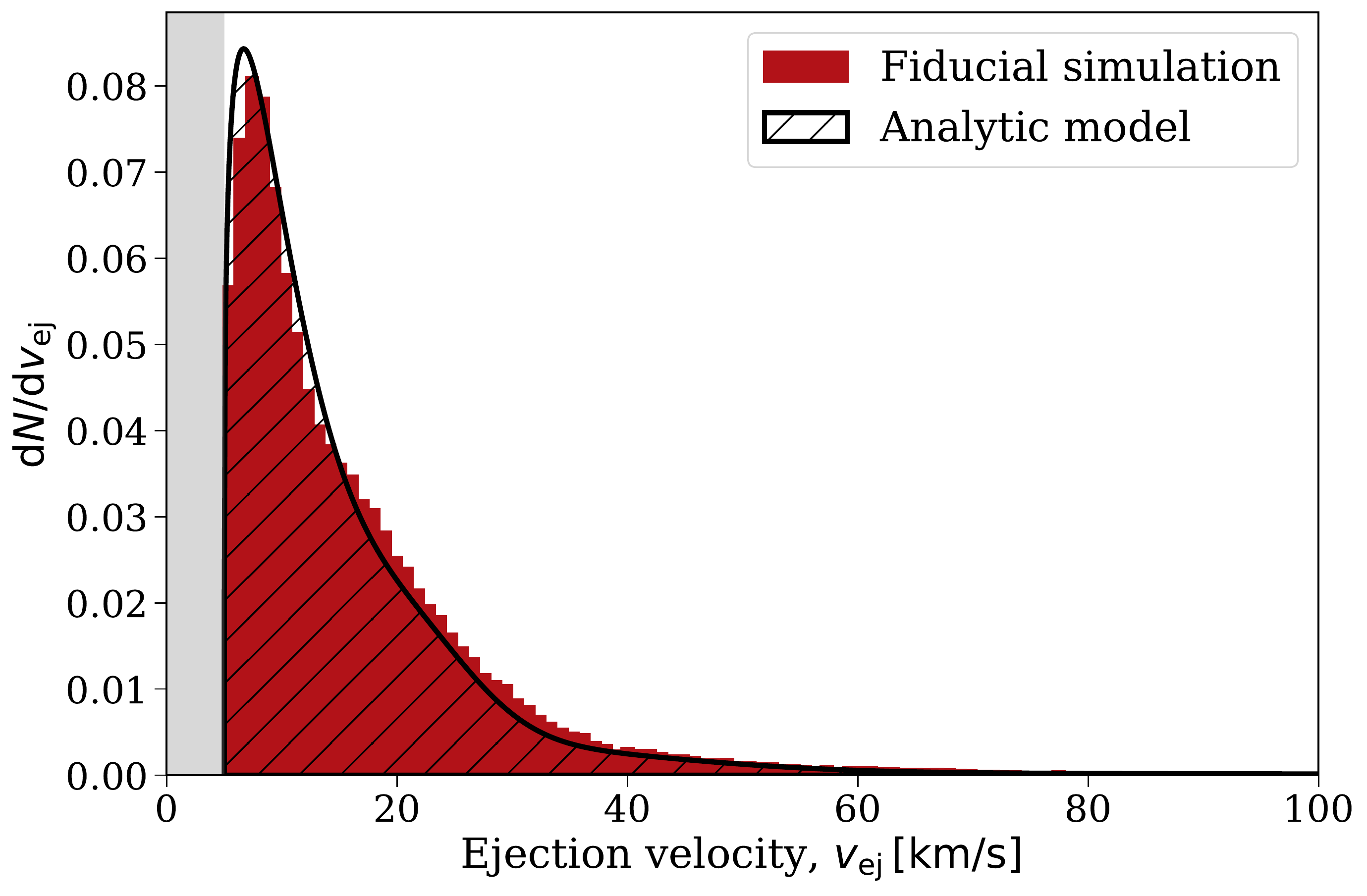}
    \includegraphics[width=\columnwidth]{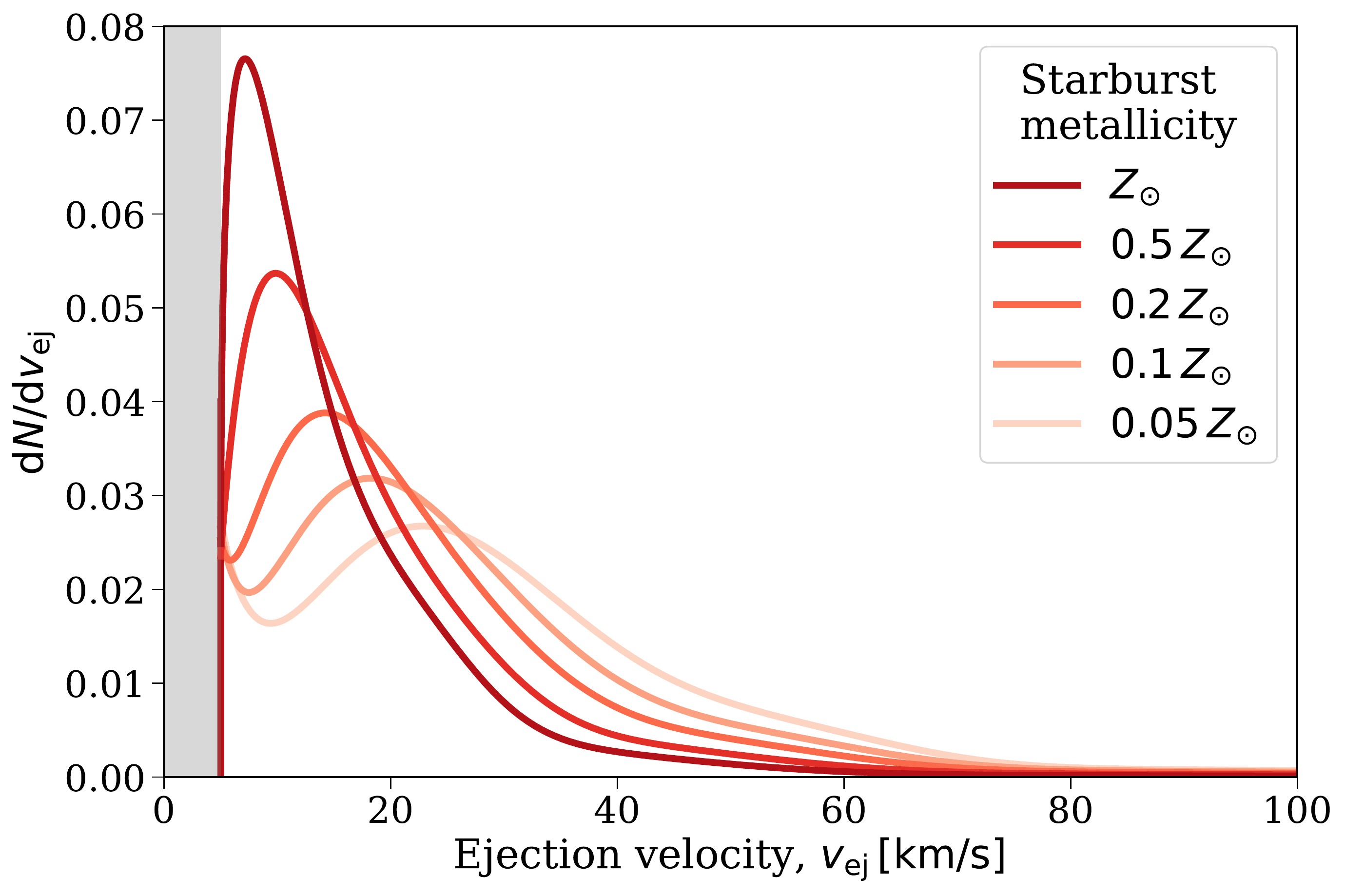}
    \caption{\textbf{Top:} A comparison of our analytic model for the distribution of ejection velocities (Eq.~\ref{eq:vel-fit-ejected}, shown in black) to our fiducial simulation for the subset of SN progenitors that are ejected with $v_{\rm ej} \ge 5\unit{km}{s^{-1}}$ (shown in red, with the lower velocity region shaded in grey). \textbf{Bottom:} The metallicity dependence of our analytic model for the ejection velocity distribution. Lower metallicity binaries typically produce higher ejection velocities. (\href{https://www.tomwagg.com/html/interact/binary-supernova-feedback.html\#fig9-10}{\faLaptopCode{} Interactive figure available.})}
    \label{fig:vel-fit}
\end{figure}

\paragraph{Ejected progenitor velocity distribution:} For progenitors ejected from their binaries, there is an additional component to the velocity beyond the initial cluster velocity dispersion, which is shown in the red histogram in the top panel of Figure~\ref{fig:vel-fit} for our fiducial model. The distribution peaks quickly around $8\unit{km}{s^{-1}}$ before declining gradually up to $100\unit{km}{s^{-1}}$.

We model the ejection velocity distribution as a mixture of four physically-motivated distributions that are distinguished by what type of mass transfer the star experienced prior to ejection; mass transfer significantly alters the pre-SN orbital separation of the system, and hence the orbital velocity and ejection speed of a companion. 
With this assumed mixture model, the probability distribution for the velocity of an ejected SN progenitor at a given metallicity can be written as
\begin{equation}\label{eq:vel-fit-ejected}
    \begin{split}p_{\rm eject}(v | Z) &=  f_{\rm no MT}(Z) \, p_{\rm no MT}(v| Z)\\
    &+ f_{\rm MT,A}(Z) \, p_{\rm MT,A}(v | Z) \\
    &+ f_{\rm MT,B/C}(Z) \, p_{\rm MT,B/C}(v | Z) \\
    &+ f_{\rm CE}(Z) \, p_{\rm CE}(v | Z)\end{split}
\end{equation}
where $v$ is the ejection velocity in $\unit{km}{s^{-1}}$, $Z$ is the metallicity, $f_{\rm no MT}$ and $p_{\rm no MT}(v | Z)$ are the fraction of ejected stars that had no mass transfer before ejection and the velocity distribution of those progenitors, respectively, while $f_{\rm MT,A}$ and $p_{\rm MT, A}(v | Z)$ are the same for stars that experienced case A mass transfer, $f_{\rm MT, B/C}$ and $p_{\rm MT, B/C}(v | Z)$ are the same for stars that experienced case B/C mass transfer, and $f_{\rm CE}$ and $p_{\rm CE}(v | Z)$ are the same for stars that experienced a common envelope. We fit the fractions of stars in each subpopulation to our five metallicity variations and find
\begin{align}\label{eq:f_noMT}
    f_{\rm no MT}(Z) &= 0.14 - 0.12 \feh,\\
    \label{eq:f-MT-A}
    f_{\rm MT, A}(Z) &= 0.12 + 0.035 \feh,\\
    \label{eq:f-MT-BC}
    f_{\rm MT, B}(Z) &= 0.67 + 0.12 \feh,\\
    \label{eq:f-CE}
    f_{\rm CE} &= 1 - f_{\rm no MT} - f_{\rm MT,A}-f_{\rm MT,B/C} .
\end{align}
The metallicity dependence in these fractions arises because the lower radial expansion of stars at low metallicity \citep[e.g.,][]{Brunish+1982:1982ApJS...49..447B, Xin+2022:2022MNRAS.516.5816X, Klencki+2022:2022A&A...662A..56K} alters when (and whether) Roche-lobe overflow occurs.

We develop fitting formulae for the distribution of ejection velocities from these subsets of binaries.

The distribution of velocities for stars that did not experience mass transfer before ejection, $p_{\rm no MT}$, follows a truncated power law distribution as
\begin{equation}\label{eq:vel-fit-nomt}
    p_{\rm no MT}(v | Z) = \begin{cases} A v^{-1.8 + 0.5 \sqrt{|\feh|}} & v_{\rm min} \le v \le v_{\rm max} \\ 0 & {\rm else} \end{cases}\end{equation}
where $A$ is a normalisation constant to ensure the distribution integrates to unity. The use of a power law in this case follows naturally from the assumption that the initial orbit period distribution also follows a power law, because no mass transfer occurs to drastically alter the shape of the orbital period distribution before ejections.

For stars that experienced case A mass transfer, the distribution of velocities, $p_{\rm MT, A}$, is restricted to moderate values, and is well-fit by a normal distribution as
\begin{equation}\label{eq:vel-fit-mt-a}
    p_{\rm MT,A}(v|Z) = \mathcal{N}(22 - 8\feh, 6 - 3\feh).
\end{equation}
Case A mass transfer doesn't produce slow ejections since that would require wider binaries, which result in case B/C mass transfer. Similarly faster ejections don't occur since the shorter orbital period required most often leads to a merger instead.

The distribution of velocities for stars that experienced case B or case C mass transfer before ejection, $p_{\rm MT, B/C}(v|z)$, is well-fit by a scaled beta distribution \citep[e.g.,][]{Beyer+1987:1987chms.book.....B} that evolves with metallicity as
\begin{equation}\label{eq:vel-fit-mt-bc}
    p_{\rm MT, B/C}(v|Z) = \frac{\beta(x, \alpha_{\rm B/C}, \beta_{\rm B/C})}{v_{\rm max}},
\end{equation}
where
\begin{equation}
    x = \frac{v - v_{\rm min}}{v_{\rm max}},\quad v_{\rm min} = 5,\quad v_{\rm max} = 100,
\end{equation}
\begin{equation}
    \alpha_{\rm B/C} = 1.5 - 1.5 \feh,\quad \beta_{\rm B/C} = 18 + 5 \feh.
\end{equation}

For stars that experiences a common-envelope, the velocity distribution also well-fit by a scaled beta distribution, but skewed to higher velocities as a result of how common-envelopes shrink binaries.
\begin{equation}\label{eq:vel-fit-ce}
    p_{\rm CE}(v|Z) = \frac{\beta(x, \alpha_{\rm CE}, \beta_{\rm CE})}{v_{\rm max}},
\end{equation}
where
\begin{equation}
    \alpha_{\rm CE} = 5 - 4 \feh,\quad \beta_{\rm B/C} = 10.
\end{equation}

The analytic model for runaway star ejection velocities accurately reproduces the distribution we find in our fiducial simulation. In the top panel of Figure~\ref{fig:vel-fit} we compare the model to the distribution of velocities for all stars that are ejected from their binary with a velocity greater than $5\unit{km}{s^{-1}}$. The fraction of stars with velocities greater than $10, 30$, and $50\unit{km}{s^{-1}}$ are each reproduced within 3\%. In the bottom panel of Figure~\ref{fig:vel-fit}, we show how the model changes as a function of metallicity. Similarly, at lower metallicity, the fraction of stars with velocities greater than $10, 30$, and $50\unit{km}{s^{-1}}$ are each reproduced within 4\%.

Overall, with these two distributions one can write the full SN progenitor velocity distribution as follows
\begin{equation}\label{eq:vel-fit}
    \begin{split} p(v | Z, t) &=f_{\rm eject}(t) p_{\rm eject}(v | Z)\\&+ (1 - f_{\rm eject}(t)) p_{\rm unejected}(v). \end{split}
\end{equation}

\subsection{Joint distribution}

\begin{figure}
    \centering
    \includegraphics[width=\columnwidth]{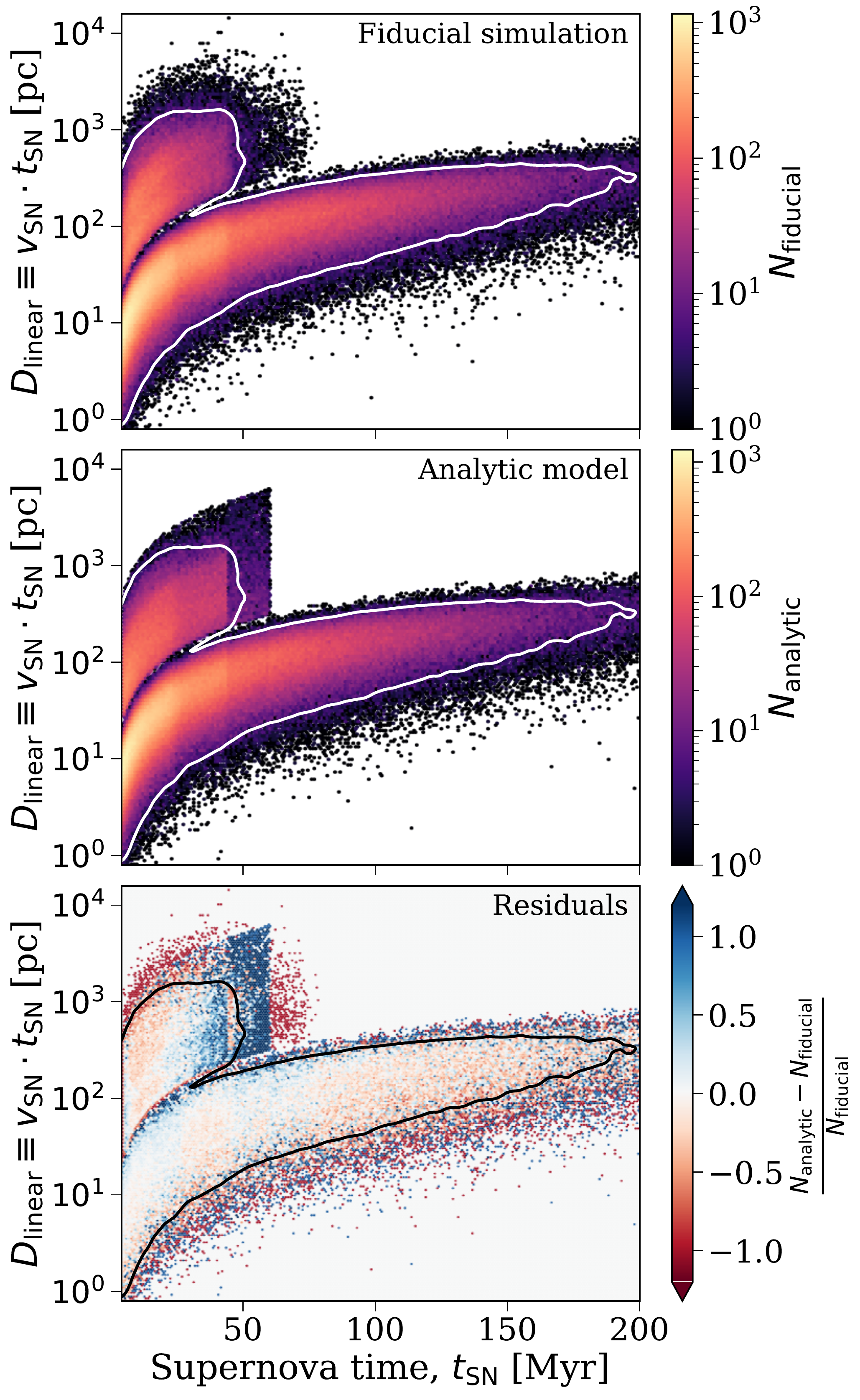}
    \caption{A comparison of the joint distribution of our analytic model to the fiducial simulation, which shows good agreement. Each panel shows a 2D histogram with the supernova time on the $x$-axis and the maximum distance travelled (i.e.\ the product of the SN time and velocity) on the $y$-axis. The top panel shows the distribution of the fiducial simulation, while the middle shows the analytic model and the bottom gives the fractional difference in each bin from the top two panels. The solid contour lines in each panel show the 98\% regime for the fiducial simulation.}
    \label{fig:2d-model}
\end{figure}

Our analytic model reproduces the joint distribution from the fiducial simulation well. In Figure~\ref{fig:2d-model} we compare the joint distribution of SN times and the linear distance an SN progenitor could travel (the product of the velocity and time). The top panel shows the distribution for the fiducial simulation, which is similar to Figure~\ref{fig:2d-dists} but with a skew to higher distances since this proxy for distance does not account for how the potential may alter the progenitor's velocity. In the middle panel we show the same distribution for a random sample from our analytic model with the same number of SNe. From the bottom panel, one can note that the model closely reproduces the distribution from the fiducial simulation, particularly within the 98\% contour of the fiducial simulation (shown with the black line).

\section{Discussion}\label{sec:discussion}

In this Section, we summarise key takeaways  (Section~\ref{sec:discuss_findings}), consider the potential implications of our results for galaxy evolution (Section~\ref{sec:implications}), and highlight potential limitations of our analysis (Section~\ref{sec:limitations}).

\subsection{Key takeaways of our results}\label{sec:discuss_findings}

\subsubsection{High-level differences between binary and single star feedback models}
We have demonstrated that binary interactions significantly shift the timing and spatial distribution of SNe when compared to an equivalent single star population. For our fiducial model, we find that the median SN occurs $22\unit{Myr}$ after a star formation event and at a distance of $35\unit{pc}$ from its parent cluster. These quantities are 29\% and 52\% larger, respectively, than for a single star population ($17\unit{Myr}$ and $23\unit{pc}$ respectively).

We also find that the total number of SN is ${\sim}11\%$ higher for the binary fiducial model than in the equivalent model including only single stars. Including binary interactions allows lower mass stars (i.e.\ those initially below the threshold for a SN) to accrete enough material, or merge, to attain the requisite mass to achieve core collapse.

In addition to lengthening the time and length scales for SNe energy injection, the inclusion of binary interactions also introduces long tails in the distributions to late times and long distances. For our fiducial model we find that $\flate = 25\%$ and $\ffar = 13\%$, compared to $\flate = 0\%$ and $\ffar = 1\%$ for the single star model.

These differences in feedback, particularly the increased spatial spread of SNe, have the potential to significantly affect the SNe energy injection in the ISM, which we will discuss in further detail in Section~\ref{sec:implications}. Given that the vast majority of core-collapse progenitors are formed in interacting binary systems \cite[e.g.,][]{Sana+2012:2012Sci...337..444S}, these differences should be accounted for in models of feedback, motivating the analytic prescription provided above in Section~\ref{sec:fits}.

\begin{figure}
    \centering
    \includegraphics[width=\columnwidth]{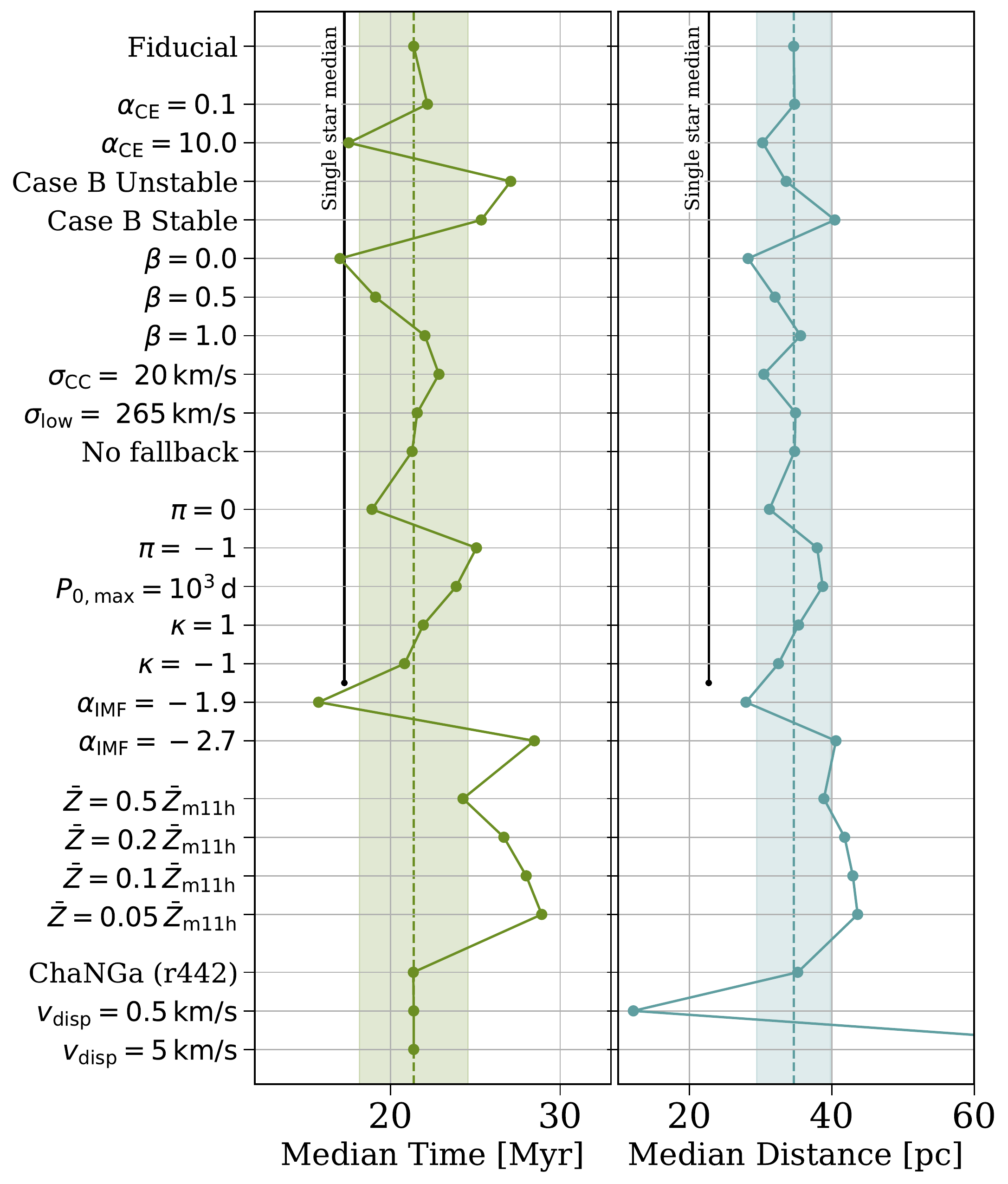}
    \caption{The median time and distance at which SNe occur for each variation (variations are outlined in Section~\ref{sec:variation-descriptions}). Dotted lines and shaded regions show the fiducial value and a $\pm15\%$ region around this value. The solid black lines show the median value of a population of single stars equivalent to our fiducial model. The lines don't continue for all variations, since the later ones would alter a single star population as well. Given the extreme physics variations that we consider, the relatively small differences in these medians demonstrates an encouraging robustness of our results to uncertainties. The exact values for each variation are given in Table~\ref{tab:percentiles}. (\href{https://www.tomwagg.com/html/interact/binary-supernova-feedback.html\#fig12-14}{\faLaptopCode{} Interactive figure available.})}
    \label{fig:trends-medians}
\end{figure}

\subsubsection{The robustness of binary SN feedback models} 
Through varying many input parameters, we find that these predictions for SN timing and distance are remarkably robust to a wide range of extreme variations to binary physics, initial conditions and galaxy parameters. In Figure~\ref{fig:trends-medians}, we summarise how the median time and distance of SNe changes under the variations that we consider. The majority of variations retain a median value within 15\% of the fiducial variation. The exceptions to this for timing are variations of metallicity, the initial mass function, and the stability of mass transfer. For SN distances, changes to the initial velocity dispersion can significantly shift the distribution. 

\begin{figure}
    \centering
    \includegraphics[width=\columnwidth]{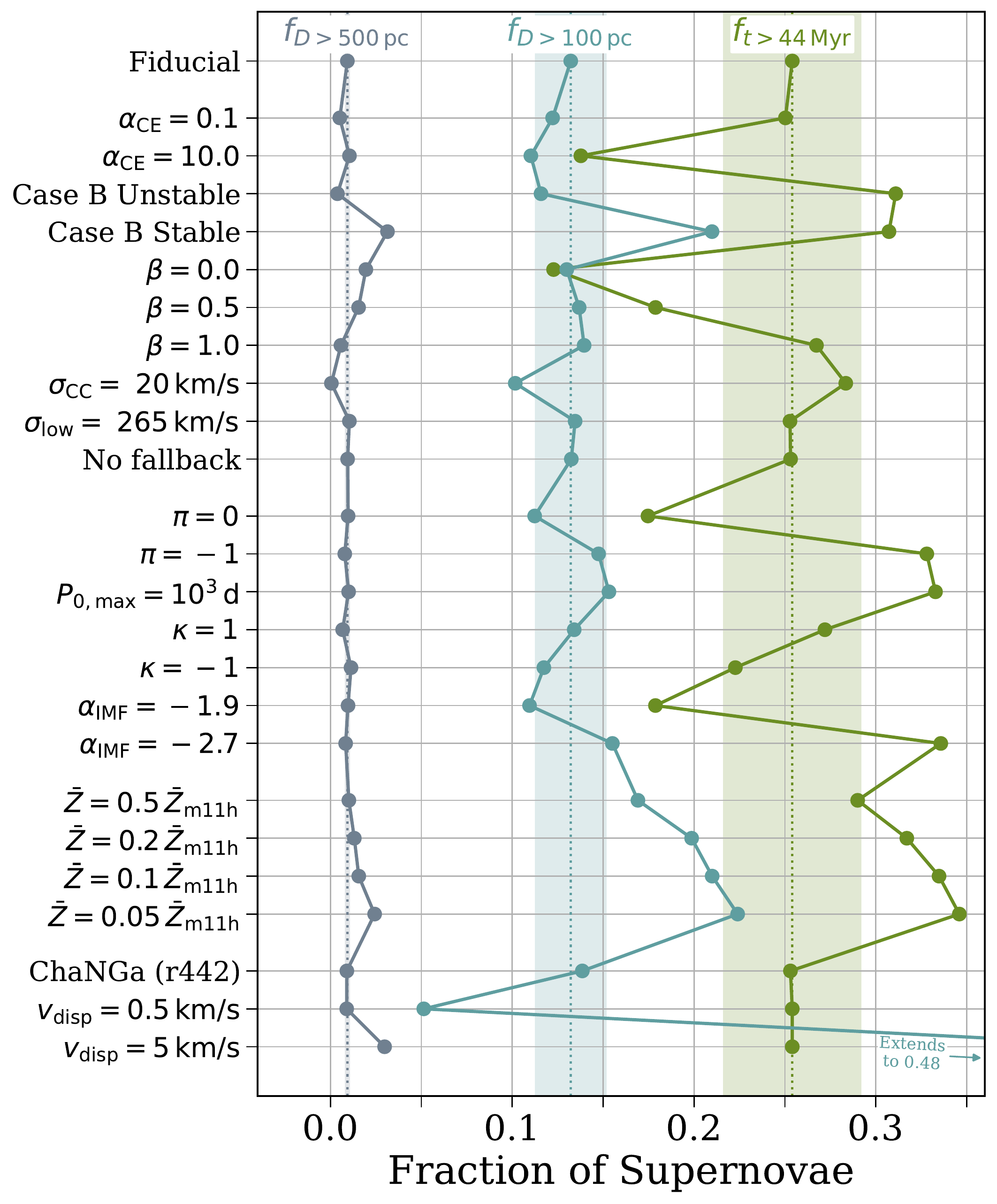}
    \caption{The fraction of the total population in the tails of the SNe timing and distance distributions for each variation (variations are outlined in Section~\ref{sec:variation-descriptions}). Scatter points indicate the \flate (green), \ffar (light blue), and \fdistant (slate blue) fractions. Note that $44\unit{Myr}$ is the assumed final core-collapse SN time in \fire-3. Dotted lines and shaded regions show the fiducial value and a $\pm15\%$ region around this value. The exact values for each variation are given in Table~\ref{tab:totals_and_stats}. (\href{https://www.tomwagg.com/html/interact/binary-supernova-feedback.html\#fig12-14}{\faLaptopCode{} Interactive figure available.})}
    \label{fig:trends-tails}
\end{figure}

In Figure~\ref{fig:trends-tails}, we summarise how the tails of the timing and distance distribution change for variations that we consider. The tails of the distance distribution are insensitive to most changes, only significantly shifting by more than $15\%$ for variations for metallicity, the stability of mass transfer, and initial cluster velocity dispersion. With the exception of the low velocity dispersion variation, all variations predict that at least $10\%$ of SNe occur more than $100\unit{pc}$ from their parent cluster.

In contrast, the timing tail does change by more than 15\% with several variations to binary physics, initial conditions and galaxy parameters. The late time tail is most sensitive to changes in metallicity, initial masses, and orbital periods, as well as mass transfer physics. Yet even for the model predicting the fewest late SNe (in which mass transfer is fully non-conservative, $\beta = 0.0$), $13\%$ of SNe still occur at times later than $44\unit{Myr}$, the assumed limit from the single star evolution models used for feedback prescriptions in current hydrodynamical codes \citep[e.g.,][]{Hopkins+2023:2023MNRAS.519.3154H}.

Given the suprisingly mild variations seen in Figures~\ref{fig:trends-medians}~\&~\ref{fig:trends-tails}, we expect that our analytic model for core-collapse SN feedback from binary progenitors (Section~\ref{sec:fits}) should be robust to significant parameter uncertainties. The model reproduces the medians, tails and overall normalisation of the timing and velocity distribution of the fiducial model, and also accounts for possible variations in metallicity and initial velocity dispersion, which produce the most significant change in the medians of the timing and distance distributions.

\begin{figure}
    \centering
    \includegraphics[width=\columnwidth]{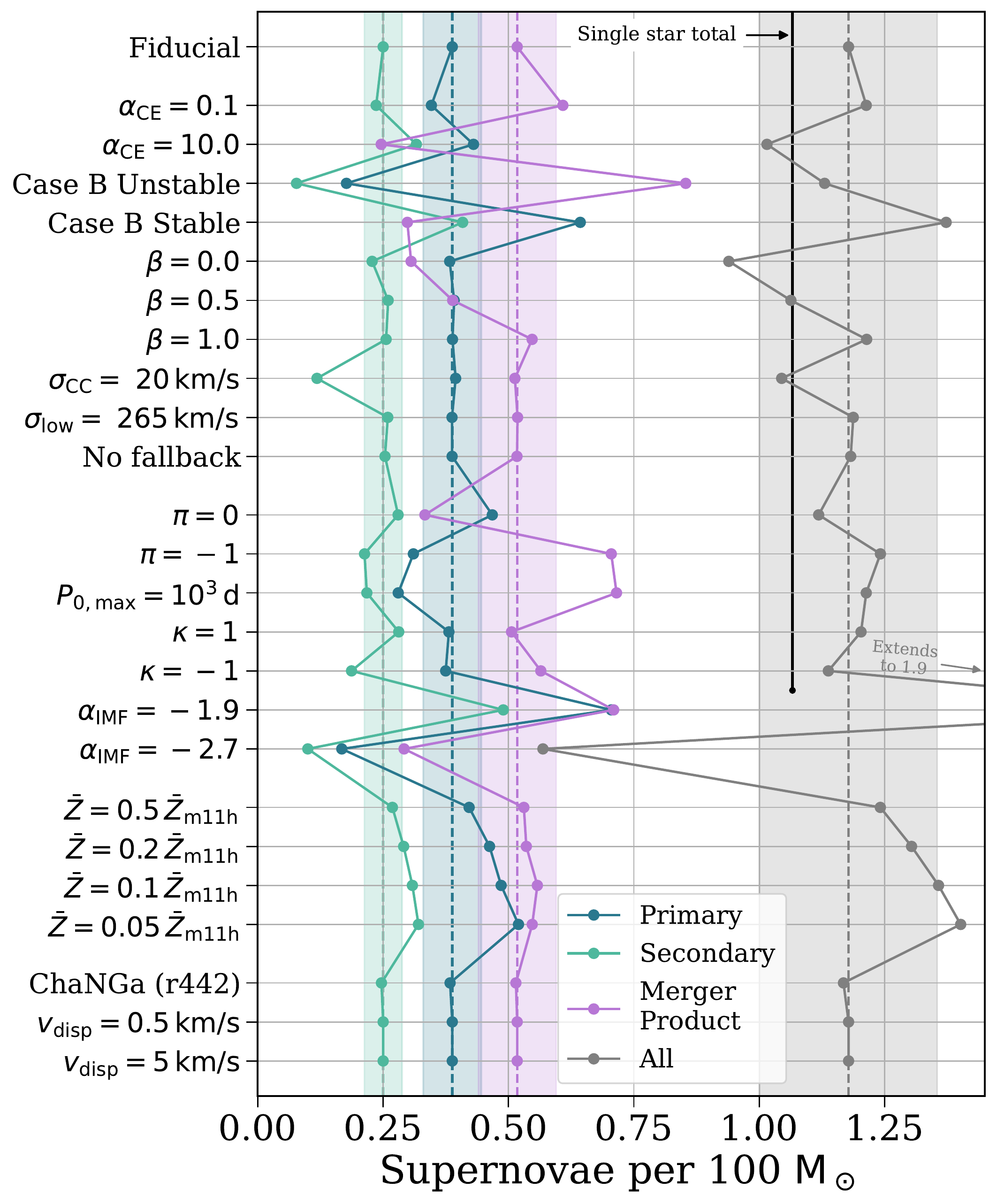}
    \caption{The total number of SNe that occur for each variation, separated by progenitor type (variations are outlined in Section~\ref{sec:variation-descriptions}). Dotted lines and shaded regions show the fiducial value and a $\pm15\%$ region around this value. The exact values for each variation are given in Table~\ref{tab:totals_and_stats}. The solid black line shows the total for a population of single stars that is equivalent to our fiducial model, indicating that binary models typically increase the total SN feedback in a population. The line doesn't continue for all variations, since the later ones would alter a single star population as well. (\href{https://www.tomwagg.com/html/interact/binary-supernova-feedback.html\#fig12-14}{\faLaptopCode{} Interactive figure available.})}
    \label{fig:trends-totals}
\end{figure}

A natural concern about implementing SNe feedback based on binary models is that there is much larger uncertainty compared to single star evolution. We show that the resulting time and spatial distributions of SNe are actually not nearly as variable as one might expect, especially when one uses fitting functions that take metallicity into account. Overall, given the robustness of our findings, our model for the timing and location of SN feedback represents a significant improvement upon the existing implicit assumption of single star evolution in hydrodynamical simulations.

\subsection{Implications for galaxy evolution}\label{sec:implications}



\subsubsection{Additional late core-collapse SN feedback}

Binary evolution may lead to more significant impacts on feedback through producing an extended tail of late-time supernova (see Figure~\ref{fig:sn-times-dists}).
This tail is not insignificant, with $\sim$25\% of SNe happening after the oldest SNe expected for single star models.

There are a number of possible changes that could result from the significant increase in late-time SNe that binary evolution produces following a star formation event.
First, the extended population of SNe will make the resulting feedback more gradual, and less impulsive. One can think of this difference as akin to changing from slamming on the brakes to slowing to a stop. This change could result in less short-term burstiness on small spatial scales, where more extended feedback could either interrupt a ``hard-stop, fast-start'' pattern of star formation, or, cause longer pauses before star formation can restart. 
Other works have previously explored the impact on ``burstiness'' from variations in stellar feedback's efficiency and energy/momentum injection and found that these variations have no strong effect \citep[e.g.,][]{Orr+2018:2018MNRAS.478.3653O, Chan+2018:2018MNRAS.478..906C, El-Badry+2018:2018MNRAS.473.1930E, Hopkins+2023:2023MNRAS.525.2241H}. However, adding in \textit{extended} stellar feedback may well produce different results, and should be tested numerically.

A second possible impact of extended SNe would be to change the susceptibility of the ISM to feedback. The longer the timescale for SNe, the more likely it is that the local ISM will have already been changed by earlier SNe and ionization from massive stars, which may make the impact of the late-time SNe qualitatively different (e.g., feedback into a dense molecular cloud versus feedback into an already expanding hot bubble). Even in the absence of immediate post-burst feedback, increasing the timescale of SNe allows more time for the ISM to ``revert to the mean'' through cooling and turbulent mixing, again changing the receptiveness of the local environment to feedback. As with assessing the impact on burstiness, full numerical tests are needed to understand the net impact.

Although we focus on core-collapse SNe in this work, we briefly consider Type Ia SNe, which are another source of binary-driven late stellar feedback. These SNe arise from slowly-evolving low-mass progenitors, have delay times that can range from ${\sim}50\unit{Myr}$ to several billion years \citep[e.g.,][]{Liu+2023:2023RAA....23h2001L, Ruiter+2025:2025A&ARv..33....1R} and may contribute to the driving of winds in dwarf galaxies \citep[e.g.,][]{Hu+2019:2019MNRAS.483.3363H}. The long tail of core-collapse SNe that we find may coincide with the earliest type Ia SNe, and therefore they could have a cumulative effect on the clustering of SNe and the driving of superbubbles.

\subsubsection{Spatially-extended feedback}

Binary stellar evolution changes where SNe occur, both through ejecting runaway stars via binary interactions, and increasing the impact of cluster dissolution through extending SNe progenitor lifetimes.
Both of these effects can shift feedback away from the higher-density ISM that hosts star formation, and into low-density environments. In all of our models, at least 12$-$15\% of all SNe occur more than $0.1\unit{kpc}$ from the centre of the clustered star formation (Figure~\ref{fig:trends-tails}. This separation is more than the size of even quite massive molecular clouds \citep[e.g.,][]{Chevance+2020:2020SSRv..216...50C}, and should be sufficient to move a significant fraction of SNe to lower density environments.

Past numerical simulations have found that more distributed, low-density feedback --- comparable to what is expected here --- may increase the likelihood of driving galactic outflows \citep[e.g.,][]{Ceverino+2009:2009ApJ...695..292C,Ceverino+2014:2014MNRAS.442.1545C,Zolotov+2015:2015MNRAS.450.2327Z,Hu+2017:2017MNRAS.471.2151H,Andersson+2020:2020MNRAS.494.3328A,Steinwandel+2023:2023MNRAS.526.1408S}. However, other works have argued that the effect on outflows is weaker \citep{Kim+2017:2017ApJ...846..133K,Andersson+2023:2023MNRAS.521.2196A} and that earlier results may have stemmed from insufficient resolution of SNe in dense gas \citep{Kim+2020:2020ApJ...900...61K}. Our updated models with a more detailed characterisation of the temporal and spatial location of SNe may affect these results and thus further numerical simulations are necessary to test this.

\paragraph{Spatially-distributed SNe in dwarfs and at high-redshift}
An underappreciated aspect of having more spatially-distributed SNe is their effect on small galaxies. The characteristic velocities at which runaways are released and the increase in SNe progenitor lifetimes are set by processes internal to the binary population, and as such, any characteristic length scale becomes increasingly important when considering physically smaller systems such as dwarf galaxies.

As an example, for galaxies with small effective stellar or gas radii (say, $R_\mathrm{eff}<1\unit{kpc}$), the $\sim12-15\%$ of SNe that take place more than $0.1\unit{kpc}$ from their birth cluster will presumably travel a substantial fraction of the effective radius, making any SNe feedback automatically a galaxy-wide event. 

We note several reasons beyond ``size'' why the possibilities for distributed or even escaping SNe may become more important for dwarfs. First, our fiducial simulation used a galaxy with a stellar mass of $4 \times 10^{9} \unit{M_\odot}$, which is technically a dwarf, but not particularly low mass compared to the entire galaxy population. Compared to this fiducial simulation, binaries evolving in galaxies with even shallower potentials may lead to SNe at even larger distances, and may lead to complete escape of SNe from the galaxy.

The second possible increased impact is due to the strong metallicity dependence of the fraction of large-distance SNe. Most dwarf galaxies are also low metallicity, which can nearly double the number of stars in the $D>100\unit{pc}$ tail (Figure~\ref{fig:trends-tails}), increasing to over 20\% of stars with $D>0.1\unit{kpc}$ for $Z\le0.1Z_{\odot}$. 

This effect may be particularly pronounced at high redshifts, where galaxies are systematically far more compact,  intrinsically lower metallicity, and (more speculatively) likely have high characteristic velocity dispersions in star forming regions. While we cannot know the exact statistical properties of binary populations at these redshifts, it seems unlikely that they would conspire to counteract these trends, given the robustness of the distance distributions (Figures~\ref{fig:trends-medians}~\&~\ref{fig:trends-tails}). 

In support of the speculations above, we note that the potentially enhanced role of runaways in dwarfs has already been identified in existing galaxy simulations, \citep[e.g.,][]{Steinwandel+2023:2023MNRAS.526.1408S}, which have shown that runaway stars have the potential to drive galactic outflow rates \citep[e.g.,][]{Hu+2017:2017MNRAS.471.2151H}, impact the overall star formation rate, and boost the energy loading factor, \citep[e.g.,][]{Steinwandel+2023:2023MNRAS.526.1408S}, albeit under the unrealistic assumption of fixed ejection velocities for all runaways.
These methods do not fully capture the more complex, long-tailed velocity distributions, or their metallicity dependence. Therefore, our models motivate future work to investigate the effect of altering the spatial distribution of the feedback in hydrodynamical simulations in this way.

\subsubsection{Stellar wind feedback}

We have previously limited our analysis to considering how binary interactions impact core-collapse SN feedback. However, binary interactions likely also have a significant impact on the feedback from stellar winds. In particular, binary interactions produce more stripped stars, which contribute a large amount of ionising emission \citep[e.g.,][]{Gotberg+2019:2019A&A...629A.134G}. 
While a quantitative analysis of the impact of stripped stars and wind-driven mass loss is outside the scope of the paper, we briefly comment on some qualitative expectations here.

In practice, we think that the spatial distribution of wind-driven feedback may not be as strongly affected by binarity as SNe feedback. The stripped stars produced via binary interactions are most likely to be primary, rather than secondary stars. These strong contributors to the feedback would therefore rarely be ejected from their parent cluster by the earlier evolution of a more massive companion. Moreover, any self-stripping from strong stellar winds is most prevalent in the most massive stars, which also have the shortest lifetimes and thus much less time to travel far from their parent cluster.

In total, the main effect of binary interactions on stellar wind feedback is likely on its magnitude (increased relative to single stars), rather than its spatial distribution. Nevertheless, future work should model the impact of these winds in detail.

\subsection{Limitations}\label{sec:limitations}

Although we have taken pains throughout this paper to widely explore the possible parameter space, there are some limitations that remain. We briefly discuss some of the most significant of these here, but note that many of these limitations are currently inherent in any analysis of this type, and may affect single-star evolution models as well. 

\paragraph{Population synthesis treatment of binary physics} Our results are computed using rapid population synthesis, which relies on parametric prescriptions for stellar evolution that mimic more detailed simulations \citep{Hurley+2000:2000MNRAS.315..543H, Hurley+2002:2002MNRAS.329..897H, Breivik2020}. In this work, we explicitly consider the impact of varying our assumptions in these parameter prescriptions for a variety of binary physics (see Section~\ref{sec:variations}), though we only vary one parameter at a time whilst some choices may be correlated. Moreover, this does not account for the fact that the parametric prescriptions themselves may not adequately describe the underlying binary physics. For example, the $\alpha$-$\lambda$ common-envelope prescription used in population synthesis is a relatively simplistic energy based prescription for a complex three-dimensional mass transfer event, which likely fails to capture all of the details of the process \citep{Webbink+1984:1984ApJ...277..355W, deKool+1990:1990ApJ...358..189D, Ivanova+2013:2013A&ARv..21...59I,Ivanova+2020:2020cee..book.....I,Ropke+2023:2023LRCA....9....2R}. This also highlights that population synthesis does not model processes on a thermal timescale and as such requires prescriptions for systems that are out of thermal equilibrium.
Further observational constraints of massive stars and these rapid evolutionary phases are necessary to better model the binary physics and improve population synthesis models. 

\paragraph{Reliance on pre-computed stellar tracks} 

Rapid population synthesis models such as \cosmic rely heavily on the choices made in the underlying stellar evolution models. These choices are then embedded into any binary population synthesis model that uses them and limit the possible variations one can consider.

For example, the original stellar tracks used by \cosmic \citep{Pols+1998:1998MNRAS.298..525P} assumed one specific model for convective boundary mixing and overshooting, which are processes that can impact the final core mass and pre-supernova structure of the star \citep[e.g.,][]{Ugliano+2012:2012ApJ...757...69U, Kaiser+2020:2020MNRAS.496.1967K}.

Some of these limitations may be particularly pronounced for aspects of stellar rotation.
The underlying stellar tracks used by \cosmic do not account for rotation, however rotation can significantly influence the evolution of massive stars \citep[e.g.,][]{Ekstrom+2012:2012A&A...537A.146E}. For example, the efficiency of mass transfer can play an important role in the timing and location of SNe, and this may be limited by rotation-enhanced mass loss, which is particularly important for the critically-rotating accretors \citep[e.g.,][though it remains debated whether this is physical, e.g., \citealt{Paczynski+1991:1991ApJ...370..597P, Popham+1991:1991ApJ...370..604P}]{Langer+1998:1998A&A...329..551L,Petrovic+2005:2005A&A...435.1013P,Renzo+2021:2021ApJ...923..277R} that appear to be common \citep[e.g.,][]{Bastian+2017:2017MNRAS.465.4795B}. 
A fuller treatment of stellar rotation could therefore impact our results. 

\paragraph{Main sequence core evolution and rejuvenation}

Most \texttt{BSE}-based population synthesis codes (including \cosmic, which we use in this work) rely upon single star evolution models for computing the core structure of stars \citep{Hurley+2000:2000MNRAS.315..543H, Hurley+2002:2002MNRAS.329..897H}. However, many works have shown that binary interactions and accretion can alter the core structure and lead to rejuvenation of accretors \citep{Hellings1983,Braun+1995,Cantiello+2007,Staritsin+2019,Renzo+2023,Lau+2024:2024arXiv240109570L, Wagg+2024:2024A&A...687A.222W}. As a result of the implicit assumption of single star evolution in core structures, the core of an accretor star that experienced case A mass transfer will have a core size commensurate with its final mass, without accounting for the core growth prior to mass transfer. Thefore, our simulations may overpredict the delay to the star's core collapse time. However, this is primarily relevant for case A mass transfer, which we find only $9\%$ of stars in our fiducial simulation experience, and therefore we expect the impact of this effect will be small.

An additional consideration is that previous rejuvenation after accreting matter from a companion can change the binding energy of a star's envelope and affect its chance of surviving a common-envelope event \citep[e.g.,][]{Renzo+2023, Landri+2025:2025ApJ...979...57L}. Thus the number of merger progenitors for SNe that we predict could be affected by including this effect.

Future updates to population synthesis prescriptions for main sequence core evolution and rejuvenation are necessary to improve upon this analysis.

\paragraph{Explodability of massive stars} In our population synthesis models we use the \citet{Fryer+2012:2012ApJ...749...91F} remnant mass prescription to infer the final core structure of the star and assume that all stars with a non-zero ejecta mass explode as SNe. However, it is possible that many of these stars may instead implode, failing to produce a SN as a result. Many works have considered the specific requirements for a massive star to end its life in a SN \citep[e.g.,][]{O'Connor+2011:2011ApJ...730...70O,Sukhbold+2014:2014ApJ...783...10S, Sukhbold+2016:2016ApJ...821...38S, Ertl+2016:2016ApJ...818..124E, Ertl+2020:2020ApJ...890...51E,Laplace+2021:2021A&A...656A..58L,Laplace+2024:2024arXiv240902058L,Ugolini+2025:2025arXiv250118689U}. In particular, recent work has shown that the explodability depends on a variety of factors beyond CO core mass, including composition and rotation. \citep[e.g.,][]{Patton+2020:2020MNRAS.499.2803P}. Yet it is also important to consider that cores that fail to explode may still produce ejecta and impact their galactic surroundings \citep[e.g.,][]{Piro+2013:2013ApJ...768L..14P, Lovegrove+2013:2013ApJ...769..109L, Antoni+2022:2022MNRAS.511..176A, Antoni+2023:2023MNRAS.525.1229A}, which, for example, may be achieved through disk outflows from massive collapsars \citep{Siegel+2022:2022ApJ...941..100S}. Future work should consider how applying updated prescriptions for pre-supernova core structure in population synthesis could impact the total SN feedback produced by massive stars.

\paragraph{Dynamical cluster ejections} During the early evolution of clusters, stars can be ejected at high velocities as a result of dynamical N-body encounters \citep{Poveda+1967}. Our simulations do not account for dynamical ejections, focusing only on creation of runaway stars via binary ejections \citep{Blaauw+1961,Boersma+1961}. This means that the distance distributions for SNe that we present are a lower limit on the true distribution. In reality, a fraction of these stars will receive additional dynamical kicks, which would result in more extended tails. This would additionally mean that primary star progenitors could also reach core collapse far from star forming regions, though we expect that the progenitors of the most distant SNe would remain secondary accretor stars. In principle,
given mass-dependent (of the star and cluster) ejection velocity distributions from N-body simulations \citep[e.g.,][]{Oh+2016:2016A&A...590A.107O}, these additional velocity kicks could be applied in our simulations at a minimal extra cost.

\paragraph{Higher-order stellar multiples} Progenitors of core-collapse SNe are expected to often be found in higher order stellar multiples. Indeed it has been predicted that the \textit{majority} of O-type stars are formed in triples and higher-order stellar multiples \citep{Offner+2023:2023ASPC..534..275O}. These systems could produce similar effects to dynamical cluster ejections (see above) by ejecting companions, and hence they should be considered in future work. Our simulations cannot evolve higher-order stellar multiples but there are several codes that address this problem, such as \texttt{TRES}, \texttt{MSE} and \texttt{TSE} \citep{Toonen+2016:2016ComAC...3....6T, Hamers+2021:2021MNRAS.502.4479H, Stegmann+2024:2024JOSS....9.7102S, Preece+2024:2024arXiv241214022P}. These codes could be coupled to galactic dynamics (as \cogsworth does with \cosmic and \gala) and applied in future work.

\paragraph{Self-consistency of post-processing hydrodynamical simulations} In this work we have used the star formation history of \fire \texttt{m11h} (and the \changa galaxy \texttt{r442} as a variation) to seed star formation in \cogsworth simulations. However, we did not adjust the feedback in the underlying hydrodynamical simulations to account for how binary physics may affect it and therefore change the star formation history of the galaxy. Nevertheless, given that the effects on feedback that we discuss are dependent on stellar evolution and binary interactions we do not expect that changes to the overall star formation rate would impact our results significantly.

\section{Comparison to previous work}\label{sec:comparison_previous_work}


This paper presents a thorough exploration of the impacts of binary evolution models on key aspects of SNe feedback. However, this work builds upon a large number of papers that have also explored some of these same issues, using a variety of techniques and physical assumptions.  Here, we place our results in the context of this existing body of work, exploring points of agreement and seeking insight from areas of potential tension.

\subsection{Core-collapse SN timing}

The impact of binaries on the delay time distribution of core-collapse SNe has been considered in several previous works \citep[e.g.,][]{DeDonder+2003:2003NewA....8..817D, Xiao+2015:2015MNRAS.452.2597X, Zapartas+2017:2017AA...601A..29Z}. These works adopt a series of different assumptions regarding initial conditions and binary physics, as well as use different simulation codes.

The earliest work exploring these effects systematically is \citealt{DeDonder+2003:2003NewA....8..817D} (hereafter \citetalias{DeDonder+2003:2003NewA....8..817D}), which used the population synthesis code described in \citet{Vanbeveren+1998:1998NewA....3..443V} to explore the galactic evolution of SN rates. Similar to our work, they make predictions for the rate of SNe after a fixed starburst at solar metallicity.

\citetalias{DeDonder+2003:2003NewA....8..817D} generally find a significantly higher rate of late SNe than our work.
We use the data from their plots, which was digitized by  \citet{Zapartas+2017:2017AA...601A..29Z}, to compute that for a $100\%$ binary population, $52\%$ of their SNe occur after $44\unit{Myr}$. This fraction is double the rate of late SNe that we find in our fiducial model (${\sim}25\%$).

The discrepancy between our results can be somewhat explained by differences in our choices regarding initial conditions. In their simulations, \citetalias{DeDonder+2003:2003NewA....8..817D} assume a bottom-heavy IMF with $\alpha_{\rm IMF} = -2.7$, thus their result is more comparable to our variation in which we find $\flate=33\%$ (see Figure~\ref{fig:trends-tails}). Moreover, they assume an upper orbital period limit of $10^3\unit{days}$. For this limit we also find $\flate = 33\%$. It is likely that a combination of this upper limit and IMF produces a much higher fraction of late SNe. Their assumed orbital period distribution is different as well, further complicating a precise comparison with work we present here. An additional consideration is that modern prescriptions of stellar wind mass loss are generally more conservative, which may lead more of our massive stars to reach core collapse on shorter timescales. Overall, the difference between our results are understandable given the various updates to binary evolution models in the past two decades.

\citealt{Zapartas+2017:2017AA...601A..29Z} (hereafter \citetalias{Zapartas+2017:2017AA...601A..29Z}) is a more recent work that investigated the delay time distribution of core-collapse SNe, exploring a series of variations similar to ours. Their physical assumptions are closer to those we adopt as well, making a comparison more straightforward than with \citetalias{DeDonder+2003:2003NewA....8..817D}. They do use a different binary synthesis code, \texttt{binary\_c} \citep{Izzard+2004:2004MNRAS.350..407I,Izzard+2006:2006A&A...460..565I,Izzard+2009:2009A&A...508.1359I,Izzard+2018:2018MNRAS.473.2984I,Izzard+2023:2023MNRAS.521...35I}, but it is based on the same underlying fitting formulae as \cosmic \citep{Pols+1998:1998MNRAS.298..525P, Hurley+2000:2000MNRAS.315..543H, Hurley+2002:2002MNRAS.329..897H,Breivik2020}.

Our results for the rate of late SNe are close to agreement with \citetalias{Zapartas+2017:2017AA...601A..29Z}. In their model assuming a binary fraction of $100\%$, they find that $20\%$ of core-collapse SNe occur at least $50\unit{Myr}$ after a star formation event. For most binaries in their simulation (those with $m_1 < 15 \unit{M_\odot}$) they assume a power law slope for their initial orbital periods of $\pi = 0$ following \citet{Opik+1924:1924PTarO..25f...1O}. For our $\pi = 0$ simulation, we find that $15\%$ of core-collapse SNe occur after $50\unit{Myr}$. The remaining discrepancy in our results can be understood through a difference in the upper period limit of binaries. In particular, we assume that upper orbital period limit is $10^{5.5} \unit{days}$, whilst \citetalias{Zapartas+2017:2017AA...601A..29Z} assumes a limit of $10^{3.5} \unit{days}$. As such, we will produce more wide, non-interacting binaries, which skew our results to having fewer late SNe (see Section~\ref{sec:ini_orb_period}).

We are in broad agreement with \citetalias{Zapartas+2017:2017AA...601A..29Z} on the trends associated with variations in initial conditions and binary physics. This agreement is evident from a comparison of our Table~\ref{tab:totals_and_stats} to Table 2 of \citetalias{Zapartas+2017:2017AA...601A..29Z}. We note that the absolute values are shifted given the difference in our assumptions regarding the fiducial binary fraction and upper orbital period limit.

One point of apparent difference is that we find an opposite trend with metallicity.  We find that the rate of late SNe increases with decreasing metallicity, whilst \citetalias{Zapartas+2017:2017AA...601A..29Z} find the opposite trend. We have traced this difference to different choices for the definition of ``late''. For our models, we use the \fire-3 limit of $44\unit{Myr}$, while \citetalias{Zapartas+2017:2017AA...601A..29Z} use the time of the final SN from a single star at each different metallicity. These alternate definitions lead to our outwardly conflicting claims regarding the trend with metallicity. We explain this reasoning in more detail, and confirm that using the same criteria in our simulations results in an agreement with \citetalias{Zapartas+2017:2017AA...601A..29Z}, in Appendix~\ref{app:manos_compare}.

\subsection{Spatial distribution of core-collapse SNe}

Earlier works have considered the rates, ejection velocities of massive runaway stars and the impact on the spatial distribution of core-collapse SNe \citep[e.g.,][]{DeDonder+1997:1997A&A...318..812D, Eldridge+2011:2011MNRAS.414.3501E, Boubert+2018:2018MNRAS.477.5261B, Renzo+2019:2019A&A...624A..66R}. We compare directly to a subset of these works which considered the spatial distribution of massive stars relative to star forming regions.

\citet{Eldridge+2011:2011MNRAS.414.3501E} (hereafter \citetalias{Eldridge+2011:2011MNRAS.414.3501E}) investigates the spatial distribution of different types of SNe. They simulate binary populations with the Cambridge \texttt{STARS} code, a detailed 1D stellar evolution code \citep{Eggleton+1971:1971MNRAS.151..351E, Pols+1995:1995MNRAS.274..964P, Eldridge+2004:2004MNRAS.348..201E, Eldridge+2008:2008MNRAS.384.1109E} and explore the fraction of binaries that are disrupted and the ejection velocities of secondary stars. They make predictions for the distances that stars will travel before exploding, but they do not account for the impact of a gravitational potential, instead multiplying the ejection velocity of the star by its remaining lifetime after ejection.

We find a higher fraction of binaries are disrupted, likely as a result of our initial orbital period distribution. \citetalias{Eldridge+2011:2011MNRAS.414.3501E} find that 70\% of binaries are disrupted after the first SN, whilst we find a fraction of 85\%. They use an upper initial orbital separation limit of $10^{4}\rsun$, while our limit is closer to $10^{5.1}\rsun$. This means that our simulations contain more weakly bound binaries, which are easier to disrupt and may account for the difference in our results.

Our results for the fraction of SNe at moderate distances are in reasonable agreement, but  \citetalias{Eldridge+2011:2011MNRAS.414.3501E} predict a factor of 4 more SNe at large distances. 
They find that $\ffar=14\%$, and $\fdistant=4\%$ travel more than $500\unit{pc}$. Their results are most comparable to our $\pi = 0$ variation in which we assume a flat-in-log distribution for the initial orbital period. In this variation, we find $\ffar=11\%$ and $\fdistant=1\%$.

The significantly higher distant SN fraction is likely due to the lack of a galactic gravitational potential in their models. In their work, they assume stars travel in a straight line without the influence of a potential, which allows them to travel much further. It is also worth nothing that their lower initial orbital separation limit (which reduces the number of disruptions on average) and higher initial mass limit of $5\msun$ compared to our $4\msun$ (which translates to shorter lifetimes after ejection) both mean that their distant SN fractions in an equivalent simulation would likely be even higher.

\citet{Renzo+2019:2019A&A...624A..66R} (hereafter \citetalias{Renzo+2019:2019A&A...624A..66R}) also investigates massive runaway stars and their spatial distribution using assumptions that are closer to our own, using \texttt{binary\_c} \citep{Izzard+2004:2004MNRAS.350..407I,Izzard+2006:2006A&A...460..565I,Izzard+2009:2009A&A...508.1359I,Izzard+2018:2018MNRAS.473.2984I,Izzard+2023:2023MNRAS.521...35I}, which as we noted above is based on the same underlying fitting formulae as \cosmic \citep{Pols+1998:1998MNRAS.298..525P, Hurley+2000:2000MNRAS.315..543H, Hurley+2002:2002MNRAS.329..897H,Breivik2020}.

We are in strong agreement on the fraction of binaries that are disrupted. \citetalias{Renzo+2019:2019A&A...624A..66R} finds that $86_{-22}^{+10}\%$ of binaries are disrupted after the first SN. The uncertainties on this estimate come from several population synthesis parameter variations, which include extreme choices for SN natal kicks. Our fiducial finding of 85\% is in excellent agreement with this result. We additionally concur that the most impactful parameters for the disruption fraction are the natal kick magnitude and metallicity.

We find a distribution of ejection velocities that is similar to that of \citetalias{Renzo+2019:2019A&A...624A..66R}'s fiducial model, with a slightly extended tail to higher velocities. Their distribution peaks at a velocity of $6\unit{km}{s^{-1}}$, in close agreement to our finding of $7 \unit{km}{s^{-1}}$. However, we find that 4.3\% of secondaries are ejected at more than $60\unit{km}{s^{-1}}$, compared to $0.2\%$ reported by \citetalias{Renzo+2019:2019A&A...624A..66R}. We speculate that this difference arises because of differences between our codes in the treatment of mass transfer and most importantly orbital evolution, which determines the ejection velocity of runaway stars.


Despite the similarity in our distribution of ejection velocities, we find a significantly lower tail of long distances for runaway star SNe. \citetalias{Renzo+2019:2019A&A...624A..66R} calculates the \textit{maximum} distance travelled as the product of the ejection velocity and stellar lifetime, neglecting the effect of a galactic potential. They find that on average main-sequence runaway stars travel $126\unit{pc}$, which is very close to our finding of $120\unit{pc}$. However, they find that ${\sim}35\%$ of these stars travel more than $100\unit{pc}$, in contrast to the ${\sim}11\%$ we find in our fiducial model. These comparisons indicate that neglecting a galactic potential may be acceptable for reproducing the bulk properties of the distribution but fails to accurately predict the long distance tails.

\section{Conclusions}\label{sec:conclusions}

We present predictions for the impact of binary interactions on the timing and location of core-collapse SNe feedback. We used self-consistent population synthesis and galactic dynamics simulations to trace the time and location of each SN from the recent star formation in the \fire \texttt{m11h} galaxy. We compared our results to an equivalent simulation that considers only single stellar evolution. We additionally repeated our simulations for a wide-range of variations in initial conditions, binary physics and galactic settings to demonstrate the robustness of our results. Based on these simulations, we designed an analytic model for core-collapse SN feedback that accounts for binary interactions. We additionally consider how the impact of binary interactions on SN feedback could affect galaxy evolution. Lastly, we compared our findings to earlier works that have previously explored the impact of binary interactions on producing late SNe and runaway stars. Our main findings can be summarised as follows:

\begin{enumerate}
    \item \textbf{Binary interactions can produce  late core-collapse SN feedback}\\We predict that ${\sim}25\%$ of core-collapse SN feedback occurs long after a star formation event, beyond the $44$ Myr cutoff typically adopted in hydrodynamical simulations. The progenitors of these SNe are primarily the products of stellar mergers.
    \item \textbf{Supernovae from binaries are displaced from their parent clusters}\\We find that ${\sim}14\%$ of core-collapse feedback occurs at least $100\unit{pc}$ from the progenitor's parent cluster. The progenitors of these SNe are primarily secondary stars that were ejected from their binary after the primary star's SN.
    \item \textbf{The distributions of times and locations are robust to changes in model parameters}\\We demonstrate that the overall distributions are surprisingly robust to variations across binary physics, initial conditions, and galactic settings. The median of the distribution of SN delay times remains within 7 Myr of the fiducial model across all variations. Similarly the distance median remains within 25 pc of the fiducial model.
    \item \textbf{An analytic model accurately reproduces joint distributions of SN times and progenitor velocities, allowing binary evolution to be seamlessly included in SNe feedback models in hydrodynamical simulations}\\We develop a metallicity-dependent analytic model for the timing of SNe and the velocities at which their progenitors travel. This physically-motivated model reproduces the timing distribution to within $0.5\%$ and the runaway ejection velocity distribution to within $4\%$ (Figures~\ref{fig:time-fit}--\ref{fig:vel-fit}).
    \item \textbf{The tails of the spatial distribution are sensitive to changes in binary mass transfer}\\The ejection velocities of secondary stars depend on their pre-supernova orbital velocity and thus are sensitive to binary mass transfer variations which alter orbital parameters. In particular, the fraction of feedback beyond $100$ pc varies from $10$--$21$\% and the fraction beyond $500$ pc varies from $0$--$3$\% (Figure~\ref{fig:trends-tails}).
    \item \textbf{The fraction of late and displaced SNe increases at low metallicity}\\
    Decreasing metallicity by a factor of 10 increases \flate to $34\%$ and $\ffar$ to $21\%$, as well as increasing the overall number of SNe by $17\%$. These trends are a result of the metallicity dependence of the assumed minimum core mass to reach core-collapse, in addition to the reduced radial expansion at low metallicity (Figure~\ref{fig:var-z} and Section~\ref{sec:metallicity_variation}).
\end{enumerate}

Overall, we have shown that core-collapse SN feedback is significantly different for binary star progenitors, yet is robust to many of the uncertainties in binary physics. Our analytic model for feedback enables future simulators to move beyond single star models at a low cost and include the more realistic impact of binaries on galaxy evolution. 

We expect that the most substantial differences will be evident in low-metallicity, high-redshift environments, in which the extended time and spatial distributions of feedback from binaries are most prominent. Future investigations of binaries should be particularly directed towards better understanding the initial mass function, orbital period distribution and binary mass transfer physics, since these drive most of the remaining uncertainties in our models for core-collapse SN feedback.

\section*{Acknowledgements}
We thank Manos Zapartas, Dimitris Souropanis, Chris Hayward, Selma de Mink, Stephen Justham, Mike Grudi\'{c}, Tobin Wainer and Lachlan Lancaster for helpful and insightful discussions related to this work. We additionally thank Manos Zapartas for making the digitized data from \citet{DeDonder+2003:2003NewA....8..817D} available to us for our comparisons to their work. T.W. thanks the Simons Foundation, Flatiron Institute and Center for Computational Astrophysics for running the pre-doctoral program during which this work was initiated. We thank the Lehman Garrison, Dylan Simon and the whole Scientific Computing Core at the Flatiron Institute for their help in running these simulations. The Flatiron Institute is funded by the Simons Foundation. T.W., K.B. and E.C.B. acknowledge support from NASA ATP grant 80NSSC24K0768.

\software{This research made use of \texttt{cogsworth} and its dependencies \citep{Wagg+2025:2025JOSS...10.7400W, Wagg+2025:2025ApJS..276...16W, Breivik2020, COSMIC_15164778, gala_JOSS, gala_13377376}, as well as the following software packages: \texttt{astropy} \citep{astropy:2013, astropy:2018, astropy:2022}, \texttt{Jupyter} \citep{2007CSE.....9c..21P, kluyver2016jupyter}, \texttt{matplotlib} \citep{Hunter:2007}, \texttt{numpy} \citep{numpy}, \texttt{pandas} \citep{mckinney-proc-scipy-2010, pandas_8301632}, \texttt{python} \citep{python}, \texttt{scipy} \citep{2020SciPy-NMeth, scipy_8259693}, \texttt{Cython} \citep{cython:2011}, \texttt{h5py} \citep{collette_python_hdf5_2014, h5py_7560547}, \texttt{pynbody} \citep{pynbody, pynbody_10276404}, \texttt{schwimmbad} \citep{schwimmbad}, \texttt{seaborn} \citep{Waskom2021}, and \texttt{tqdm} \citep{tqdm_8233425}. This research has made use of NASA's Astrophysics Data System. Software citation information aggregated using \texttt{\href{https://www.tomwagg.com/software-citation-station/}{The Software Citation Station}} \citep{software-citation-station-paper, software-citation-station-zenodo}. We use simulations from the \fire-2 public data release \citep{Wetzel2023}. The \fire-2 cosmological zoom-in simulations of galaxy formation are part of the Feedback In Realistic Environments (\fire) project, generated using the Gizmo code \citep{Hopkins2015} and the \fire-2 physics model \citep{Hopkins2018a}. The \changa simulation \texttt{r442} was
run using resources made available by the Flatiron Institute. The Flatiron Institute is a division of the
Simons Foundation.}

\bibliographystyle{aasjournal}
\bibliography{bibs/paper, bibs/software, bibs/cosmic}{}

\restartappendixnumbering

\allowdisplaybreaks
\appendix



\section{Dependence of late SN fraction on metallicity}\label{app:manos_compare}

In Section~\ref{sec:comparison_previous_work}, we highlighted that our results for the trend of the late SN fraction as a function of metallicity seem to disagree with \citet{Zapartas+2017:2017AA...601A..29Z} upon immediate inspection. In this Appendix, we demonstrate that this discrepancy is in fact due to a difference in our definitions of ``late'' SNe and that, when using the same definition, our results are in agreement.

\citet{Zapartas+2017:2017AA...601A..29Z} found that a decrease in metallicity also leads to a decrease in the fraction of late SNe, which the authors point out is a result of lower metallicity stars being more compact and therefore interacting later or avoiding interaction entirely. Later interactions (or a lack of interactions) will have a lesser effect on the core mass of the star, which determines its time until core collapse. In contrast, we find the opposite trend, that decreasing metallicity increases \flate. We argue this is primarily a results of lower metallicity decreasing the mass required to reach core collapse, which results in more (slowly evolving) low mass stars being able to reach core collapse.

The difference in our results is a result of our choices of the time beyond which SNe are considered ``late''. We use the \fire-3 limit of $44\unit{Myr}$, while \citet{Zapartas+2017:2017AA...601A..29Z} uses the time of the final SN from a single star population at each different metallicity.

\begin{figure}
    \centering
    \includegraphics[width=\columnwidth]{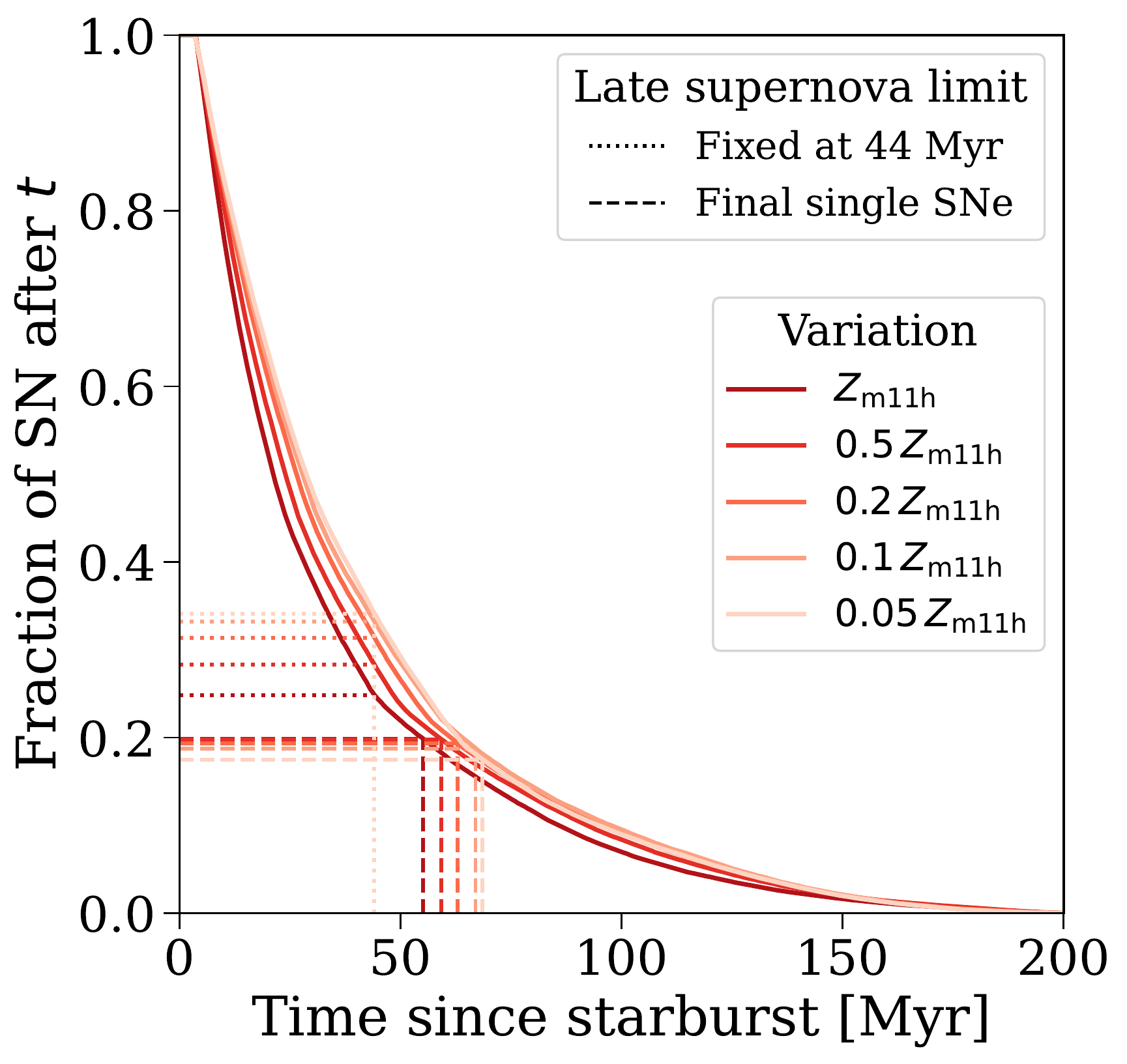}
    \caption{The fraction of SNe that occur after a given time for the difference metallicity variations we consider are shown in coloured lines. The dotted lines indicate the fraction of late SNe when using a fixed 44 Myr limit. The dashed lines show the same fraction when using a limit set by the time of the last single star SN. The fixed limit results in a negative correlation of the late SNe fraction with metallicity, while the variable limit gives a positive correlation.}
    \label{fig:manos-metallicity}
\end{figure}

We apply both limit choices to our simulations in Figure~\ref{fig:manos-metallicity}. Our choice of a fixed limit results in a negative correlation of the late SNe fraction with metallicity, while the variable limit used by \citet{Zapartas+2017:2017AA...601A..29Z} gives a positive correlation. This is because the variable limit results in the additional SNe from lower mass stars reaching core collapse no longer being counted towards the ``late'' SNe. Instead, as \citet{Zapartas+2017:2017AA...601A..29Z} explains, the more compact stars at low metallicity have later and fewer interactions, which reduces the number of delayed SNe and hence the late SN fraction. We note that the time of the last single star reaching core collapse is dependent upon the underlying stellar tracks used in population synthesis \citep{Pols+1998:1998MNRAS.298..525P}.

Overall, our results are therefore in agreement with \citet{Zapartas+2017:2017AA...601A..29Z} when using the same definition of ``late'' SN. Our fixed limit corresponds to the limit currently applied in \fire-3 subgrid feedback models and hence is most applicable for the purposes of this work.

\section{Sampling routine flowchart}\label{app:flowchart}

\begin{figure}[htbp]
    \centering
    \begin{tikzpicture}
        \node (start) [startstop] {Start};
        \node (timesample) [process, below of=start, yshift=-0.5cm, text width=3.5cm] {Sample $t_{\rm SN}$ (Eq.~\ref{eq:time_fit})};
        \node (runaway) [decision, below of=timesample, yshift=-1.25cm, text width=2.5cm] {Progenitor ejected? (Eq.~\ref{eq:f-eject})};
        \node (mt) [decision, below of=runaway, yshift=-2cm, text width=2cm] {Experienced MT? (Eq.~\ref{eq:f_noMT})};
        \node (mtType) [decision, below of=mt, yshift=-3cm, text width=2cm] {Type of MT? (Eq.~\ref{eq:f-MT-A}--\ref{eq:f-CE})};
        \node (unejected) [process, right of=runaway, xshift=3.0cm, text width=2cm] {Sample $v_{\rm SN}$ from cluster velocity dispersion (Eq.~\ref{eq:vel-fit-cluster})};
        \node (ejectedNoMT) [process, below of=unejected, yshift=-2cm, text width=2cm] {Sample $v_{\rm SN}$ from Eq.~\ref{eq:vel-fit-nomt}};
        \node (ejectedCaseA) [process, below of=ejectedNoMT, yshift=-1cm, text width=2cm] {Sample $v_{\rm SN}$ from Eq.~\ref{eq:vel-fit-mt-a}};
        \node (ejectedCaseBC) [process, below of=ejectedCaseA, yshift=-1cm, text width=2cm] {Sample $v_{\rm SN}$ from Eq.~\ref{eq:vel-fit-mt-bc}};
        \node (ejectedCE) [process, below of=ejectedCaseBC, yshift=-1cm, text width=2cm] {Sample $v_{\rm SN}$ from Eq.~\ref{eq:vel-fit-ce}};
        \node (stop) [startstop, below of=ejectedCE, xshift=-2cm,yshift=-1cm] {Return $t_{\rm SN}, v_{\rm SN}$};

        \draw [arrow] (start) -- (timesample);
        \draw [arrow] (timesample) -- (runaway.north);
        \draw [arrow] (runaway.east) -- node[anchor=south] {No} (unejected.west);
        \draw [arrow] (runaway.south) -- node[anchor=east] {Yes} (mt.north);
        \draw [arrow] (mt.east) -- node[anchor=south] {No} (ejectedNoMT.west);
        \draw [arrow] (mt.south) -- node[anchor=east] {Yes} (mtType.north);
        \draw [arrow] (mtType.east) |- node[anchor=east] {Case A} (ejectedCaseA.west);
        \draw [arrow] (mtType.east) -- node[anchor=south, text width=1cm, text centered] {Case B/C} (ejectedCaseBC.west);
        \draw [arrow] (mtType.east) |- node[anchor=east, text width=1.5cm, text centered] {Common envelope} (ejectedCE.west);
        \draw [arrow] (unejected.east) -- +(0.5,0) -- +(0.5,-11)  -- (stop.east);
        \draw [arrow] (ejectedNoMT.east) -- +(0.5,0);
        \draw [arrow] (ejectedCaseA.east) -- +(0.5,0);
        \draw [arrow] (ejectedCaseBC.east) -- +(0.5,0);
        \draw [arrow] (ejectedCE.east) -- +(0.5,0);
        \draw [arrow, line width=2pt] (unejected.east) +(1.25,0) -- node[rotate=90, yshift=0.2cm] {Increasing average velocity} +(1.25,-9);
    \end{tikzpicture}
    \caption{A flowchart of the sampling routine for our analytic model of SN feedback, where $t_{\rm SN}$ is the time of the SN and $v_{\rm SN}$ is the velocity at which its progenitor moves away from its parent cluster. Based on the sampled time, we draw a progenitor velocity, $v_{\rm SN}$. The distribution for this velocity changes based on (a) whether the progenitor was ejected from its binary and, if so, (b) whether it experienced mass transfer before doing so, and, if so, (c) what type of mass transfer it experienced. The velocity distributions are sorted in this flowchart from top to bottom as slowest to fastest (where unejected progenitors proceed slowest on average, whilst stars ejected after a common-envelope are typically fastest).}
    \label{fig:flowchart}
\end{figure}
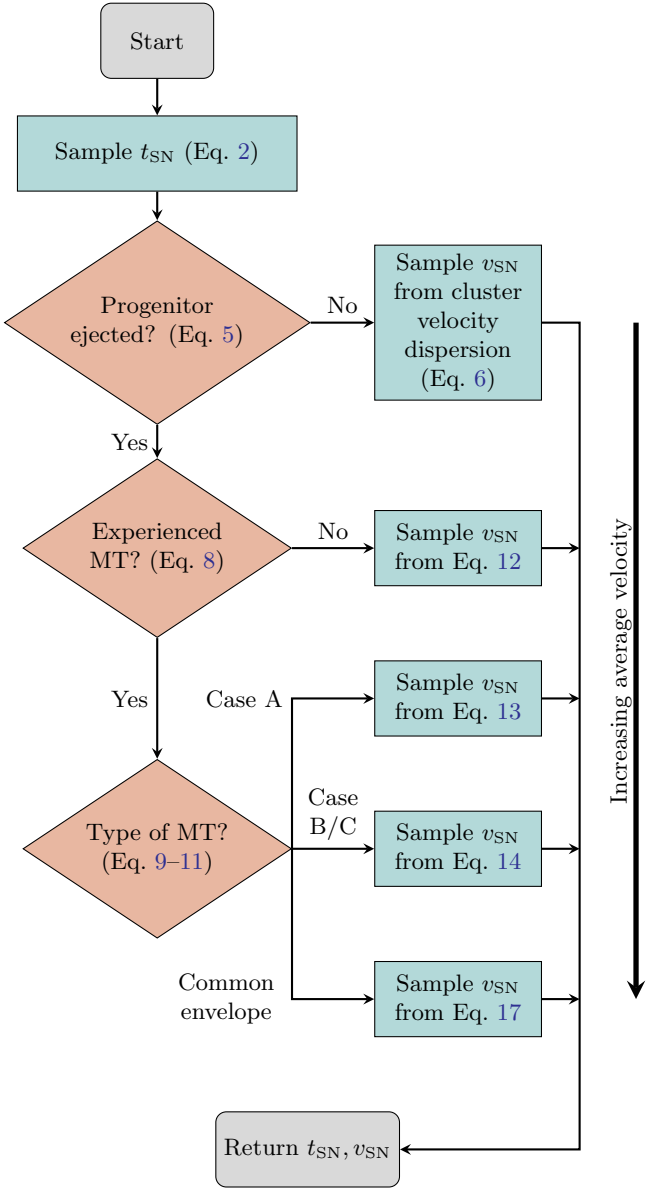

This appendix contains Figure~\ref{fig:flowchart}, which illustrates the sampling routine for the analytic model that we present in Section~\ref{sec:fits}.

\section{SN rate and distribution data}\label{app:supple}

This section contains supplementary data for this work. Table~\ref{tab:totals_and_stats} list the total number of SNe that occur in each simulation per $100\msun$, separated by progenitor type, as well as the fractions of SNe in the tails of the timing and distance distributions. Table~\ref{tab:percentiles} details the data for the total distributions and SN times and distances. 

\begin{table*}
    \centering
    \begin{tabular}{l cccc |ccc}
        \hline\hline
        \multirow{2}{*}{Model variation} & \multicolumn{4}{c}{Supernovae per $100 \msun$} & \multicolumn{3}{c}{Distribution tails} \\
        \cline{2-5} \cline{6-8}
        & All & P & S & MP & \flate & \ffar & \fdistant \\
        \hline
        \textbf{Fiducial} & 1.18 & 0.39 & 0.25 & 0.52 & 25.4\% & 13.2\% & 0.9\% \\
        \textbf{Binary physics} &&&&&&& \\
        \quad Common envelope, $\alpha_{\rm CE} = 0.1$ & 1.21 & 0.35 & 0.24 & 0.61 & 25.0\% & 12.2\% & 0.5\% \\
        \quad Common envelope, $\alpha_{\rm CE} = 10.0$ & 1.02 & 0.43 & 0.32 & 0.25 & 13.8\% & 11.0\% & 1.0\% \\
        \quad Case B Unstable & 1.13 & 0.18 & 0.08 & 0.85 & 31.1\% & 11.6\% & 0.4\% \\
        \quad Case B Stable & 1.37 & 0.64 & 0.41 & 0.30 & 30.7\% & 21.0\% & 3.1\% \\
        \quad Mass transfer efficiency, $\beta = 0.0$ & 0.94 & 0.38 & 0.23 & 0.31 & 12.3\% & 13.0\% & 2.0\% \\
        \quad Mass transfer efficiency, $\beta = 0.5$ & 1.06 & 0.39 & 0.26 & 0.39 & 17.9\% & 13.7\% & 1.6\% \\
        \quad Mass transfer efficiency, $\beta = 1.0$ & 1.21 & 0.39 & 0.26 & 0.55 & 26.7\% & 14.0\% & 0.6\% \\
        \quad Supernova kicks, $\sigma_{\rm CC} = 20 \, {\rm km/s}$ & 1.04 & 0.39 & 0.12 & 0.51 & 28.3\% & 10.2\% & 0.1\% \\
        \quad Supernova kicks, $\sigma_{\rm low} = 265 \, {\rm km/s}$ & 1.19 & 0.39 & 0.26 & 0.52 & 25.3\% & 13.5\% & 1.0\% \\
        \quad Supernova kicks, No fallback & 1.18 & 0.39 & 0.25 & 0.52 & 25.3\% & 13.3\% & 0.9\% \\
        \textbf{Initial conditions} &&&&&&& \\
        \quad Singles, $f_{\rm bin} = 0.0$ & 1.07 & 0.00 & 0.00 & 0.00 & 1.3\% & 0.9\% & 0.0\% \\
        \quad Initial mass function slope, $\alpha_{\rm IMF} = -1.9$ & 1.95 & 0.71 & 0.49 & 0.71 & 17.9\% & 11.0\% & 1.0\% \\
        \quad Initial mass function slope, $\alpha_{\rm IMF} = -2.7$ & 0.57 & 0.17 & 0.10 & 0.29 & 33.6\% & 15.5\% & 0.8\% \\
        \quad Orbital period slope, $\pi = 0$ & 1.12 & 0.47 & 0.28 & 0.33 & 17.5\% & 11.2\% & 1.0\% \\
        \quad Orbital period slope, $\pi = -1$ & 1.24 & 0.31 & 0.21 & 0.71 & 32.8\% & 14.8\% & 0.8\% \\
        \quad Initial upper orbital period limit, $P_{\rm 0, max} = 10^{3} \, {\rm d}$ & 1.21 & 0.28 & 0.22 & 0.72 & 33.3\% & 15.3\% & 1.0\% \\
        \quad Mass ratio slope, $\kappa = 1$ & 1.20 & 0.38 & 0.28 & 0.51 & 27.2\% & 13.4\% & 0.7\% \\
        \quad Mass ratio slope, $\kappa = -1$ & 1.14 & 0.37 & 0.19 & 0.56 & 22.3\% & 11.7\% & 1.1\% \\
        \textbf{Metallicity} &&&&&&& \\
        \quad Metallicity, $\bar{Z} = 0.5 \, \bar{Z}_{\rm m11h}$ & 1.24 & 0.42 & 0.27 & 0.53 & 29.0\% & 16.9\% & 1.0\% \\
        \quad Metallicity, $\bar{Z} = 0.2 \, \bar{Z}_{\rm m11h}$ & 1.30 & 0.46 & 0.29 & 0.54 & 31.7\% & 19.9\% & 1.3\% \\
        \quad Metallicity, $\bar{Z} = 0.1 \, \bar{Z}_{\rm m11h}$ & 1.36 & 0.49 & 0.31 & 0.56 & 33.5\% & 21.0\% & 1.6\% \\
        \quad Metallicity, $\bar{Z} = 0.05 \, \bar{Z}_{\rm m11h}$ & 1.40 & 0.52 & 0.32 & 0.55 & 34.6\% & 22.4\% & 2.4\% \\
        \textbf{Galaxy settings} &&&&&&& \\
        \quad Velocity dispersion, $v_{\rm disp} = 0.5 \ {\rm km/s}$ & 1.18 & 0.39 & 0.25 & 0.52 & 25.4\% & 5.1\% & 0.9\% \\
        \quad Velocity dispersion, $v_{\rm disp} = 5 \ {\rm km/s}$ & 1.18 & 0.39 & 0.25 & 0.52 & 25.4\% & 45.3\% & 3.0\% \\
        \quad ChaNGa (r442) & 1.17 & 0.38 & 0.25 & 0.51 & 25.3\% & 13.9\% & 0.9\% \\
    \end{tabular}
    \caption{The total numbers of SNe per $100 \msun$ for different subpopulations as well as summary statistics for the tails of the timing and distance distributions, for each model variation in our simulations. 
    Column 1 indicates the variation, column 2-5 are the total SNe per $100\msun$ for the total population, primary stars (P), secondary stars (S) and merger products (MP) respectively, column 6 is the fraction of SNe that occur after 44 Myr, columns 7 and 8 are the fractions of SNe that occur beyond 100pc and 500pc respectively.
    Each column gives the value for a different subpopulation and the total number.}
    \label{tab:totals_and_stats}
\end{table*}

\begin{table*}
    \centering
    \begin{tabular}{lccccc|ccccc}
        \hline\hline
        \multirow{2}{*}{Model variation} & \multicolumn{5}{c}{Supernova time [Myr]} & \multicolumn{5}{c}{Distance from parent cluster [pc]} \\
        \cline{2-11}
        & 2.5 & 25 & 50 & 75 & 97.5 & 2.5 & 25 & 50 & 75 & 97.5 \\
        \hline
        \textbf{Fiducial} & 4.4 & 10.8 & 21.4 & 44.6 & 143.5 & 4.6 & 16.4 & 34.7 & 66.8 & 242.1 \\
        \textbf{Binary physics} &&&&&&& \\
        \quad Common envelope, $\alpha_{\rm CE} = 0.1$ & 4.4 & 11.0 & 22.2 & 44.0 & 139.9 & 4.6 & 16.7 & 34.8 & 65.3 & 208.7 \\
        \quad Common envelope, $\alpha_{\rm CE} = 10.0$ & 4.3 & 9.5 & 17.5 & 31.5 & 105.6 & 4.4 & 14.7 & 30.3 & 58.8 & 250.5 \\
        \quad Case B Unstable & 4.3 & 12.1 & 27.1 & 54.2 & 142.5 & 4.4 & 15.8 & 33.6 & 63.8 & 197.8 \\
        \quad Case B Stable & 4.5 & 12.1 & 25.4 & 58.0 & 185.5 & 4.9 & 18.5 & 40.4 & 85.1 & 556.7 \\
        \quad Mass transfer efficiency, $\beta = 0.0$ & 4.2 & 9.0 & 17.0 & 31.1 & 92.2 & 4.1 & 13.4 & 28.3 & 59.2 & 412.8 \\
        \quad Mass transfer efficiency, $\beta = 0.5$ & 4.3 & 10.0 & 19.1 & 36.3 & 99.3 & 4.4 & 15.1 & 32.0 & 64.6 & 345.7 \\
        \quad Mass transfer efficiency, $\beta = 1.0$ & 4.4 & 11.0 & 22.0 & 47.1 & 143.3 & 4.6 & 16.8 & 35.6 & 68.9 & 237.5 \\
        \quad Supernova kicks, $\sigma_{\rm CC} = 20 \, {\rm km/s}$ & 4.3 & 10.7 & 22.9 & 50.2 & 147.7 & 4.3 & 14.6 & 30.5 & 59.3 & 177.4 \\
        \quad Supernova kicks, $\sigma_{\rm low} = 265 \, {\rm km/s}$ & 4.4 & 10.8 & 21.6 & 44.4 & 143.3 & 4.6 & 16.5 & 34.9 & 67.4 & 253.8 \\
        \quad Supernova kicks, No fallback & 4.4 & 10.7 & 21.3 & 44.5 & 143.4 & 4.6 & 16.4 & 34.8 & 67.0 & 244.1 \\
        \textbf{Initial conditions} &&&&&&& \\
        \quad Singles, $f_{\rm bin} = 0.0$ & 4.2 & 8.9 & 17.3 & 29.1 & 43.0 & 3.9 & 11.8 & 22.7 & 40.1 & 83.7 \\
        \quad Initial mass function slope, $\alpha_{\rm IMF} = -1.9$ & 4.0 & 8.0 & 15.8 & 34.2 & 125.1 & 3.8 & 12.5 & 28.0 & 58.2 & 230.0 \\
        \quad Initial mass function slope, $\alpha_{\rm IMF} = -2.7$ & 5.1 & 14.2 & 28.5 & 60.1 & 157.4 & 5.5 & 20.4 & 40.6 & 74.6 & 249.0 \\
        \quad Orbital period slope, $\pi = 0$ & 4.3 & 9.9 & 18.9 & 36.1 & 122.4 & 4.3 & 14.9 & 31.2 & 60.5 & 237.4 \\
        \quad Orbital period slope, $\pi = -1$ & 4.5 & 11.8 & 25.1 & 58.7 & 157.5 & 4.8 & 17.9 & 37.9 & 72.1 & 237.7 \\
        \quad Initial upper orbital period limit, $P_{\rm 0, max} = 10^{3} \, {\rm d}$ & 4.6 & 11.7 & 23.9 & 58.6 & 153.0 & 4.9 & 18.4 & 38.7 & 73.6 & 250.7 \\
        \quad Mass ratio slope, $\kappa = 1$ & 4.4 & 10.8 & 21.9 & 48.3 & 151.4 & 4.6 & 16.7 & 35.3 & 67.8 & 229.9 \\
        \quad Mass ratio slope, $\kappa = -1$ & 4.4 & 10.7 & 20.8 & 40.9 & 125.1 & 4.5 & 15.6 & 32.5 & 62.7 & 241.5 \\
        \textbf{Metallicity} &&&&&&& \\
        \quad Metallicity, $\bar{Z} = 0.5 \, \bar{Z}_{\rm m11h}$ & 4.5 & 12.2 & 24.3 & 49.0 & 154.1 & 4.7 & 17.9 & 38.9 & 76.2 & 289.9 \\
        \quad Metallicity, $\bar{Z} = 0.2 \, \bar{Z}_{\rm m11h}$ & 4.6 & 13.3 & 26.7 & 52.4 & 148.1 & 4.9 & 18.9 & 41.8 & 83.9 & 352.7 \\
        \quad Metallicity, $\bar{Z} = 0.1 \, \bar{Z}_{\rm m11h}$ & 4.6 & 13.9 & 28.0 & 54.9 & 146.9 & 4.9 & 19.3 & 43.0 & 87.2 & 385.7 \\
        \quad Metallicity, $\bar{Z} = 0.05 \, \bar{Z}_{\rm m11h}$ & 4.6 & 14.3 & 28.9 & 55.8 & 146.3 & 4.9 & 19.5 & 43.6 & 91.1 & 492.3 \\
        \textbf{Galaxy settings} &&&&&&& \\
        \quad Velocity dispersion, $v_{\rm disp} = 0.5 \ {\rm km/s}$ & 4.4 & 10.8 & 21.4 & 44.6 & 143.5 & 1.9 & 5.8 & 12.1 & 26.2 & 181.2 \\
        \quad Velocity dispersion, $v_{\rm disp} = 5 \ {\rm km/s}$ & 4.4 & 10.8 & 21.4 & 44.6 & 143.5 & 12.5 & 43.7 & 89.0 & 169.4 & 537.4 \\
        \quad ChaNGa (r442) & 4.4 & 10.8 & 21.3 & 44.5 & 142.4 & 4.6 & 16.7 & 35.2 & 68.1 & 251.6 \\
    \end{tabular}
    \caption{The distributions for (a) the times at which SNe occur and (b) the distances at which they occur relative to their parent cluster, for each model variation in our simulations. Each column gives the value for a different percentile, which is listed in the heading.}
    \label{tab:percentiles}
\end{table*}

\end{document}